\RequirePackage[colorlinks, linkcolor=blue, urlcolor=blue, citecolor=blue]{hyperref}
\RequirePackage[table,xcdraw]{xcolor}
\documentclass[draft]{agujournal2019}
\usepackage{lineno}
\usepackage[inline]{trackchanges} 
\usepackage{soul}

\usepackage{amssymb}
\usepackage{amsthm}
\usepackage{amsmath}
\usepackage{setspace}
\usepackage{listings}
\usepackage{color}

\usepackage{multirow}

\usepackage{url}

\newcommand{\be}{\begin{equation}}
\newcommand{\ee}{\end{equation}}
\newcommand{\bes}{\begin{equation*}}
\newcommand{\ees}{\end{equation*}}
\newcommand{\ba}{\begin{eqnarray}}
\newcommand{\ea}{\end{eqnarray}}
\newcommand{\bas}{\begin{eqnarray*}}
\newcommand{\eas}{\end{eqnarray*}}

\newcommand{\bfu}{{\bf u}}

\newcommand{\bft}{{\boldsymbol{\tau}}}

\nolinenumbers
\draftfalse

\journalname{Journal of Advances in Modeling Earth Systems (JAMES)}

\begin{document}




\title{Implementation and Evaluation of a Machine Learned Mesoscale Eddy Parameterization into a Numerical Ocean Circulation Model}


\authors{Cheng Zhang\affil{1}, Pavel Perezhogin\affil{2}, Cem Gultekin\affil{2}, Alistair Adcroft\affil{1}, Carlos Fernandez-Granda\affil{2,3}, Laure Zanna\affil{2,3}}

\affiliation{1}{Program in Atmospheric and Oceanic Sciences, Princeton University, Princeton, NJ 08542, USA}
\affiliation{2}{Courant Institute of Mathematical Sciences, New York University, New York, NY 10012, USA}
\affiliation{3}{Center for Data Science, New York University, New York, NY 10011, USA}

\correspondingauthor{Cheng Zhang}{cheng.zhang@princeton.edu}

\begin{keypoints}
\item A stochastic-deep learning model is implemented in an ocean circulation model, MOM6
\item We evaluate the online performance of a stochastic-deep learning model as a subgrid parameterization
\item We identify certain limitations of the machine learned parameterization which otherwise has the potential to improve specific metrics.
\end{keypoints}

\begin{abstract}

We address the question of how to use a machine learned parameterization in a general circulation model, and assess its performance both computationally and physically.
We take one particular machine learned parameterization \cite{Guillaumin1&Zanna-JAMES21} and evaluate the online performance in a different model from which it was previously tested.
This parameterization is a deep convolutional network that predicts parameters for a stochastic model of subgrid momentum forcing by mesoscale eddies.
We treat the parameterization as we would a conventional parameterization once implemented in the numerical model.
This includes trying the parameterization in a different flow regime from that in which it was trained, at different spatial resolutions, and with other differences, all to test generalization.
We assess whether tuning is possible, which is a common practice in general circulation model development.
We find the parameterization, without modification or special treatment, to be stable and that the action of the parameterization to be diminishing as spatial resolution is refined.
We also find some limitations of the machine learning model in implementation:
1) tuning of the outputs from the parameterization at various depths is necessary;
2) the forcing near boundaries is not predicted as well as in the open ocean;
3) the cost of the parameterization is prohibitively high on CPUs.
We discuss these limitations, present some solutions to problems, and conclude that this particular ML parameterization does inject energy, and improve backscatter, as intended but it might need further refinement before we can use it in production mode in contemporary climate models.

\end{abstract}

\nolinenumbers

\section*{Plain Language Summary}

This paper discusses how machine learning can be used to make climate models more accurate.
Specifically, we import an existing machine learning model that predicts how small eddies (in the order of 10km to 100km) in the ocean affect larger currents.
We test this machine learning model in a different ocean circulation model than the one it was originally designed for, and found that it worked well.
However, we also found some limitations: the model works differently at different depths in the ocean, and it does not work as well near the coasts of the ocean.
We also found that the model takes a long time to run on normal computers.
Overall, we concluded that the model is promising, but more work is needed to make it work well in realistic situations.

\section{Introduction}\label{sec1}

The numerical global circulation models used for climate research solve the governing equations at a finite resolution and are unable to resolve all dynamical scales that influence climate.
The spatial resolution of global circulation models has been incrementally refined decade by decade, gradually resolving or admitting new processes. 
However, the closure problem of parameterizing the influence of unresolved subgrid processes will likely remain for many decades to come.
Historically, the development of subgrid parameterizations has required a synergy of theory, observations, and large-eddy simulations (LES), or even direct numerical simulations (DNS).
Many of these parametrizations have been developed by suggesting a mathematical operator which mimics the bulk effect of the subgrid processes on the large-scale flow properties \cite<e.g.,>{gent_parameterizing_1995,griffies_isoneutral_1998,anstey2017deformation,juricke2017stochastic}.  
To construct and then implement parameterizations, in production climate-simulation codes, has required teams of researchers to be funded, e.g. the Climate Process Teams \cite{legg_improving_2009,mackinnon_climate_2017}.
Despite tremendous progress in the development of such parameterizations, they continue to be a source of error in climate simulations \cite{hewitt2020resolving} and a source of uncertainty in climate projections \cite{zanna_uncertainty_2018}. 
Recently, there is growing interest in the use of machine learning to develop parameterizations directly from data, rather than building an ad-hoc mathematical operator for the bulk effect of the subgrid scales onto the large-scale \cite{krasnopolsky2010accurate, rasp2018deep, o2018using,Bolton2019, maulik2019subgrid, zanna2020data, beucler2021climate, Guillaumin1&Zanna-JAMES21,Ross-et-al2022}.
Many of these show significant skill in offline tests and online evaluation has been demonstrated in several cases, e.g. \citeA{rasp2018deep}, \citeA{brenowitz2018GRL}, \citeA{Guillaumin1&Zanna-JAMES21} and \citeA{yuval2021GRL}.
However, few machine learned (ML) parameterizations have been fully implemented in a general circulation model, nor evaluated for effectiveness in realistic simulations.
There are technical and practical hurdles that contribute to the current state of affairs, and we lay out and examine some of those issues in this study.

We set out to implement an ML parameterization in a conventional ocean circulation model, the Modular Ocean Model version 6 \cite<MOM6,>{adcroft2019gfdl}.
This study explores and documents the issues associated with implementing a pre-defined ML parameterization, as well as to further evaluate the ML parameterization beyond the assessment of the ML parameterization authors.
The ML parameterization we chose to implement and evaluate is that of \citeA{Guillaumin1&Zanna-JAMES21}, hereafter referred to as GZ21.
The ML parameterization takes the form of a stochastic-deep learning model and was designed to parameterize the upscale transfer of energy in the inverse cascade of mesoscale turbulence in the ocean.
This parameterization is of particular interest for models with eddy-permitting resolution that must account for specific physics inherent to mesoscale eddies.
Mesoscale eddies strengthen mean jet currents \cite{greatbatch2010transport} by upgradient momentum fluxes, and result in an inverse kinetic energy cascade \cite{scott2007spectral, kjellsson2017impact, balwada2022direct}.
General circulation models, at typical spatial resolutions, are missing a systematic energy exchange from subgrid to resolved scales.
Both properties underline the energizing effect of subgrid mesoscale eddies on the resolved flow.
Additionally, mesoscale eddies are responsible for large fraction of heat and salt transport \cite{delman2021global}, and thus failing to resolve or parameterize them results in significant biases in mean surface temperature and overturning circulation \cite{hewitt2020resolving}.
The deep learning model of GZ21, like the majority of other machine learning models, is developed in a high-level programming language Python.
MOM6, however, like most of large-scale scientific computation models, is written in a low-level programming language Fortran.
One approach to this language barrier is to "port" the code, translating from one language to another.
In this case, this would entail rewriting some machine learning libraries in Fortran.
While this solution works for some straightforward network architectures it is nevertheless time-consuming and not necessarily extensible.
When new and more complex deep learning architectures are invented, more porting would be needed.
There are some existing machine learning libraries in Fortran aiming to make this step easier, e.g., Neural Fortran \cite{curcic2019neuralFortran} and Fortran-Keras Bridge \cite<FKB,>{ott2020FKB}.
Such libraries are few in number and typically not up to date as their Python counterparts.
One other challenge faces ports to Fortran is that machine learning methods are computationally intensive and dedicated hardware devices are normally used for rapid computation (e.g. a graphics processing unit, GPU, or tensor processing unit, TPU).
Fortran is not widely used on such devices.
An alternative to porting code is coupling the Fortran codes and Python scripts.
There are several packages already available that facilitate interoperability between Fortran and python.
Recently, \citeA{partee2022SmartSim} describe using a turn-key package called {\it SmartSim} to implement a parameterization in a large scale ocean model at scale in a high-performance computing (HPC) environment.
{\it SmartSim} provides a client library that is compiled into the Fortran model, with put/get/run semantics to communicate with a distributed database capable of handling machine learning and data sciences services, and an infrastructure library capable of executing simulation, visualization, and analysis workloads on a variety of HPC platforms.
Forpy (\url{https://github.com/ylikx/forpy}) is a light-weight alternative to {\it SmartSim}. It is a Fortran module that enables the use of Python features in Fortran, and has been utilized for machine learning parameterizations in atmospheric global climate models \cite{liu2021gas}, atmospheric gas-phase chemistry models \cite{Espinosa_etal-GRL2022}, and computational fluid dynamics turbulence models \cite{beck2021CFDparameter}.
As a Fortran module Forpy can be compiled on any computer or clusters that have Python and Fortran installed, and all locally available Python libraries are then accessible from the Fortran application.
The demonstrated simplicity, compatibility, and versatility of this light-weight package led us to use Forpy for this study.

The paper is organized as follows.
In Section {\ref{sec2}}, the ocean model MOM6, the stochastic-deep learning model and their bridge are introduced.
To demonstrate the online performance of the system, Section {\ref{sec3}} presents an idealized case: a wind-driven double gyre.
In Section {\ref{sec4}}, some potential issues when applying the deep learning model to a numerical ocean model are highlighted, along with some potential solutions. Conclusions and ideas for future study are discussed in Section \ref{sec5}. 

\section{Methods}\label{sec2}

In this section, we describe the framework consisting of the numerical ocean model MOM6 and the stochastic-deep learning model for predicting mesoscale ocean dynamics.
The numerical ocean model MOM6 that we use for the ML-based parameterizations is described in Section \ref{sec2.1}.
In Section \ref{sec2.2}, we recapitulate the methodology of the CNN model in GZ21 and describe the inference stage in MOM6. Finally, in Section \ref{sec2.2} we provide the workflow and techniques of online implementation. 

\subsection{Ocean model description} \label{sec2.1}

The numerical model employed in this study is the Modular Ocean Model version 6 \cite<MOM6,>{adcroft2019gfdl}, a solver for ocean circulation written in Fortran used for ocean climate simulations.
We use the model in an adiabatic limit with no buoyancy forcing which simplifies the equations of motion to the stacked shallow water equations.
The layer momentum equations given in vector-invariant form are 
\be
\frac{\partial \bfu_k}{\partial t} + \frac{f+\zeta_k}{h_k} \hat{\bf{z}} \times h_k \bfu_k + \nabla K_k + \nabla M_k
=  \mathbf{F}_k
\label{eq2.1_1}
\ee
where $\bfu_k$ is the horizontal component of velocity, $h_k$ is the layer thickness, $f$ is Coriolis parameter, $\zeta_k$ is the vertical component of the relative vorticity, $K_k=(1/2) \bfu_k\cdot\bfu_k$ is the kinetic energy per unit mass in horizontal, $M_k$ is the Montgomery potential (defined in \ref{appa}) and $\mathbf{F}_k$ represents the accelerations due to the divergence of stresses including the lateral parameterizations that are not inferred from ML-based models, $\hat{\bf{z}}$ is the unit vector pointing in the upward vertical direction, $k$ is the vertical layer index with $k=1$ at the top, $\nabla$ is the horizontal gradient and $\nabla\cdot$ is the horizontal divergence.
The governing equations are discretized on C-type staggered rectangular grid with finite volume method, and the advection operator is energy-conserving (in our setup).
We take the limit of an adiabatic fluid with a single constituent so that the governing equations simplify to the stacked shallow water equations.
In the adiabatic limit used here, vertical advection of all quantities is represented by the Lagrangian motion of the model layers. \ref{appa} contains a full description of governing equations.
 
The oceanic mesoscale turbulence that interests us involves an upscale cascade of energy from small (unresolved) scales, so a finite resolution model needs a subgrid momentum forcing on account for nonlinear interactions with the unresolved eddies.
This subgrid momentum forcing can be diagnosed by
\be
\mathbf{S}_k = \left ( \begin{array}{c} S_{kx} \\ S_{ky} \end{array} \right )
= (\bar{\bfu}_k\cdot \bar{\nabla})\bar{\bfu}_k - \overline{ (\bfu_k\cdot \nabla)\bfu_k }
\label{eq2.1_2}
\ee
where the overbar is the horizontal filtering and coarse graining, and we make use of the identity $\bfu_k \cdot \nabla \bfu_k = \zeta_k \hat{\bf{z}} \times \bfu_k + \nabla K_k$.
Operator $\bar{\nabla}$ stands for the discretization of $\nabla$ on coarse grid.
In equation \eqref{eq2.1_2} we consider nonlinear interactions only due to momentum advection operator, and neglect subgrid forcing from nonlinearity in vertical transport, varying Coriolis parameter and subgrid dissipation.
The coarse resolution flow evolves according to the equation
\be
 \frac{\partial \bar{\bfu}_k}{\partial t} + \frac{f+\bar{\zeta}_k}{\bar{h}_k} \hat{\bf{z}} \times \bar{h}_k \bar{\bfu}_k + \nabla \bar{K}_k + \nabla \bar{M}_k
=  \bar{\mathbf{F}}_k + \mathbf{S}_k
\label{eq2.1_3}
\ee
and the parameterization of subgrid forcing $\mathbf{S}_k$ is function of $\bar{\mathbf{u}}_k$. GZ21 developed a deep learning model to predict the statistical moments of $\mathbf{S}_k$ that can be used in a stochastic parameterization in the coarse resolution model.

\subsection{Stochastic-deep learning model}\label{sec2.2}

The stochastic-deep learning model of GZ21 is a Fully Convolutional Neural Network (CNN) with eight convolutional layers, where the kernel size of the first two layers is $5 \times 5$ and the kernel size of the rest layers is $3 \times 3$.
Each of the convolutional layers has $128$, $64$, $32$, $32$, $32$, $32$, $32$ and $4$ filters, respectively.
The ReLU activation function is used for hidden layers and no padding is used in the convolutional layers.
The CNN architecture results in the stencil size of $21 \times 21$ for predicting the forcing on a single grid point.
In contrast to a deterministic parameterization for predicting the momentum forcing, the CNN models the mean and standard deviation of a Gaussian probability distribution of the subgrid momentum forcing.
The mean square error (MSE) loss function is replaced by a full negative Gaussian log-likelihood of the forcing.
The CNN was trained and validated with surface velocity data from the high-resolution coupled climate model CM2.6 \cite{Griffies_etal-JC15} which has a nominal resolution in the ocean model of $1/10 ^\circ$.
This resolution is considered sufficiently fine to resolve eddies in the tropics and mid-latitudes of the global ocean \cite{Hallberg-2013}.
The simulated ocean surface velocity fields from four subdomains are selected as representative of different dynamical regimes. More details about the model, training, and data can be found in Section 2 of GZ21.


The parameterization is evaluated at each time step in the ocean model using the velocity components as the inputs to the CNN model which returns the mean and standard deviation of a Gaussian probability distribution of the subgrid momentum forcing.
The stochastic subgrid momentum forcing is then generated by 
\be
S_{C,i,j}=S_{C,i,j}^{(mean)}+\epsilon_{C,i,j} \cdot S_{C,i,j}^{(std)}; \hspace{0.2in}  C = x,y;\hspace{0.1in} i=1,...,m;\hspace{0.1in} j=1,...,n  \label{eq2.2_1}
\ee
where $i$ and $j$ are the ocean model spatial indices, $C$ indicates the component of momentum forcing (zonal "$x$" or meridional "$y$"), and $\epsilon_{C,i,j}$ are random 2D fields sampled from the standard normal distribution, independent for each grid cell, zonal/meridional component, vertical layer, and time step.


\subsection{Online implementation of CNN with MOM6} \label{sec2.3}

The MOM6 ocean circulation model is exclusively written in Fortran, while the stochastic-deep learning model was developed with the machine learning package PyTorch (many deep learning practitioners favor developing machine learning models in Python, and other recent languages, since machine learning tools are readily available).
Computer language interoperability is a technical barrier that we overcome here by using the package, Forpy.
Python is an interpreted language, while Fortran is compiled.
A system call from Fortran to run a python script would require booting the Python interpreter each time the Python functions are needed.
Most approaches to embed Python in a compiled language therefore use the C-language API to call the Python run-time library directly.
This embedding method requires writing an intermediate software layer for all the possible combinations of arguments to functions and so is not readily extensible.
Forpy is a Fortran module that provides that interface to the Python library, and appears to avoid any significant overheads.
The module conveniently allows data to be passed from the calling Fortran code to functions in the python script.
In addition, Forpy allows us to use any Python libraries from Fortran, is independent of the computing environment, and does not require installing any other software that needs system privileges.
Another benefit of using the Python language for inference in MOM6 is that the network can utilize the graphical processing units (GPUs) even though MOM6 exclusively executes on central processing units (CPUs).

\begin{figure}[htbp]
\centering
\includegraphics[width=1.0\textwidth]{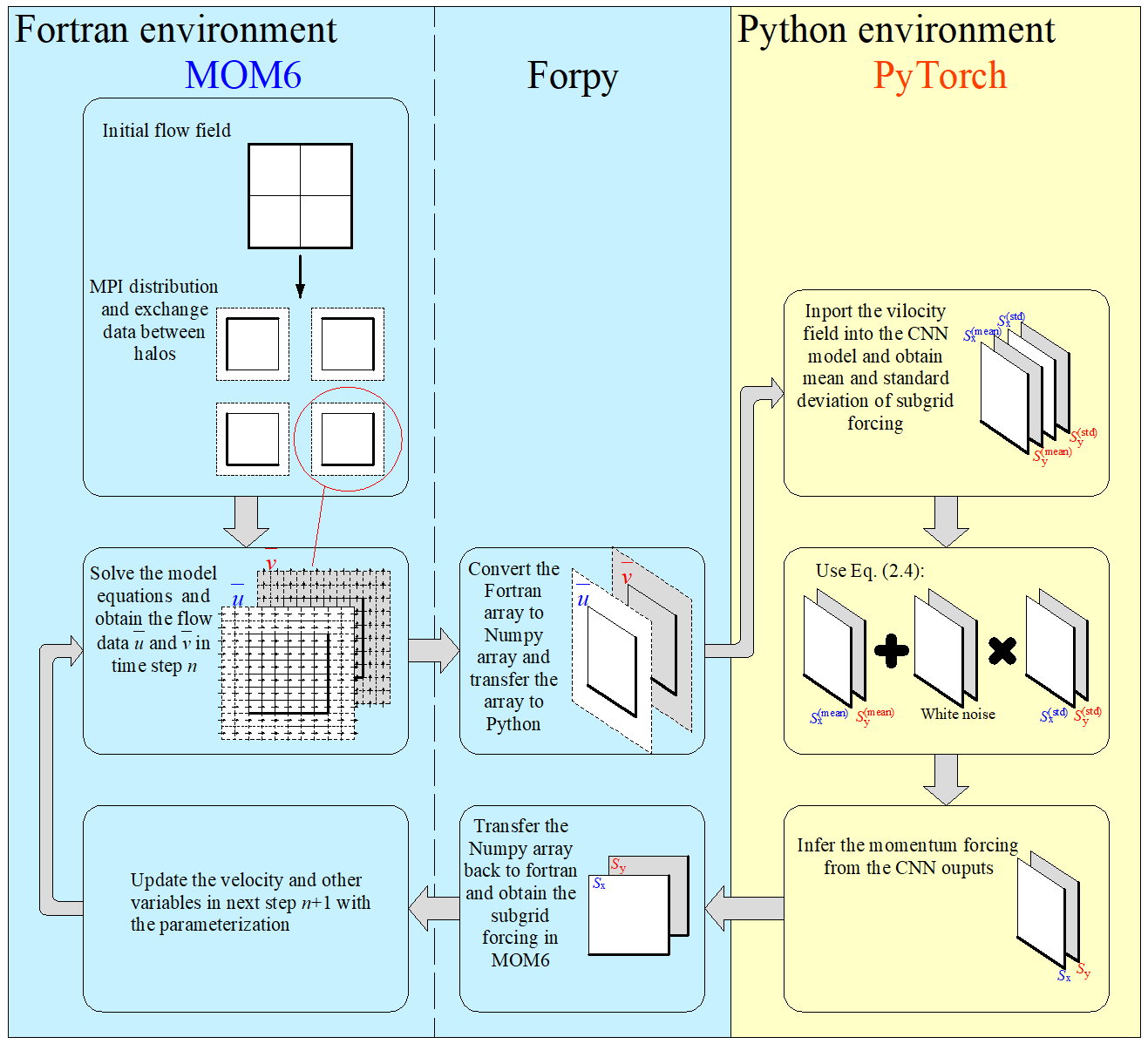}
\caption{Flowchart of the coupling system between MOM6 and the CNN model, using Forpy.}
\label{fig2.3_1}
\end{figure}

In the hybrid model consisting of MOM6 and the CNN parameterization, the velocity field is computed by MOM6 first using all available terms in equation \eqref{eq2.1_3}.
The Fortran array of the velocity is then wrapped up as a Numpy array by Forpy and transferred to Python as the input of the CNN model.
The CNN returns the moments in equation \eqref{eq2.2_1} and then random numbers are generated to yield the momentum forcing in a Numpy array. 
The momentum forcing is then transferred back to Fortran and Forpy provides an interface to read the data from the Numpy array in Fortran.
The momentum is then updated with this stochastic forcing and the hybrid model continues as would the conventional MOM6.
Figure \ref{fig2.3_1} illustrates the flowchart of the whole hybrid model.

Not only does the language barrier complicate the implementation of a CNN into an ocean model, but it also complicates how computations are distributed among computing resources.
The MOM6 ocean model utilizes data parallelism, where the computational domain is divided into subdomains with overlapping halo regions which are kept in-sync as needed by communication between adjacent processors using the MPI communications libraries.
In the conventional MOM6 model, the width of the halo region is determined by the stencil of numerical discretization and is typically on the order of $3$ or $4$ cells.
A computation involving spatial stencils generally needs to be preceded or followed by a halo synchronization (MPI exchange).
Optimal scaling of MOM6 is obtained when the costs of communication, additional memory, and extra computation, associated with the halos are minimized.
On contemporary platforms this typically leads to using the number of cores such that the width of halo is less than a quarter of the sub-domain area belonging to each core.
The CNN has a stencil of $21\times21$ cells which is far wider than any discretized terms in MOM6 and which requires expanding the width of the halos to $10$, and sometimes violating the less-than-quarter rule.
We discuss this further in Section \ref{sec4.4}

For the treatment of land, wherever the CNN parameterization would return momentum forcing on land (dry points), the velocity and forcing are set to zero.

\section{Online performance: wind-driven double gyre} \label{sec3}

GZ21 evaluated the CNN parameterization in a barotropic model and show good online performance.
Here, we test the online performance of the parameterizations in a baroclinic model, applying the closure in the ocean interior for which it was not trained.
In this paper, we focus on different metrics from those used in the network training, and evaluate the parameterization from the perspective of model large scale solution and not the details of the processes being parameterized.
We examine the effect of spatial resolution and tuning, in which the parameterization is attenuated or amplified.
We also make qualitative comparisons between parameterized coarse grid results and fine grid results.
It should be noted that in this work, the term "online" refers to the process of inferring from a trained deep learning model rather than the process of continuously updating a deep learning model as simulations progress, which is referred to as "online learning."

\subsection{Case setup} \label{sec3.1}

\begin{figure}[htbp]
\centering
\includegraphics[width=0.5\textwidth]{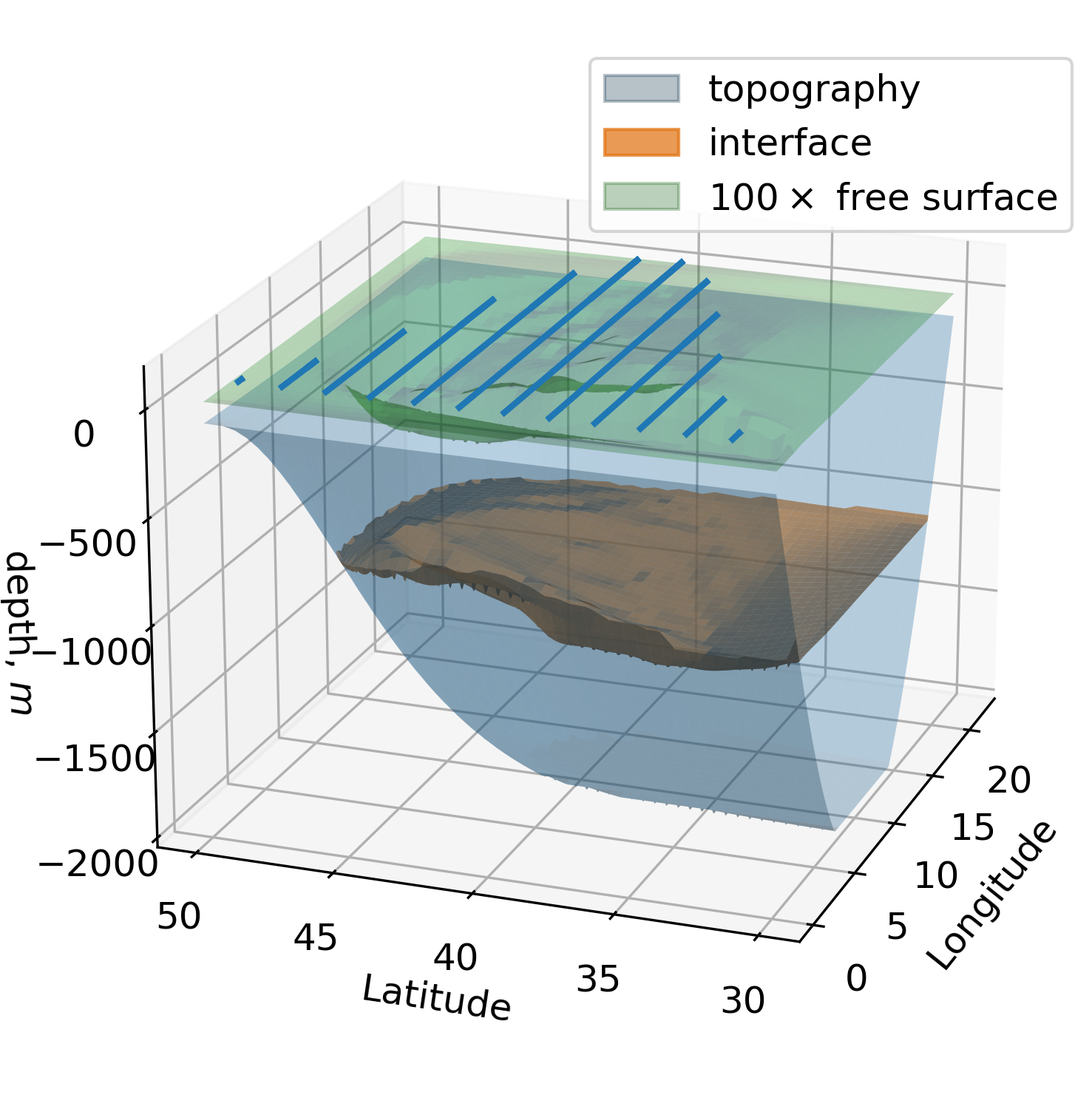}
\caption{Sketch of the wind-driven double gyre configuration: free surface (green), and layer interface (brown) are averaged over five years and bowl-shaped topography (blue/grey);. Blue lines indicate the strength and direction of imposed fixed zonal wind-stress.}
\label{fig3_1}
\end{figure}

The ocean model is configured to simulate a wind-driven double gyre in a bowl-shaped basin \cite{hallberg_rhines_2000} and a vertical wall at the southern boundary (Figure \ref{fig3_1}). The coordinate system is spherical, with computational domain ranging from $0$ degree to $22$ degree in longitude and from $30$ degree to $50$ degree in latitude.
Coriolis parameter is given by $f=2 \Omega \sin(\phi)$, where $\Omega=7.2921 \cdot 10^{-5} s^{-1}$ is planetary rotation rate and latitude $\phi$.
Although we use primitive equation ocean model, in this configuration the governing equations are simplified to two-layer shallow water model without thermodynamics (no computations involving equation of state, temperature and salinity).
The maximum depth is $2000$m and an interface between layers is initially located at the depth of $1000$m (at rest).
Let $h_1$ and $h_2$ are local fluid layer thickness, upper and lower, respectively.
The density of the upper layer is $\rho_1 = 1035 kg/m^3$, and lower layer is $\rho_2 = 1036.035 kg/m^3$, and corresponding reduced gravity for the interior interface is $g'=g (\rho_2-\rho_1)/\rho_1 = 0.0098 m/s^2$, where $g=9.8 m/s^2$.
The Rossby deformation radius is $R_d=c/f$ decreases from $30 km$ in the south to $15km$ in the north (approximate), where $c=\sqrt{g' \frac{h_1 h_2}{h_1+h_2}}$ is the gravity wave speed of the baroclinic mode \cite{gill1982atmosphere}.
The flow is forced by a constant (in time) surface wind stress $\tau_x$ and varies latitudinally with a maximum at center latitude ($\phi=40$) and zero stress at borders ($\phi=30, 50$):
\begin{equation}
    \tau_x = \tau_0 \left[ 1-\cos\left(2\pi \cdot \frac{\phi-30}{20}\right) \right] \geq 0
    \label{eq3.1_1}
\end{equation}
$\tau_0 = 0.1 N/m^2$.
The simulations last $10$ years and are initialized from rest.
The circulation and mesoscale turbulence reaches statistical equilibrium after about $5$ years.
The full specification of parameters is given in Zenodo\cite{doublegyre_setup}.
For the turbulence model we use a biharmonic friction with a Smagorinsky eddy viscosity following \citeA{griffies2000biharmonic}, where the details are in \ref{appa}.
Scale selective friction is required to remove small-scale numerical noise and stabilize the computations and is applied in both reference and parameterized simulations.
The Smagorinsky constant in all experiments here is $C_S=0.06$.
We vary the spatial grid size and time step (see Table \ref{table_resolution}) in these experiments.

\begin{table}[h!]
\centering
\begin{tabular}{  l | c | c  }
 Experiment & R4 & R32 \\ 
 \hline
 Grid spacing, $degree$ & $1/4$ & $1/32$ \\  
 Grid spacing, $km$ (approx.) & $24$ & $3$ \\
 Time step, $min$ &  $18$    & $2.25$ \\
 Cell count (Lon. $\times$ Lat.) & $88\times80$ & $704\times640$
\end{tabular}
\caption{Summary of the spatial and temporal resolution of the reference simulations used.}
\label{table_resolution}
\end{table}

Most evaluations we present will be in a model with $1/4^\circ$ horizontal resolution, hereafter referred to as R4. R4 is "eddy permitting" in that it exhibits mesoscale variability that contribute to variability of the separating boundary current.  

For the purposes of evaluating the CNN parameterization in R4, a $1/32^\circ$ model is run to obtain a "truth" run (hereafter referred to as R32). R32 is fine enough to resolve some of the mesoscale cascade. Note that R32 is also finer than the training data from the global model used to construct the CNN parameterization.

Figure \ref{fig3_2.1} shows the snapshots of the upper layer relative vorticity (normalized by the planetary vorticity) (a and b) and kinetic energy (KE, c and d), at the end of the run, and the five-year averaged sea surface height (SSH, e and f), for coarse resolution model, R4, (a, c and e) and fine resolution model, R32, (b, d and f).
The fine resolution model generates more energetic flow and finer-scale eddies.
The time-mean flow, indicated by the time-mean sea-surface displacement, of R4 has a double gyre, but fails to simulate well the boundary current extension separating the gyres (see region around $(5^\circ N,38^\circ E)$).
In this section, we will focus on the performance of the stochastic parameterization in improving the boundary current and the under-energized flow for coarse grid models.

\subsection{Results without tuning} \label{sec3.2}

The stochastic parameterization is implemented in R4, applied equally in both layers without tuning.
To take advantage of the stochastic nature of the parameterization we run a $50$ member ensemble with different random seeds.
The models are run for the same duration as R32 and R4 to permit a direct comparison between runs at the same model time since rest.
Examining and averaging an ensemble at the same model time avoids aliasing any systematic drift even though we did not find any significant long term trends.

\begin{figure}[htbp]
\centering
\includegraphics[width=1.0\textwidth]{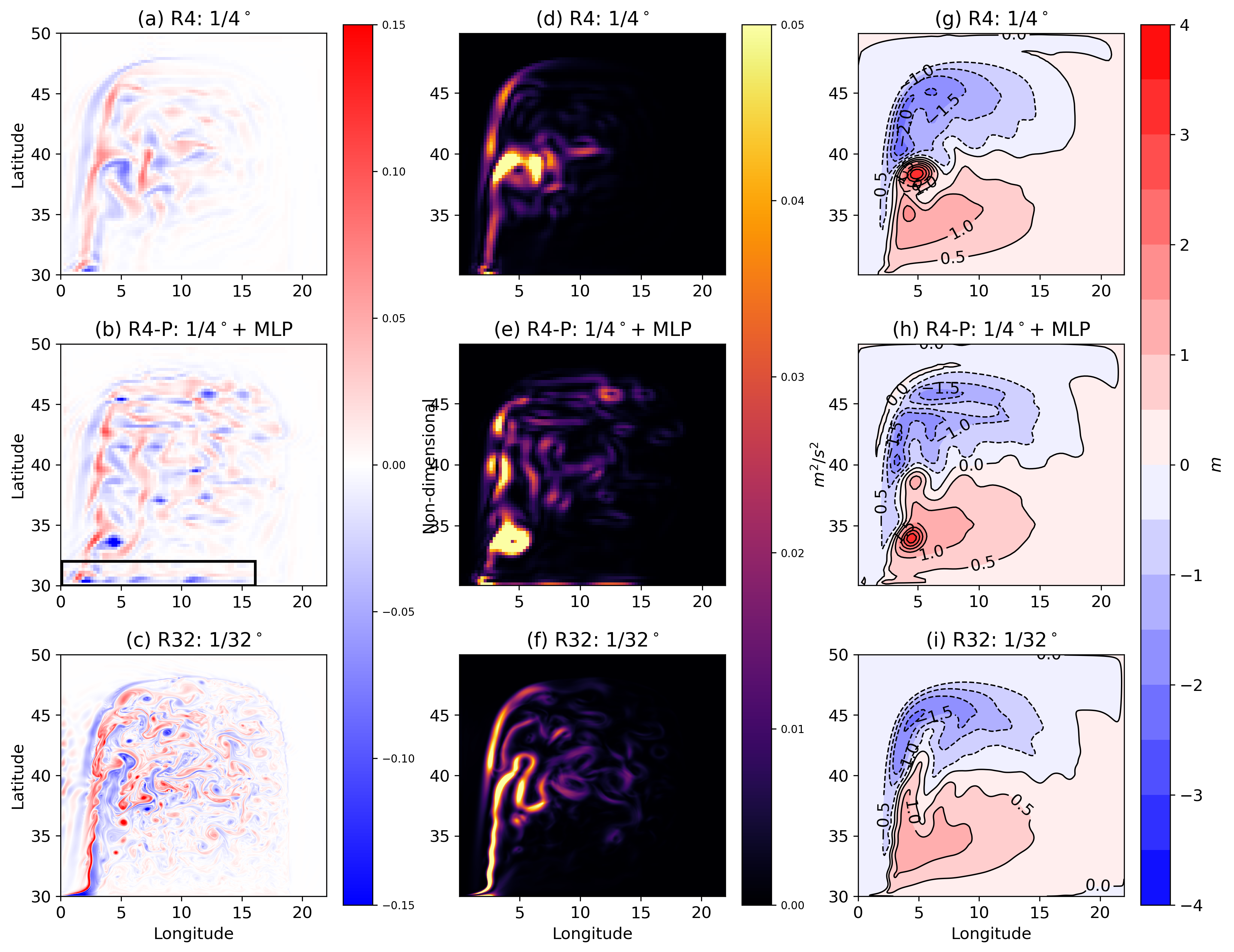}
\caption{Snapshots of the upper layer relative vorticity (normalized by the planetary vorticity, a-c) and KE in [m$^2$/s$^2$](d-f), and five-year averaged SSH in [m] (g-i), for coarse resolution model R4 (top row; a, d and g), coarse resolution model with ML parameterizations R4-P (middle row; b, e and h) and fine resolution model R32 (bottom row; c, f and i).}
\label{fig3_2.1}
\end{figure}

We use snapshots of upper layer relative vorticity and KE, shown in Figures \ref{fig3_2.1}(b) and \ref{fig3_2.1}(e) respectively, from the end of the one ensemble run, to present a qualitative assessment of the effect of the parameterization.
We illustrate by showing only one of the ensemble members, but the other ensemble members produce similar statistics.
Further details about the similarity of ensemble members are given in the Supplementary Information.
The subgrid momentum forcing from the CNN model energizes the flow and introduces some small-scale eddies, and they are perhaps more comparable to the eddies in R32 (Figure \ref{fig3_2.1}.b).
Two striking features can be observed from the vorticity and KE maps.
First, there is longitudinal stretching of some eddy features.
It is possible that this is due to a statistical bias in the structure of eddies in the training data.
Second, there are structures or artifacts on the southern boundary (highlighted by the black box, near a vertical wall) and western and northern boundaries where the topography is shallow.
On the southern wall in both the vorticity and KE maps, for all members shown, an unrealistic zonal strutcture is apparent.
We will discuss the boundary condition problem in more details in Section \ref{sec4.3}.
Figure \ref{fig3_2.1}(h) shows the SSH averaged over the last five years, for the same ensemble member.
Randomness from equation \eqref{eq2.2_1} leads to the different SSH patterns for each realization, especially in the region that we focus on (the separated boundary current).
Broadly speaking, the patterns of SSH appear to be improved by the parameterization and more similar to the pattern of the fine resolution model (Figure \ref{fig3_2.1}.f).

\begin{figure}[htbp]
\centering
\includegraphics[width=1.0\textwidth]{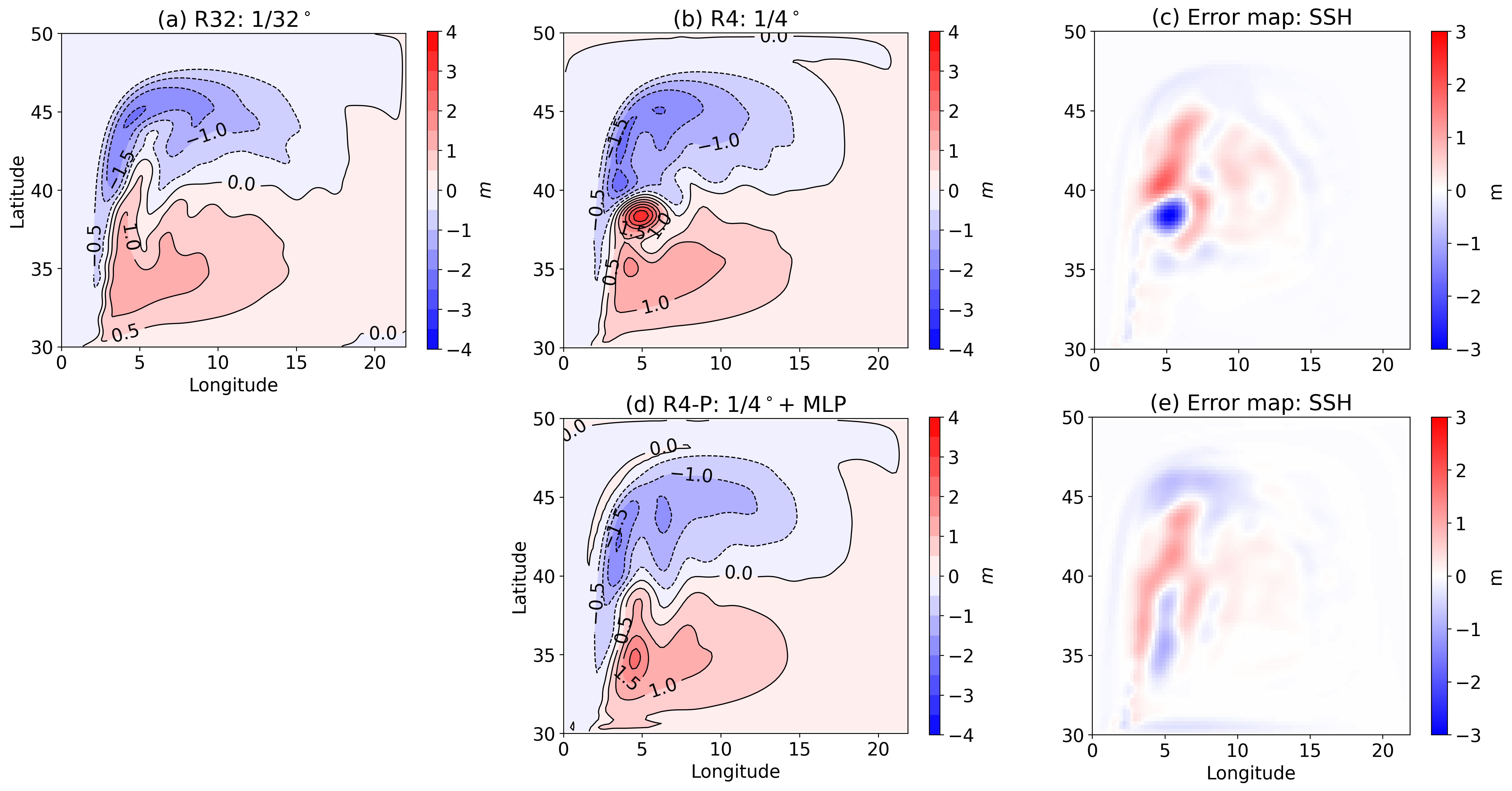}
\caption{Comparison of five-year averaged SSH between the coarse resolution model with (R4-P) and without (R4) the subgrid parameterization and the target fine resolution model (R32). The error maps (c and d) are obtained by subtracting low-resolution SSH (with or without parameterization) from coarse-grained high-resolution SSH. 
The R4-P SSH (d) is averaged from $50$ ensemble members. MLP is short for the ML parameterization.}
\label{fig3_2.5}
\end{figure}

\begin{figure}[htbp]
\centering
\includegraphics[width=0.495\textwidth]{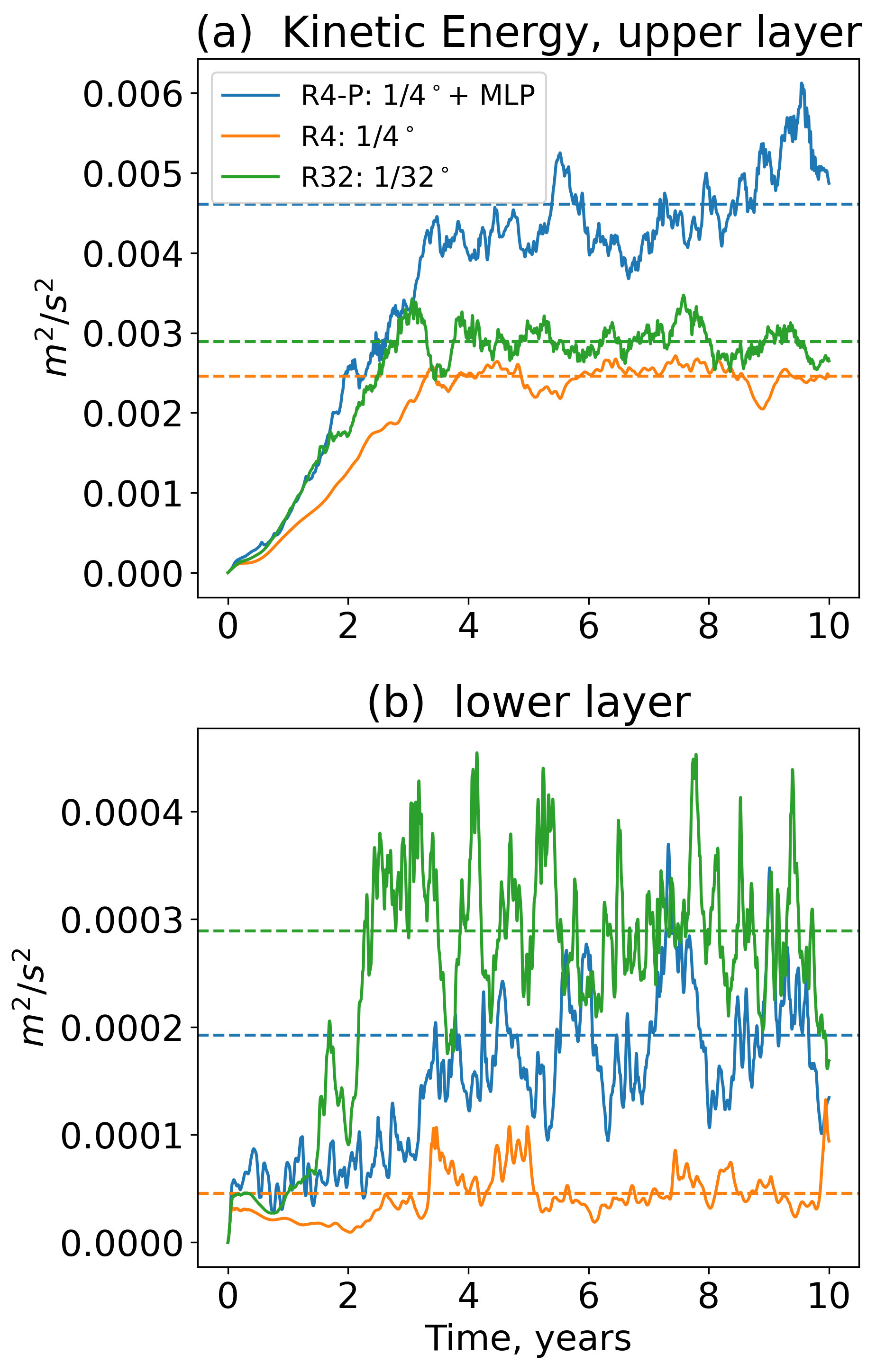}
\includegraphics[width=0.495\textwidth]{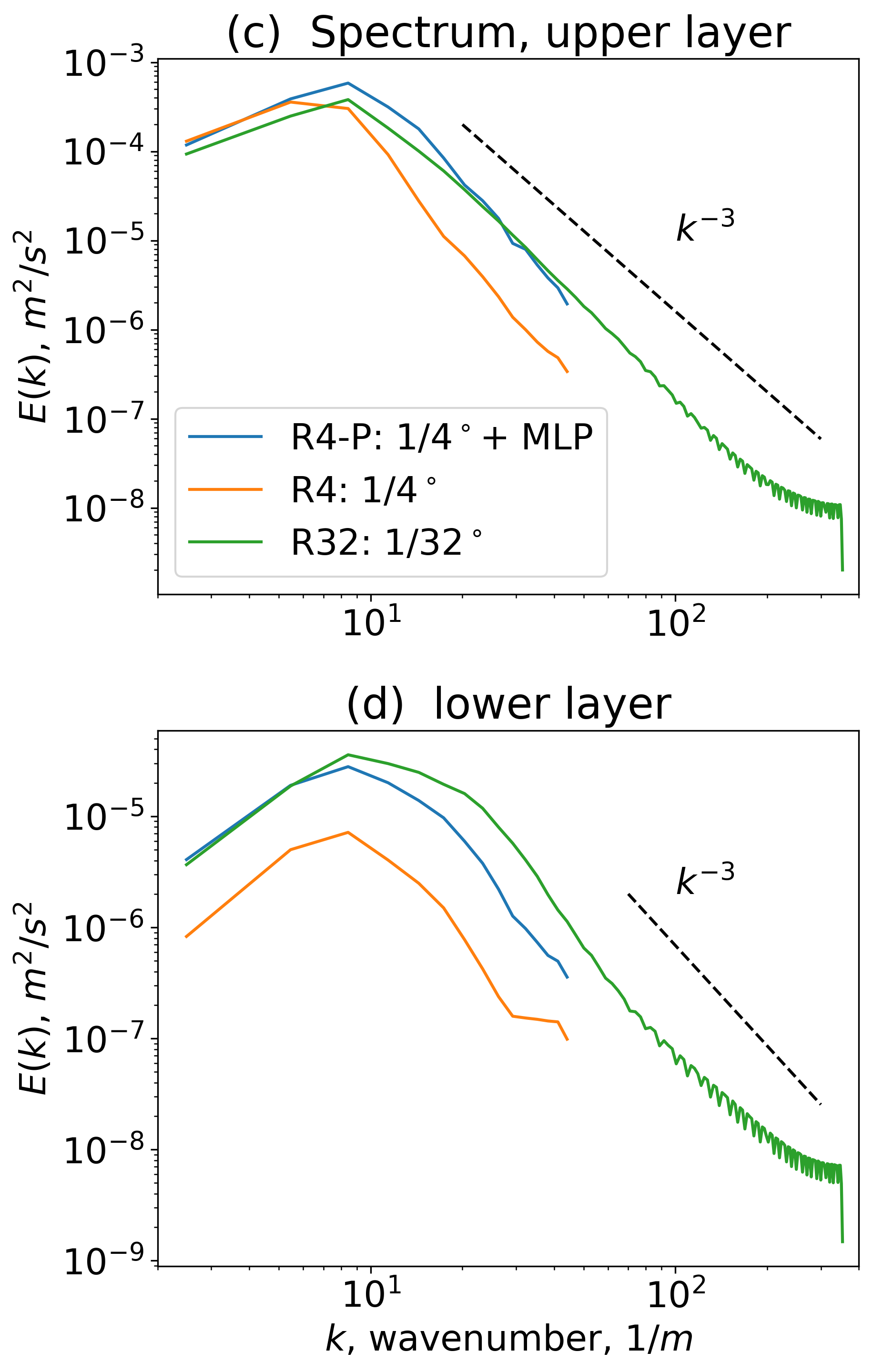}
\caption{Comparison of KE time series (a and b) and spectra (c and d) for the flow upper layer (top row) and the lower layer (bottom row) between the coarse resolution model R4 (orange), fine resolution R32 (green) and the coarse resolution model with ML parameterizations R4-P (blue). The dashed lines in (a and b) are mean values of KE over the last 5 years. The dashed lines in (c and d) are the spectral slope of kinetic energy spectrum corresponding to inertial interval of enstrophy. MLP is short for the ML parameterization.}
\label{fig3_2.6}
\end{figure}

To more quantitatively assess the impact of the subgrid parameterization, we use two metrics, errors in the five-year averaged sea surface height, and change in the kinetic energy spectra.
The metrics used when training the CNN model's offline accuracy in GZ21 are to minimize the statistical moments of the momentum forcing.
For individual realizations, a metric based on the local subgrid forcing is not meaningful.
Instead, we use metrics more amendable to model evaluation that uses the model state.
In Figure \ref{fig3_2.5}(a-c), we compare the five-year averaged SSH between R4 and the fine resolution R32. 
To make a fair comparison between the results from different resolutions, both R4 and R32 SSH are first filtered using a Gaussian kernel with the window size of $1^\circ$, and then the results of fine resolution R32 are coarsened to the grid of coarse resolution R4.
The error map shows that the largest errors appear around the region of the separated boundary current near $(5^\circ N,38^\circ E)$.
The CNN parameterization in the coarse model (hereafter R4-P) reduces the local error of the ensemble averaged SSH (Figure \ref{fig3_2.5} a,d and e).
The root mean square error (RMSE) of R4 SSH (relative to R32 SSH) is $0.2780$m and the RMSE of R4-P SSH is $0.2202$m.
The KE time series and spectra are compared between R4, R4-P and R32, in Figure \ref{fig3_2.6}.
The coarse resolution model R4 has less energetic flow then R32.

\subsection{Applying the CNN parameterization at different spatial resolutions} \label{sec3.3}

In the CNN training procedure, the velocity from the fine resolution CM2.6 $1/10 ^\circ$ ocean grid was used to generate momentum forcing on the coarse resolution $1/4 ^\circ$ grid of the CM2.5 model.
As a result, the CNN might be considered "optimized" for the R4 resolution for the double gyre tests above.
Parameterizations used in realistic ocean circulation models will likely be deployed at a range of spatial resolutions and even need to accommodate variable spatial resolutions within one model.

\begin{figure}[htbp]
\centering
\includegraphics[width=1.0\textwidth]{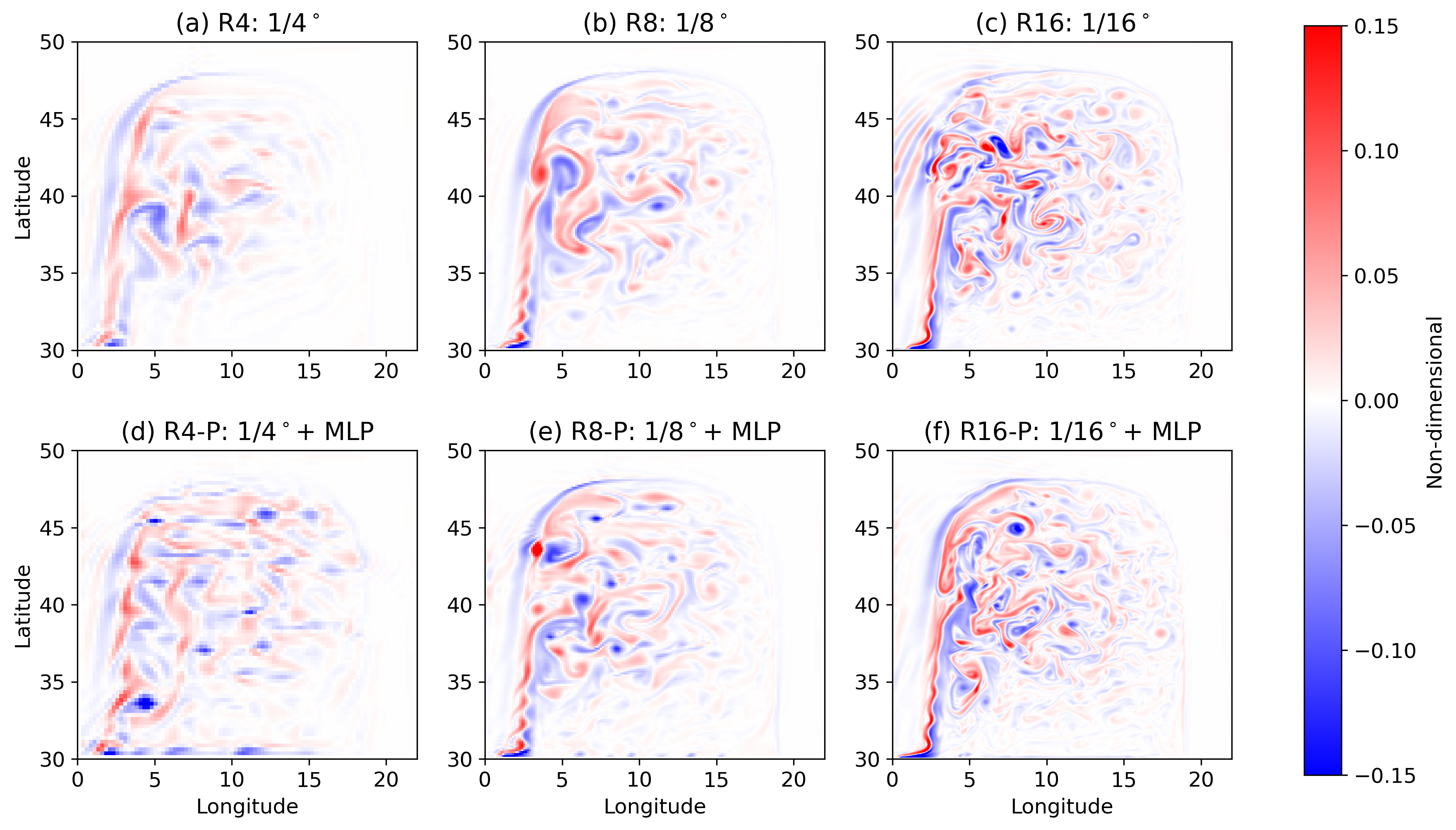}
\caption{Snapshots of upper layer relative vorticity (normalized by the planetary vorticity) without any subgrid parameterizations (a-c) or with the ML parameterization (d-f) in simulations with 3 different horizontal resolutions. The grid sizes of the simulations are $1/4 ^\circ$ (R4, a and d), $1/8 ^\circ$ (R8, b and e) and $1/16 ^\circ$ (R16, c and f).  MLP is short for the ML parameterization.}
\label{fig3.3_1}
\end{figure}

\begin{figure}[htbp]
\centering
\includegraphics[width=0.495\textwidth]{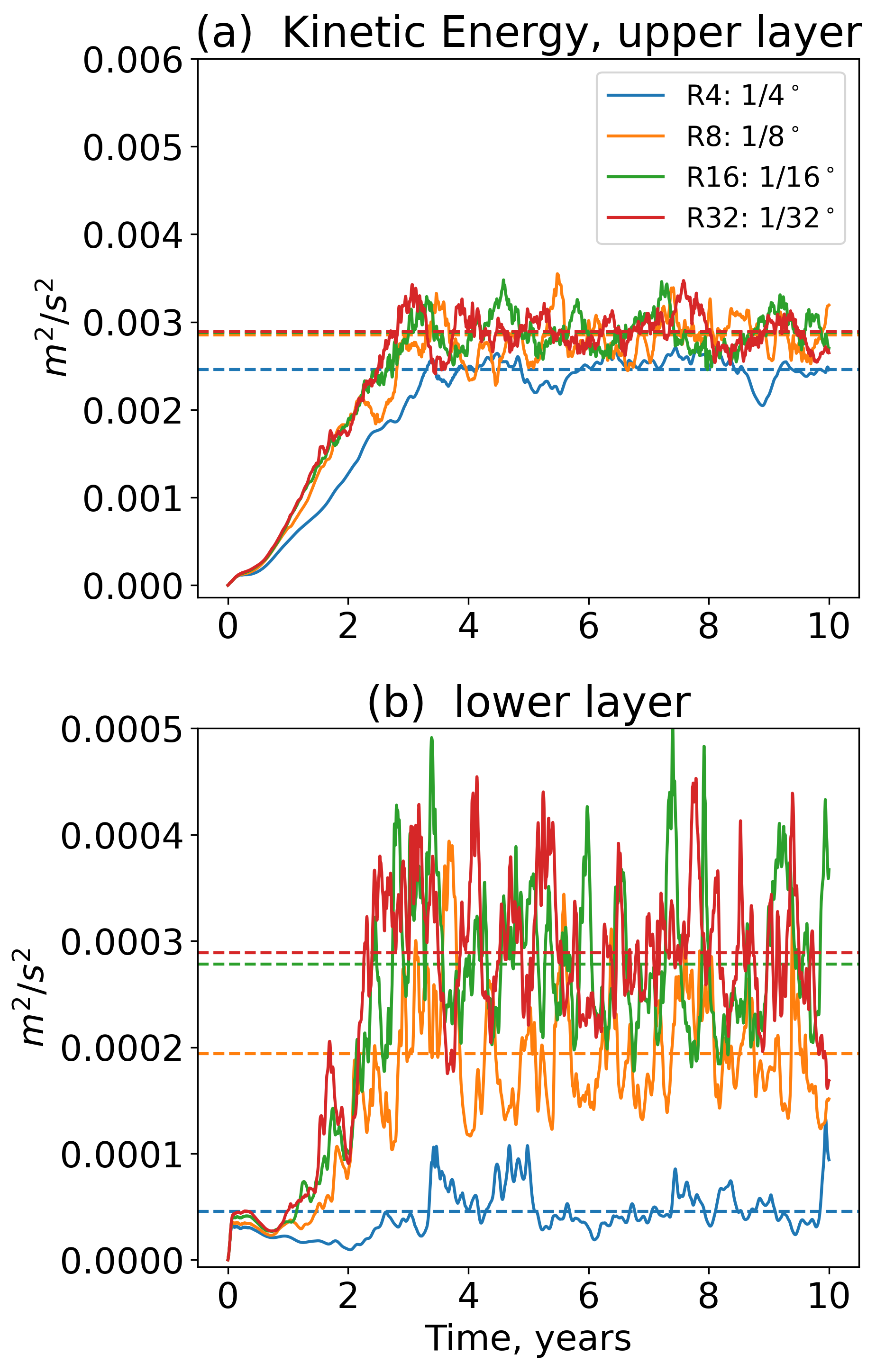}
\includegraphics[width=0.495\textwidth]{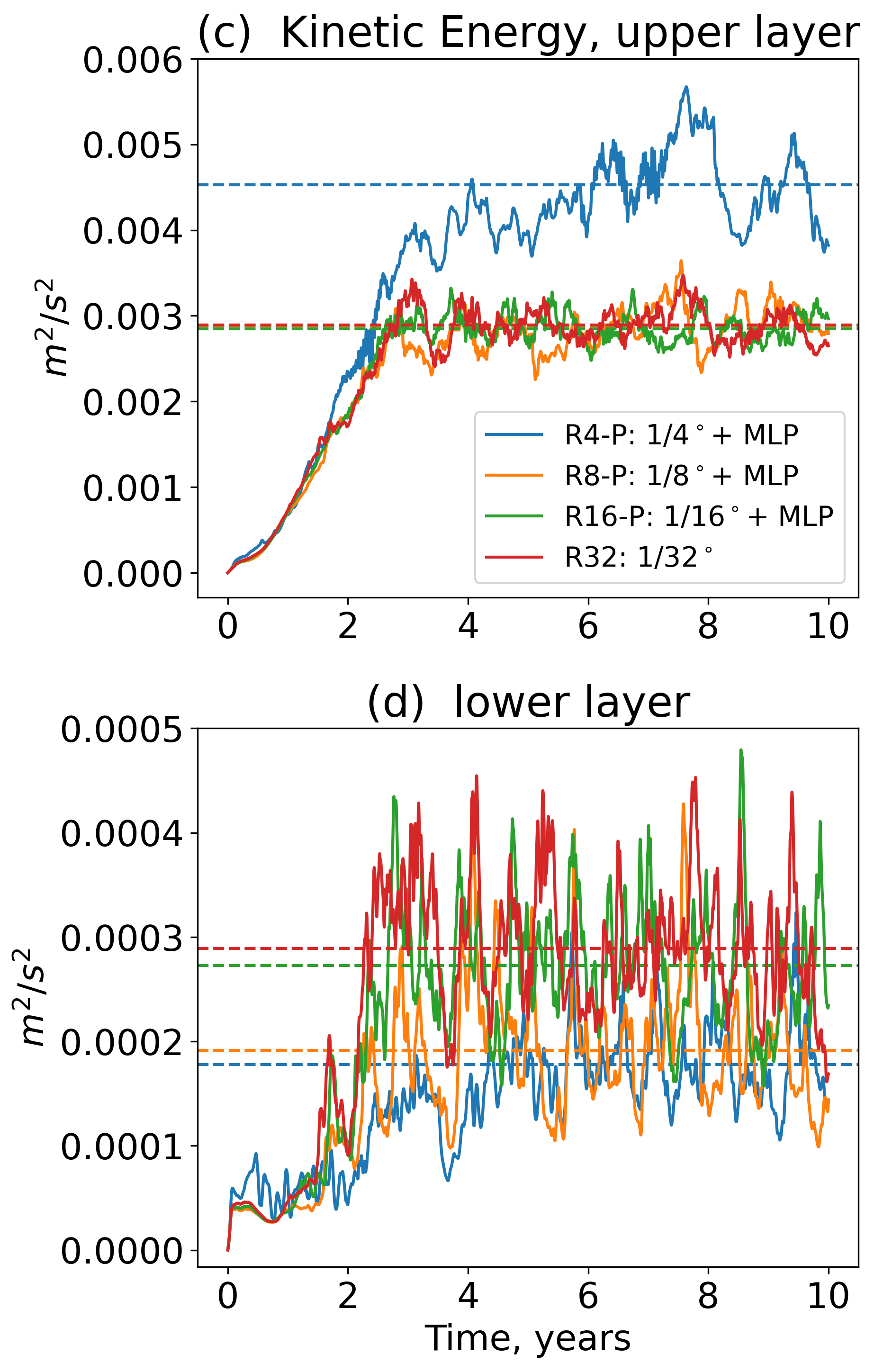}
\caption{Kinetic energy time series comparison, across different horizontal resolutions, of coarse resolution simulations against the finest (truth) 1/32$^2$ resolution (red): (a,b) coarse resolution simulations without subgrid parameterizations for the upper and lower layer; (c,d): coarse resolution simulations with subgrid ML parameterizations.  The dashed lines are mean values of KE over the last 5 years.  MLP is short for the ML parameterization.}
\label{fig3.3_2}
\end{figure}

\begin{figure}[htbp]
\centering
\includegraphics[width=0.495\textwidth]{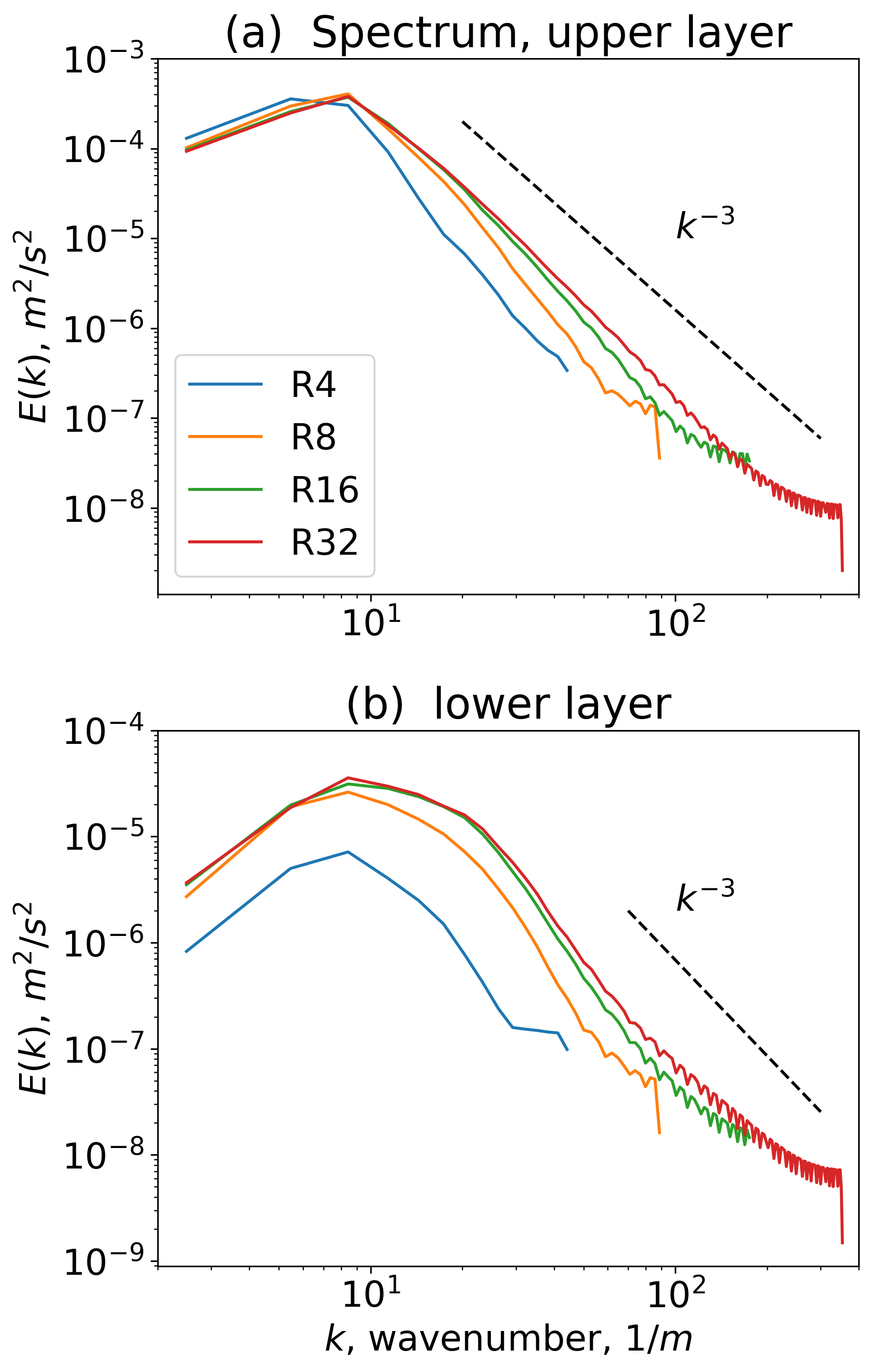}
\includegraphics[width=0.495\textwidth]{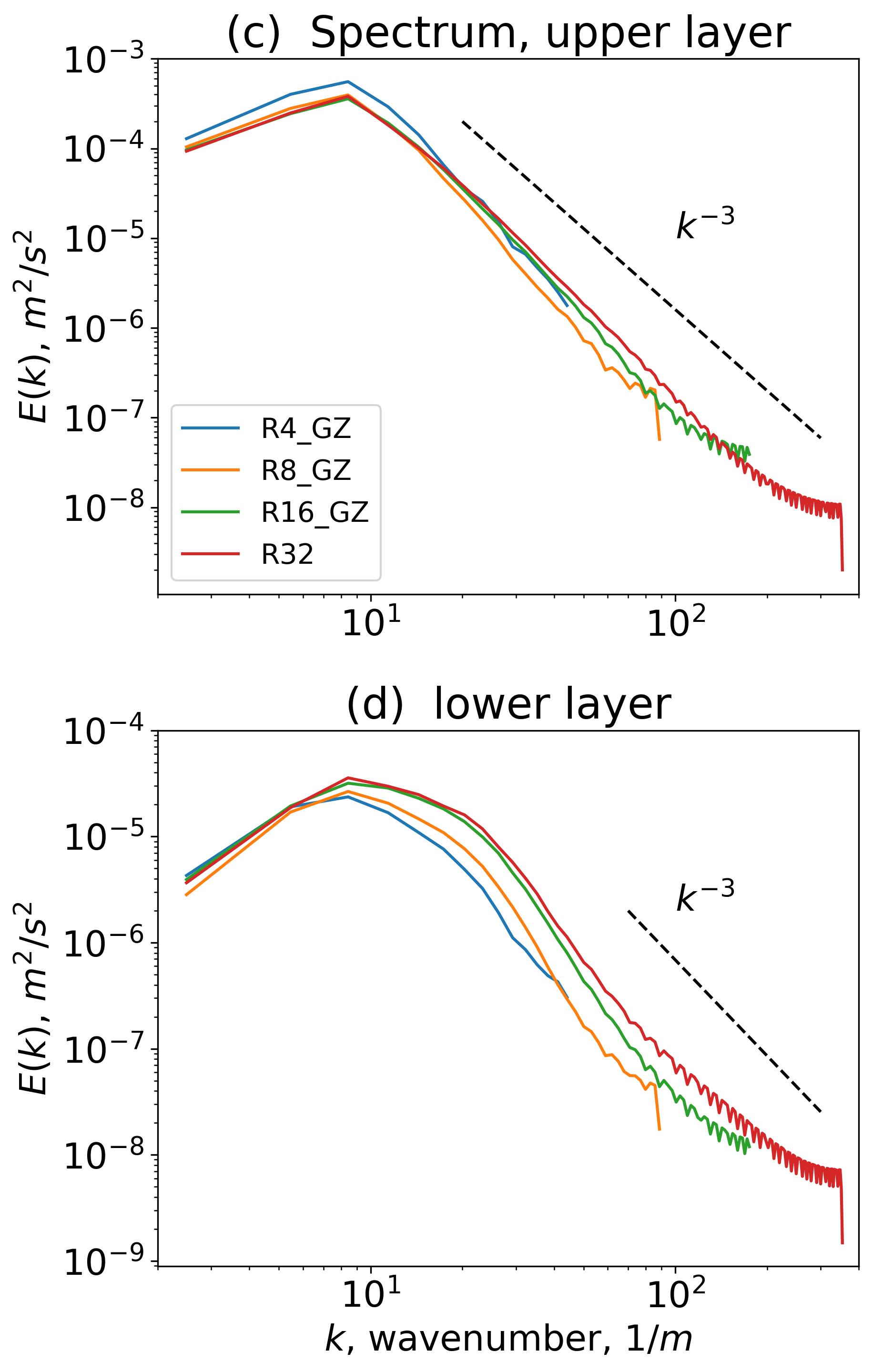}
\caption{(a and b) Comparison of KE spectra for flow without subgrid parameterizations versus target fine resolution flow; (c and d) Comparison of KE spectra for flow with subgrid parameterizations versus target fine resolution flow. The dashed lines are the spectral slope of kinetic energy spectrum corresponding to inertial interval of enstrophy.  MLP is short for the ML parameterization.}
\label{fig3.3_3}
\end{figure}

\begin{figure}[htbp]
\centering
\includegraphics[width=1.0\textwidth]{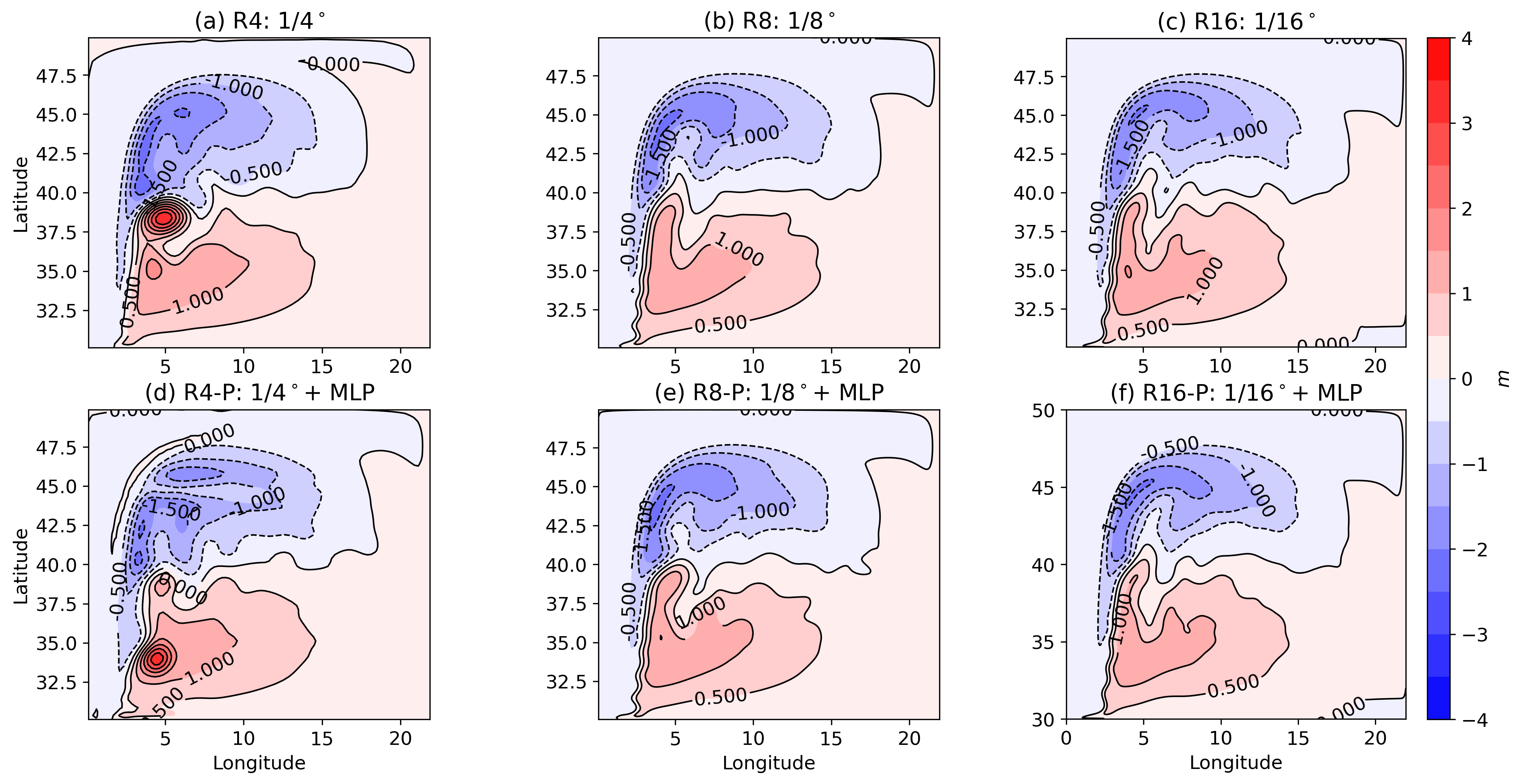}
\caption{Five-year averaged SSH maps for the flow without subgrid parameterizations (a, b and c) or with subgrid parameterizations (d, e and f). The grid sizes of the simulations are $1/4 ^\circ$ (R4, a and d), $1/8 ^\circ$ (R8, b and e) and $1/16 ^\circ$ (R16, c and f).  MLP is short for the ML parameterization.}
\label{fig3.3_4}
\end{figure}

To investigate the applicability of the CNN subgrid parameterization at different grid resolutions, we test the model against the grids ranging in size from $1/4 ^\circ$ (R4) to $1/16 ^\circ$ (R16).
Figure \ref{fig3.3_1} shows the snapshots of relative vorticity at the upper layer flow for different spatial resolutions.
The three runs in (a-c) have no parameterized momentum forcing, and the three runs in (d-f) have the stochastic CNN parameterization.
At all resolutions, small scales are qualitatively modified relative to the unparameterized counterpart.
As the spatial resolution is refined, the amplification by energy-injection appears to diminish; the CNN stochastic momentum forcing injects lots of energy in R4, but hardly any in R16.
This is more obvious in the plots of the total kinetic energy time series (Figure \ref{fig3.3_2}).
In the upper layer flow, the R4 case has significantly less KE (${\sim}17\%$) than the R32 case, and the parameterization overcompensates for this so that R4-P has almost ${\sim}50\%$ too much KE.
The intermediate resolution cases R8 and R16 have nearly identical total KE to that of R32.
In the lower layer flow, R4 also has smaller KE than R32, 
and the parameterization does increase the KE (R4-P), but in contrast to the upper layer, the parameterization does not add enough.
As for the upper layer, the parameterization has minor effects on the lower layer KE for both R8 ad R16.
The kinetic energy spectra in Figure \ref{fig3.3_3} and the five-year averaged sea surface height in Figure \ref{fig3.3_4} show the similar diminishing trend that the finer the grid resolution, the less effect the parameterization has on the flow. 

\subsection{Tuning the parameterization for online performance} \label{sec3.4}

\begin{figure}[htbp]
\centering
\includegraphics[width=1.0\textwidth]{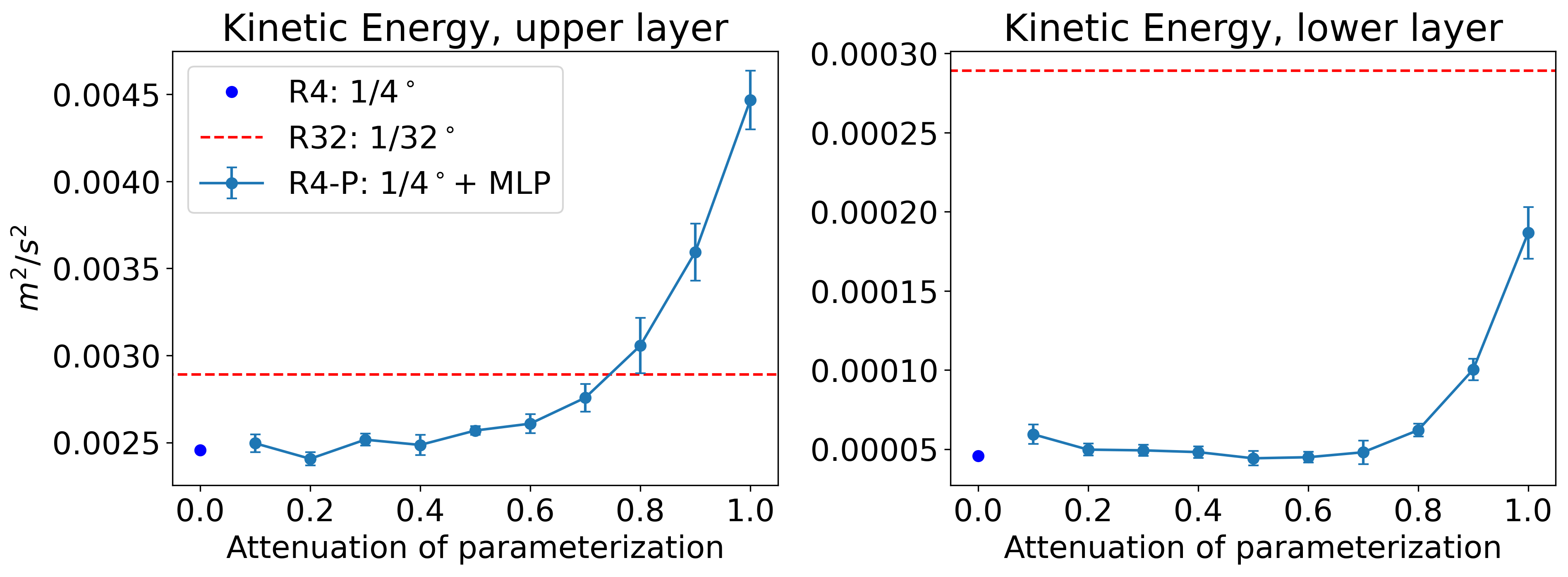}
\caption{Sensitivity of five-year time-averaged KE to vertically uniform attenuation of the momentum forcing. The error bar represents the standard deviation of KE among $10$ ensemble members for each value of attenuation.  MLP is short for the ML parameterization.}
\label{fig3.4_1}
\end{figure}

It is a common practice to tune a simulation by scaling a parameterization to optimize some metrics.
The simplest form of scaling is to multiply the parameterized accelerations by a fixed factor that will either amplify or attenuate depending on whether the scaling factor is larger or less than 1.
As shown in Figure \ref{fig3_2.6}, the momentum parameterization in R4-P over-energizes the upper layer flow, but under-energizes the lower layer flow.
We consider two strategies to tune the parameterization.
In the first strategy, we attenuate the momentum forcing by multiplying it for both layers by the same constant coefficient, ranging from $0$ to $1$, as done in \cite{zanna2020data}.
The metric we use to measure the attenuation is the integrated kinetic energy for each layer, averaged over the last five years.
Figure \ref{fig3.4_1} shows the sensitivity of the five-year averaged KE to vertically uniform attenuation of the momentum parameterization.
In general, an increase in the strength of parameterization results in more energization of the flow, i.e. this subgrid parameterization represents kinetic energy backscatter, see \citeA{frederiksen1997eddy}, \citeA{berner2009spectral}, \citeA{thuburn2014cascades}, \citeA{jansen2014parameterizing}, \citeA{zanna2017scale}, \citeA{juricke2020ocean}, and \citeA{zanna2020data}.
The sensitivity of time-averaged KE to parameterization strength appears to be different for the upper and lower layer flows.
The upper layer flow becomes more strongly sensitive to the attenuation coefficient about $0.6$, and provides optimal energization at ${\sim}0.75$, whereas the lower layer flow is relatively insensitive until the attenuation of $0.8$ and would apparently require an amplification coefficient greater than $1$.
Therefore, there is no shared value of scaling coefficient that can optimize the solution in both layers.

\begin{figure}[htbp]
\centering
\includegraphics[width=1.0\textwidth]{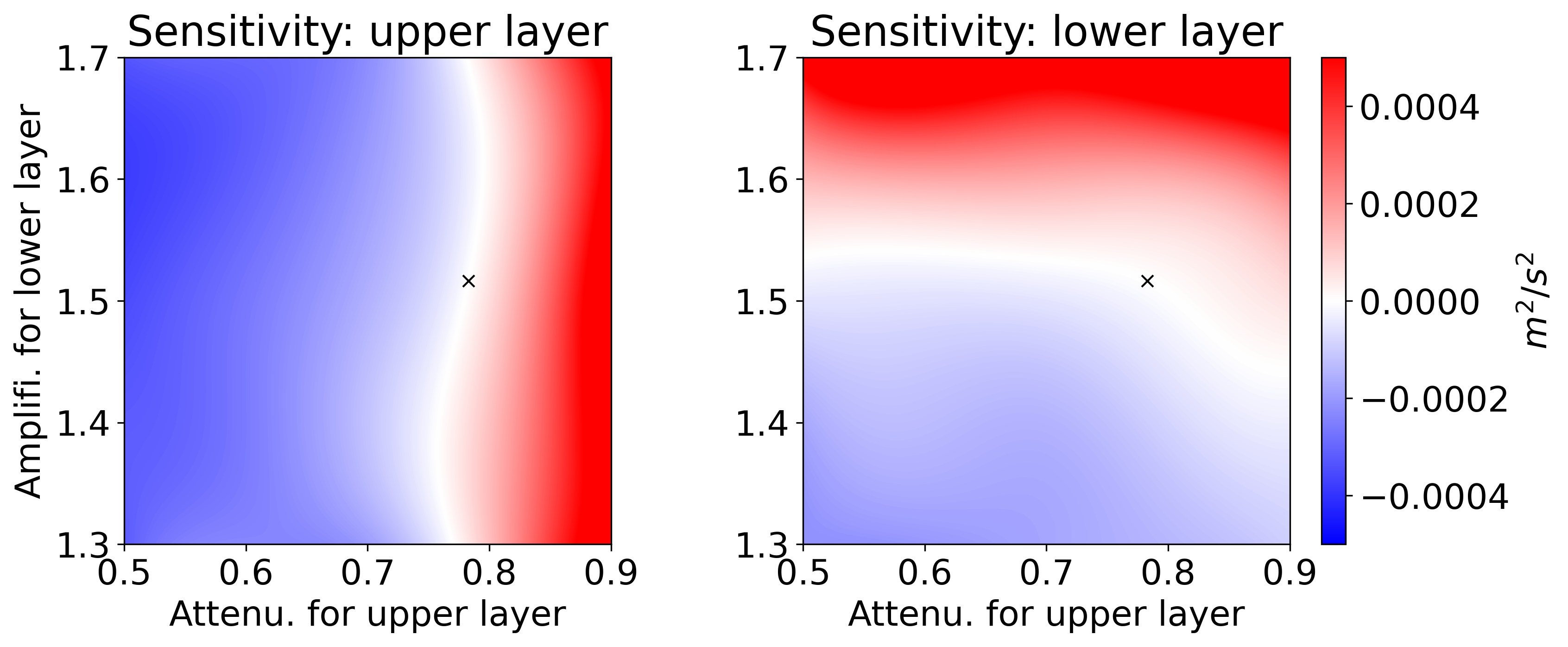}
\caption{Sensitivity of five-year time-averaged KE to vertically nonuniform scaling of the momentum forcing for the grid size of $1/4 ^\circ$. The $x$-axis is the attenuation applied to the upper layer forcing, and the $y$-axis is the amplification applied to the lower layer forcing. The kinetic energy is relative to KE for the target fine resolution case R32. The cross represents the fitting numbers that minimize relative KE in both layers.}
\label{fig3.4_2}
\end{figure}

\begin{figure}[htbp]
\centering
\includegraphics[width=0.495\textwidth]{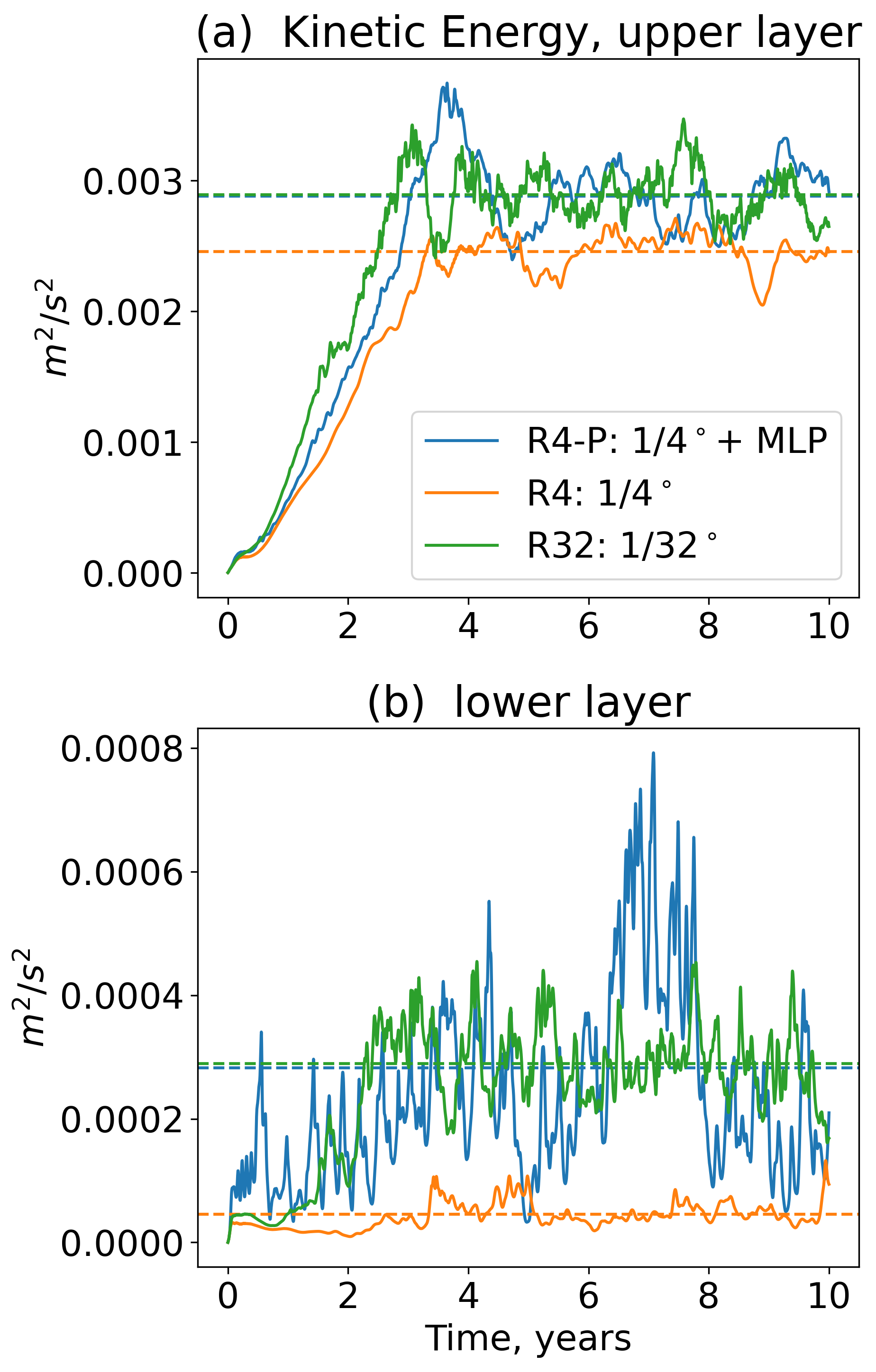}
\includegraphics[width=0.495\textwidth]{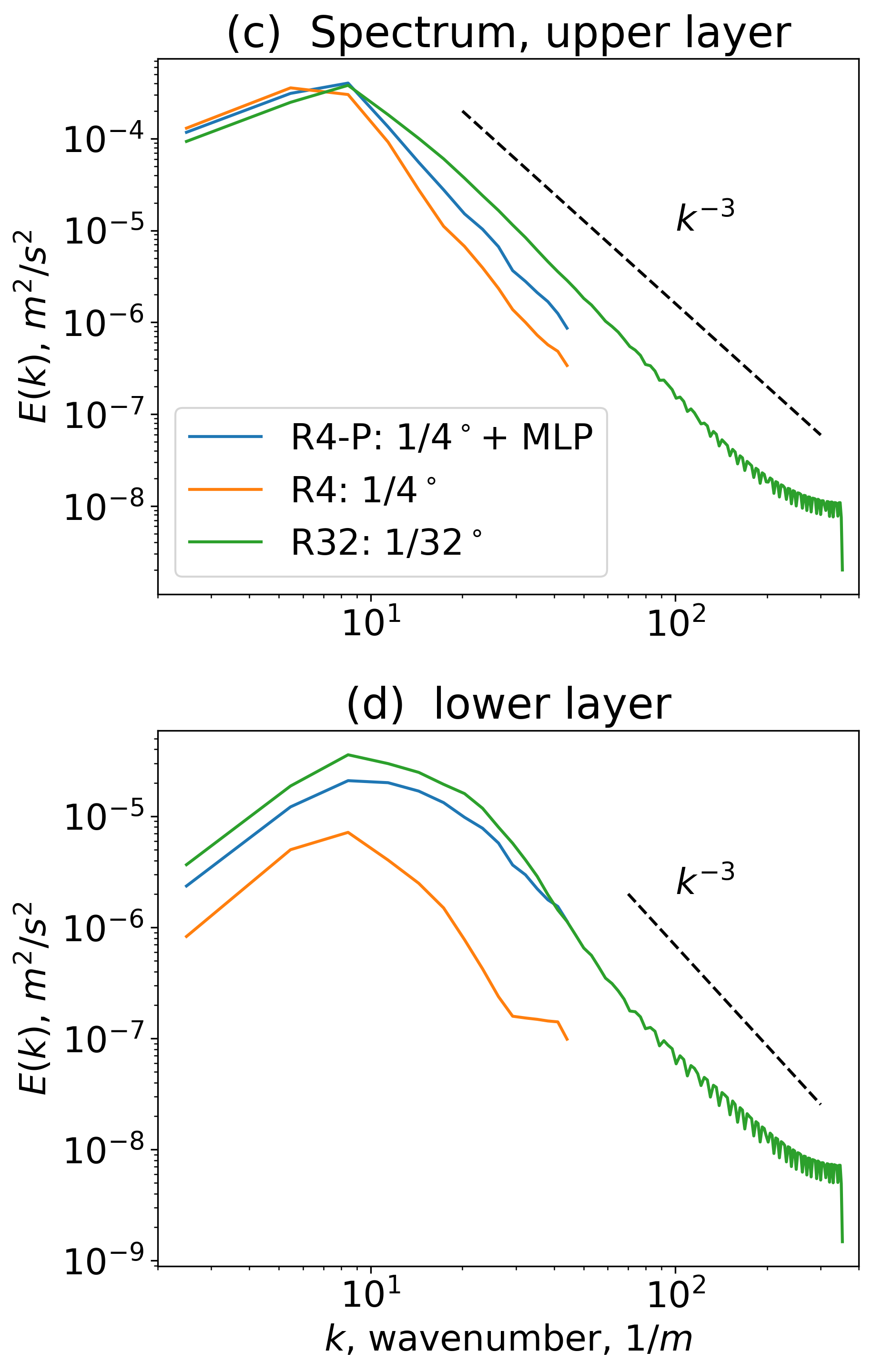}
\caption{Comparison of KE time series (a and b) and spectra (c and d) flow upper layer (top row) and the lower layer (bottom row) between the coarse resolution model R4 (orange), fine resolution R32 (green) and the coarse resolution model with the optimal scaling of momentum forcing R4-P (blue). The dashed lines in (a and b) are mean values of KE over the last 5 years. The dashed lines in (c and d) are the spectral slope of kinetic energy spectrum corresponding to inertial interval of enstrophy.  MLP is short for the ML parameterization.}
\label{fig4.1_1}
\end{figure}

The second tuning strategy we consider uses two different scaling coefficients, one for each layer.
Again, we use the time-averaged integrated kinetic energy for each layer as a metric.
The attenuation coefficient for the upper layer forcing is varied from $0.5$ to $0.9$, while the amplifying coefficient for the lower layer is varied from $1.3$ to $1.7$.
Figure \ref{fig3.4_2} shows a 2D sensitivity map where the $x$-axis is the upper layer attenuation coefficient and the $y$-axis is the lower layer amplification coefficient.
The color values are the KE difference relative to KE of R32, and we refer to it as relative KE.
The energy in both layers does not increase in a strictly linear fashion as the scaling number increases.
The energy increases in upper layer slightly slower when lower layer amplification coefficient is larger, while the energy increases in lower layer somewhat slower when the upper layer attenuation coefficient is larger.
In other words, the scaling of top layer can influence the lower flow, and vice versa.
Despite the influence between layers, the sensitivity for each layer is dominated by that layer's scaling coefficient.
For this case with the specific resolution and metric, the upper layer scaling number is $0.7827$ and the lower layer number is $1.5164$.
Using this set of scaling number, the momentum forcing parameterization vastly improves mean kinetic energy and its spatial spectrum (see Figure \ref{fig4.1_1}).
The mean of KE time series for R4-P almost exactly matches the KE mean for R32, and the KE spectra for R4-P are closer to the target.
This sensitivity analysis shows that it is possible to retroactively tune the machine learned parameterization of momentum forcing to optimize some metric of the ocean model solution.

\section{Problems and remedies} \label{sec4}

In the two-layer double gyre tests of the GZ21 parameterization (Section \ref{sec3}), we find the parameterization can be made to work well but might be limited in generality and has some artifacts at boundaries.
We will discuss distinct aspects of our results, noting challenges and suggest some remedies here or for future work.

\subsection{Parameter optimization and vertical structure} \label{sec4.1}

Without attenuation the GZ21 parameterization over-energizes the upper layer flow and under-energizes the lower layer flow.
That tuning is needed at all is not unexpected, with many conventional and machine learned parameterizations performing differently between "offline" and "online". 
All parameterizations are ultimately tuned.
In the online test by GZ21 it appears a parameterization trained on surface fields was reasonably effective for a barotropic model.
Here, we essentially tested the hypothesis that the interior momentum forcing was functionally similar to the surface momentum forcing, and whether the momentum forcing could be treated independently layer by layer, i.e. decoupled in the vertical except through correlations between the layer flows.
We find that vertical structure is needed since tuning yielded significantly different scaling coefficients for the two layers (attenuation for the upper layer, amplification for the lower layer).
Here, we could afford to find the optimal combination of just two scaling values that yield the "best" coarse resolution model with the CNN parameterization, using the time-averaged integrated kinetic energy as a metric (Figure \ref{fig3.4_2}).

\subsection{Resolution dependence and scale-awareness} \label{sec-scale}

\begin{figure}[htbp]
\centering
\includegraphics[width=1.0\textwidth]{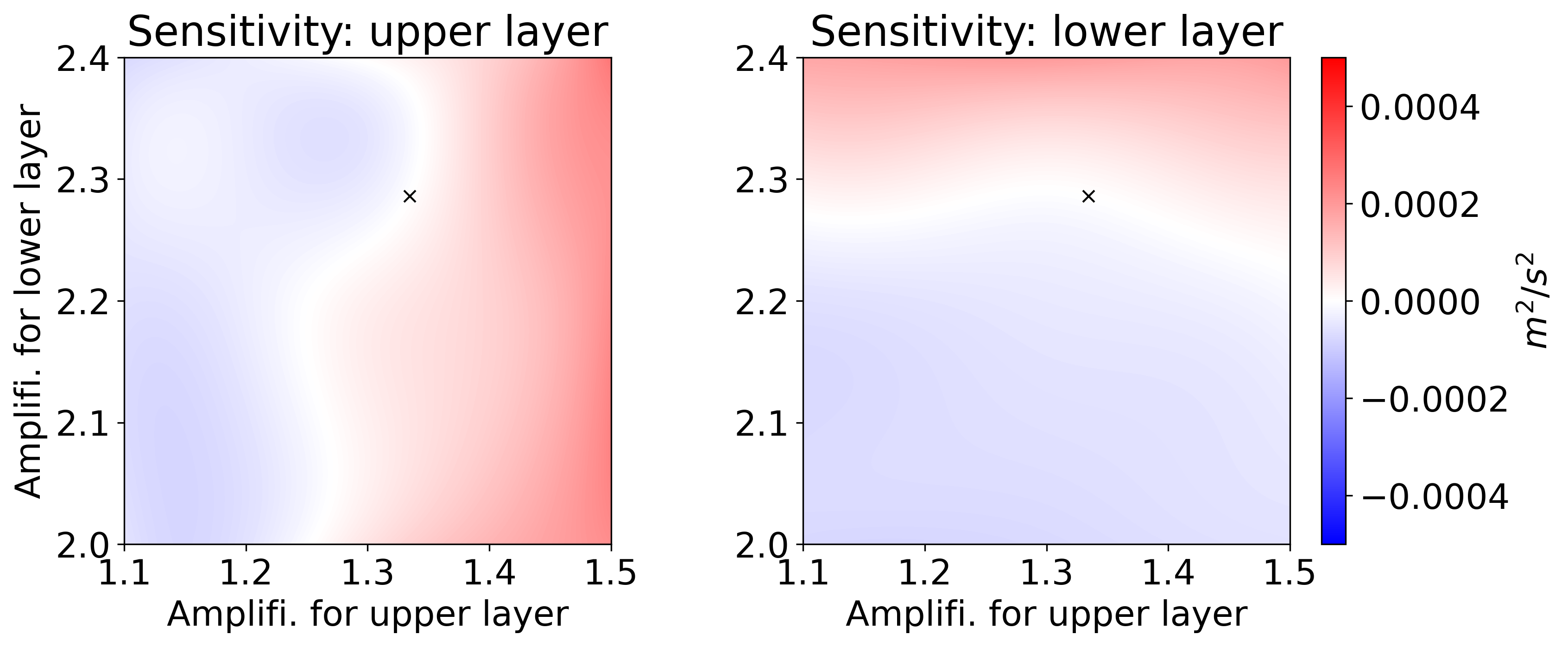}
\caption{Sensitivity of five-year averaged KE to scaling of the momentum parameterization for the grid size of $1/8 ^\circ$. The definition of $x$-axis, $y$-axis, the variable in the map and the cross is same to Figure \ref{fig3.4_2}.}
\label{fig4.1_2}
\end{figure}

\begin{figure}[htbp]
\centering
\includegraphics[width=1\textwidth]{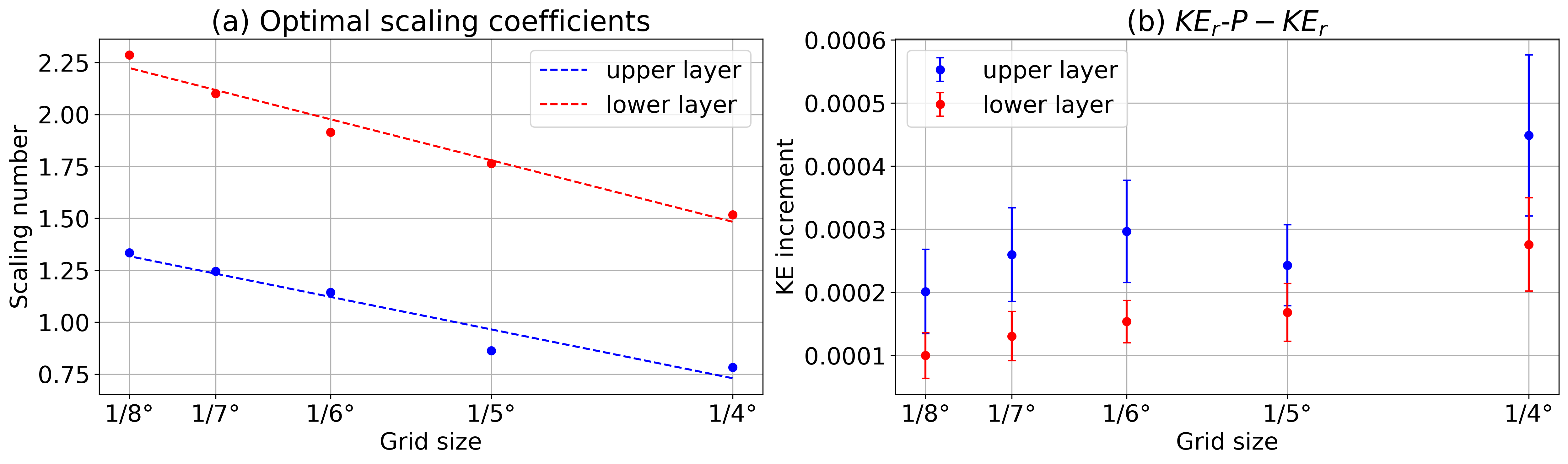}
\caption{(a) Relation between optimal scaling numbers to subgrid parameterization and grid sizes of the double gyre simulations; (b) Relation between KE increment from the optimal scaled parameterization and grid sizes; the error bar represents the standard deviation of KE among $20$ ensemble members for optimal scaling number.}
\label{fig4.1_3}
\end{figure}

The optimal tuning indicated in Figure \ref{fig3.4_2} is for the spatial resolution of $1/4 ^\circ$.
In section \ref{sec3.3} we asked if the parameterization performed well at other spatial resolutions.
We noted that at finer resolutions the parameterized momentum forcing is diminished.
This resolution dependence might be coming from the change in flow structure and amplitude at different resolved scales.
We repeat the tuning exercise for the spatial resolution of $1/8 ^\circ$, varying the layer-wise scaling coefficients to optimize the time-mean intergrated KE (Figure \ref{fig4.1_2}).
The sensitivity patterns are broadly similar to those in Figure \ref{fig3.4_2} but with smaller amplitudes, indicating less sensitivity.
The coefficients that optimize the time-mean KE of R8-P to be most similar to that of R32 (cross in Figure \ref{fig4.1_2}) are an upper layer amplification of $1.3345$ and a lower layer amplification of $2.2862$.
Here, the upper layer in R8-P needs amplification while in R4-P the upper layer needed attenuation.
If the relationship between grid size and scaling factor is assumed to be linear, the slope of the regression line for upper layer scaling numbers to the grid sizes is $-4.6987$ ($\text{scaling number} = -4.6987 \times \text{grid size} +1.9048$), while the slope of the lower layer scaling numbers to the grid sizes is $-5.9207$ ($\text{scaling number} = -5.9207 \times \text{grid size} +2.9635$).
We repeat the tuning at the spatial resolutions of $1/5 ^\circ$, $1/6 ^\circ$ and $1/7 ^\circ$, and plot the optimal scaling coefficients in Figure \ref{fig4.1_3}(a).
We find a broadly linear fit with increasing amplification as resolution is refined.
It is interesting to see that the scaling factor increases as the resolution gets finer, which contrary to what we normally expect that the parameterization impact should taper off when the resolution gets finer.
Figure \ref{fig4.1_3}(b) shows the difference between KE from the optimal scaled parameterization (KE$_r$-P) and from no parameterization(KE$_r$) which is an integral measure of how much work the parameterization has done.
The measure of work tends to decrease with finer resolution even though the scaling factor gets larger.
In comparison to the trend, the KE difference for R5 is relatively small. It could be due to a modest size of the ensemble (20 member), which was chosen to minimize computing expense.

\subsection{Role of metric in evaluation}\label{sec4.2}

\begin{figure}[htbp]
\centering
\includegraphics[width=1.0\textwidth]{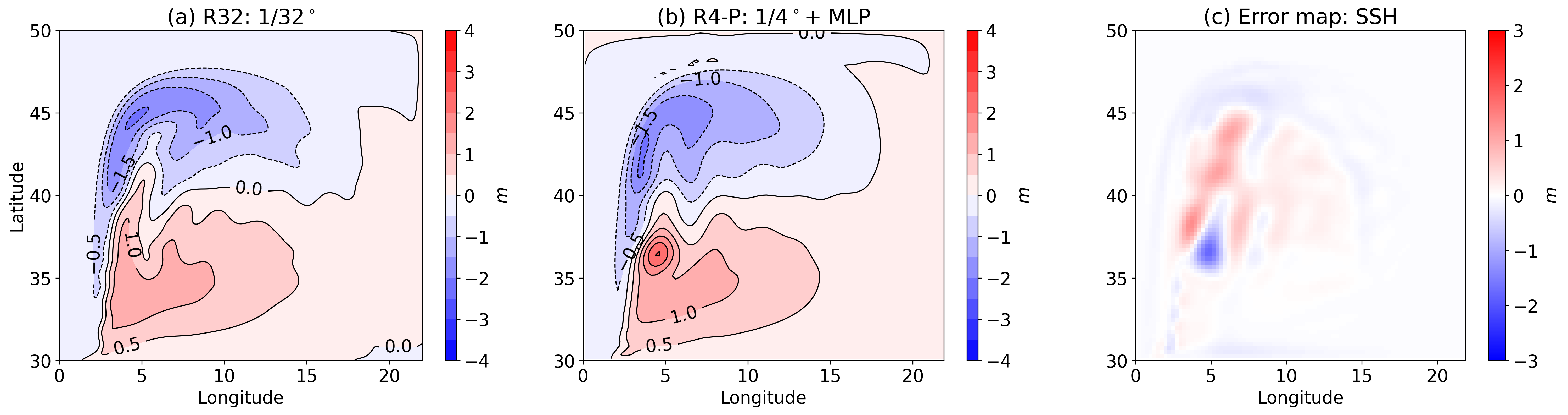}
\caption{Comparison of five-year averaged SSH between the coarse resolution model with the scaled subgrid parameterization (the upper layer scaling number is 0.7827 and the lower layer number is 1.5164, indicated by the cross in Figure \ref{fig3.4_2}) and the target fine resolution model R32. The error map is obtained by subtracting low-resolution SSH with the optimal parameterization from coarse-grained high-resolution SSH.  MLP is short for the ML parameterization.}
\label{fig4.1_4}
\end{figure}

\begin{figure}[htbp]
\centering
\includegraphics[height=5.2cm]{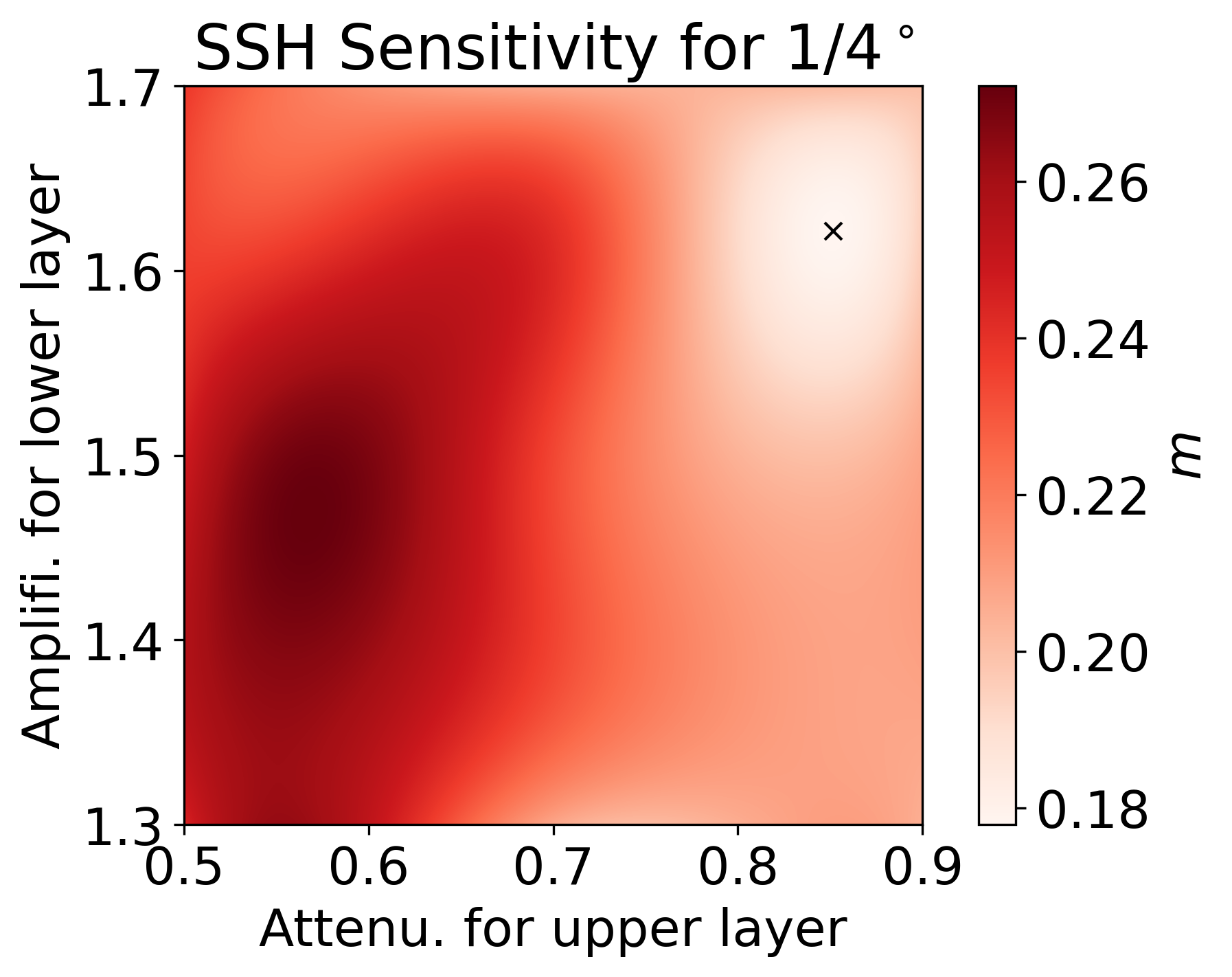}
\includegraphics[height=5.2cm]{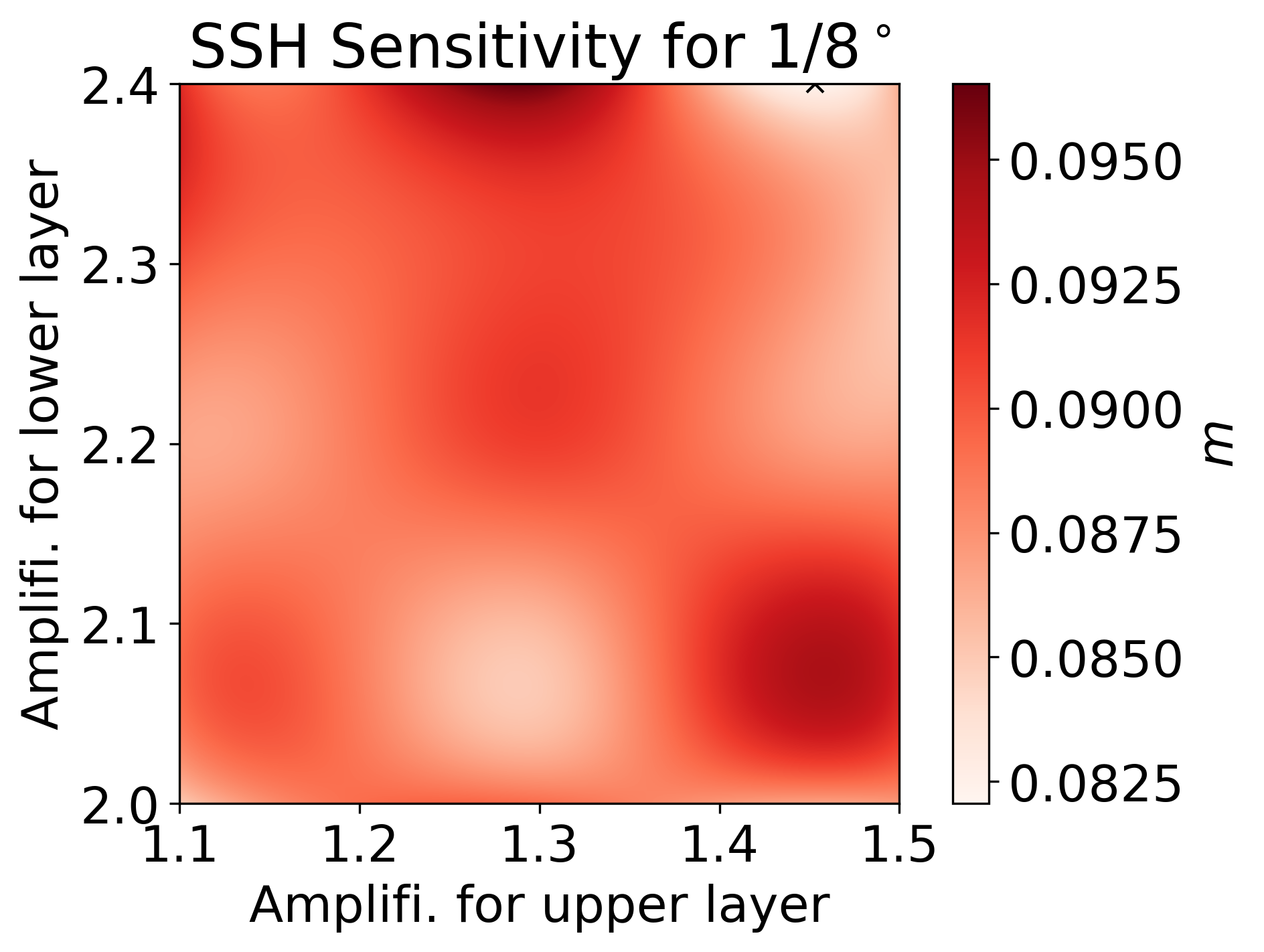}
\caption{Scaling optimization of momentum parameterization based on five-year averaged SSH. The $x$-axis is the amount of prescribed scaling for upper layer flow and the $y$-axis is the amount of the scaling for lower layer flow. The cross represents the fitting numbers that minimize RMSE of this SSH to R32 SSH.}
\label{fig4.1_5}
\end{figure}

So far we have only used the time-averaged integrated kinetic energy as a metric for tuning the scaling of momentum forcing.
Qualitatively, other aspects of the solution improve when the total KE is optimized.
Figure \ref{fig4.1_4} shows the difference between the five-year averaged SSH of R4-P and R32 using the best scaling of momentum forcing based on the optimized KE, where the upper layer scaling number is 0.7827 and the lower layer number is 1.5164 (indicated by the cross in Figure \ref{fig3.4_2}).
The scaled parameterization improves this metric if we look at RMSE of the error map where the RMSE value is now $0.2034$m, down from $0.2202$m for the parameterization without scaling (Figure \ref{fig3.4_2}(e)).
Table \ref{tab1} shows the SSH improvement based on the RMSE of error maps for the various grid sizes from $1/4^\circ$ (R4) to $1/8^\circ$ (R8).
For all resolution models, we find that the best-scaled parameterizations based on the metric of KE also improve the metric of SSH.
While the best scaling numbers for KE also improve SSH, these numbers are not the best scaling numbers for SSH.
Figure \ref{fig4.1_5} depicts the optimal scaling numbers for R4 and R8 based on another metric, i.e., RMSE of SSH deviation.
The best scaling numbers for SSH are apparently not same to the number for KE.
Furthermore, the patterns in the 2D maps are less coherent, in contrast to the patterns in KE sensitivity maps in Figures \ref{fig3.4_2} and \ref{fig4.1_2}.

\begin{table}[h!]
\begin{tabular}{|l|l|ll|ll|}
\hline
\rowcolor[HTML]{9B9B9B} 
\cellcolor[HTML]{9B9B9B} &
  No Param. &
  \multicolumn{2}{l|}{\cellcolor[HTML]{9B9B9B}Param. without scaling} &
  \multicolumn{2}{l|}{\cellcolor[HTML]{9B9B9B}Param. with best scaling} \\ \cline{2-6} 
\rowcolor[HTML]{9B9B9B} 
\multirow{-2}{*}{\cellcolor[HTML]{9B9B9B}Res.} &
  RMSE (m)&
  \multicolumn{1}{l|}{\cellcolor[HTML]{9B9B9B}RMSE (m)} &
  Improved(\%) &
  \multicolumn{1}{l|}{\cellcolor[HTML]{9B9B9B}RMSE (m)} &
  Improved(\%) \\ \hline
$1/4^\circ$ & 0.2780 & \multicolumn{1}{l|}{0.2202} & 20.7914 & \multicolumn{1}{l|}{0.2034} &  26.8345 \\ \hline
$1/5^\circ$ & 0.2093 & \multicolumn{1}{l|}{0.1827} & 12.7090 & \multicolumn{1}{l|}{0.1706} &  18.4902 \\ \hline
$1/6^\circ$ & 0.1432 & \multicolumn{1}{l|}{0.1305} &  8.8687 & \multicolumn{1}{l|}{0.1253} &  12.5000 \\ \hline
$1/7^\circ$ & 0.1167 & \multicolumn{1}{l|}{0.1088} &  6.7695 & \multicolumn{1}{l|}{0.1001} &  14.2245 \\ \hline
$1/8^\circ$ & 0.0971 & \multicolumn{1}{l|}{0.0905} &  6.7971 & \multicolumn{1}{l|}{0.0891} &   8.2389 \\ \hline
\end{tabular}
\caption{Improvement of time-mean sea surface height for scaled momentum forcing at various spatial resolutions, based on optimal scaling factors for KE.}
\label{tab1}
\end{table}

As for many conventional parameterizations, we find the parameterization of momentum forcing to be able to improve different aspects of the solution but to different degrees and not necessarily optimally together.
The parameterization injects momentum and kinetic energy so we should expect to be able to have a direct effect on total kinetic energy.
The parameterization has a more indirect control over time-mean sea-surface height (through geostrophy if any) and we find less coherent response in the RMSE SSH.
Neither metric was used in the training of the CNN in GZ21, so the result that we can optimally tune total KE, whilst observing a modest reduction in RMSE SSH, is therefore a success for the parameterization.

\subsection{Effect of lateral boundaries on CNN inference} \label{sec4.3}

\begin{figure}[htbp]
\centering
\includegraphics[width=0.5\textwidth]{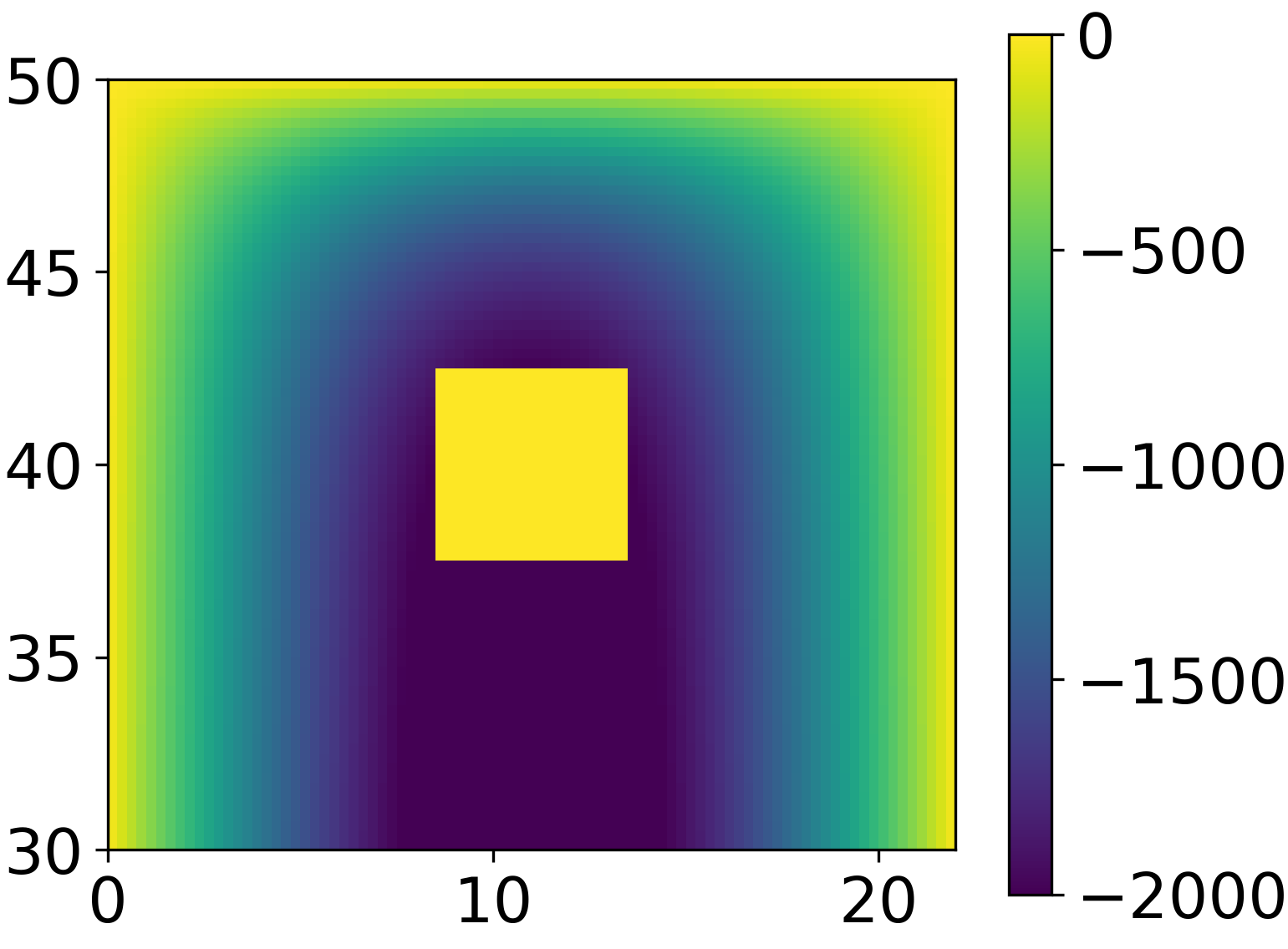}
\caption{Plan view (latitude, longitude) of the bathymetry with a box in the middle of the domain for wind-driven double gyre.}
\label{fig4.2_5}
\end{figure}

\begin{figure}[htbp]
\centering
\includegraphics[width=1.0\textwidth]{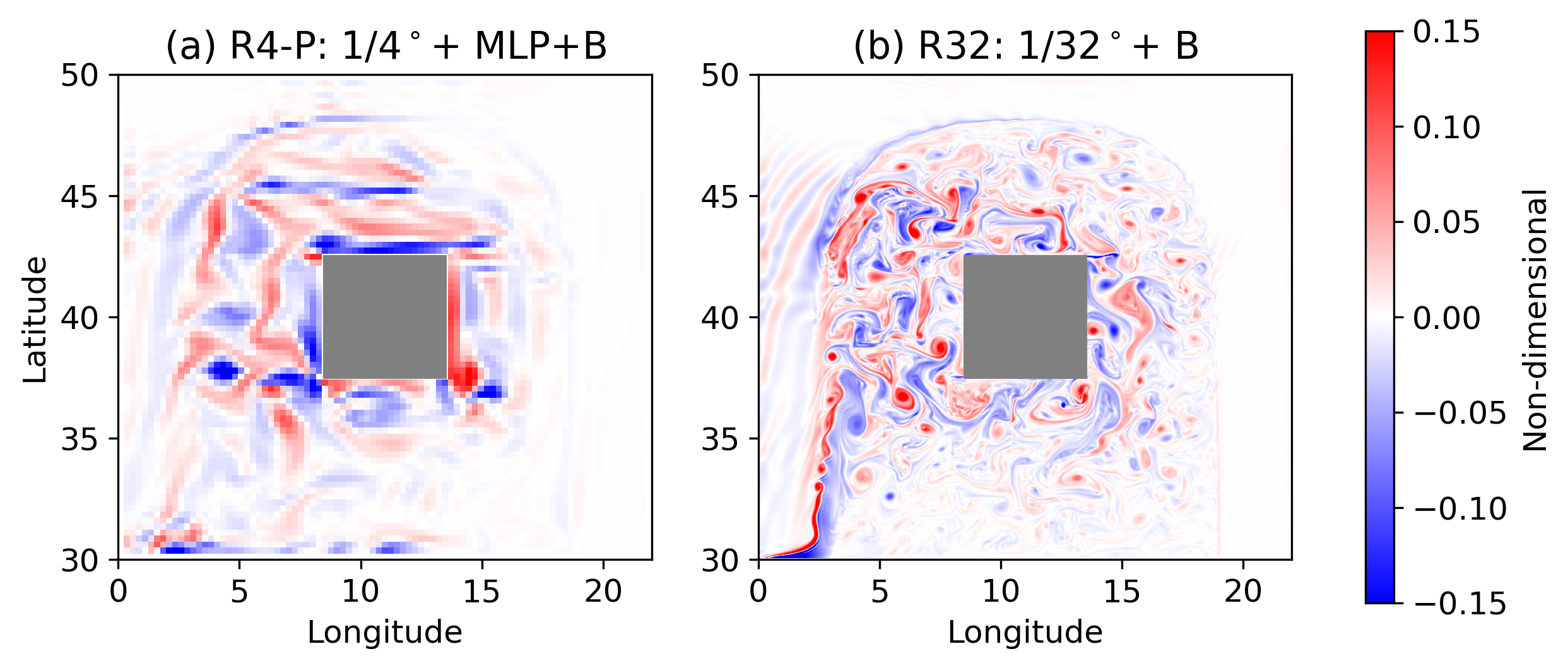}
\caption{Snapshots of the upper layer relative vorticity (normalized by the planetary vorticity) at the end of perturbed-topography simulations from the coarse resolution model with ML parameterizations R4-P (a) and the fine resolution R32 (b). The grey rectangle indicates the region where the unrealistic eddies are generated. MLP is short for the ML parameterization and B is short for the box (grey rectangle).}
\label{fig4.2_6}
\end{figure}

\begin{figure}[htbp]
\centering
\includegraphics[width=0.495\textwidth]{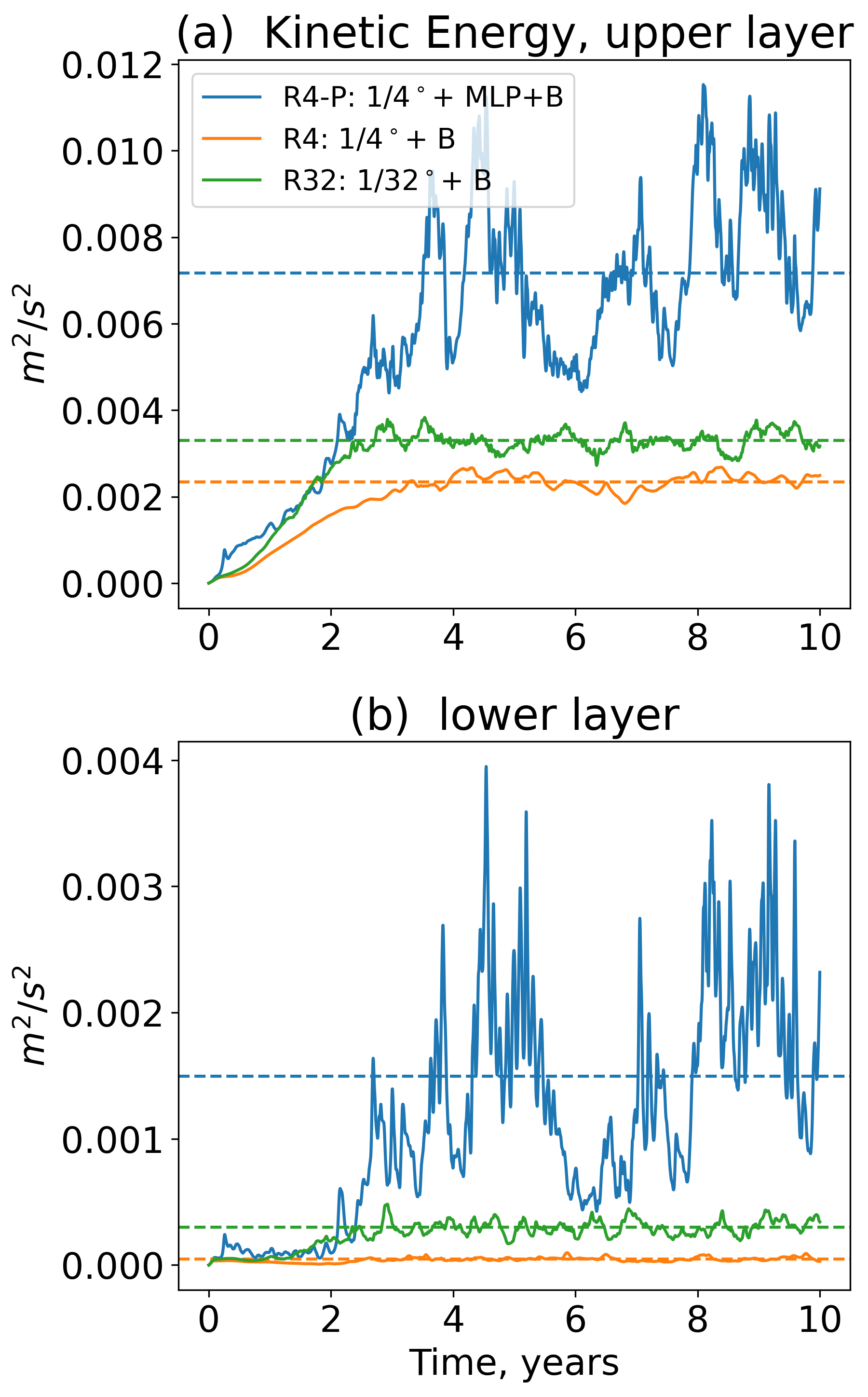}
\includegraphics[width=0.495\textwidth]{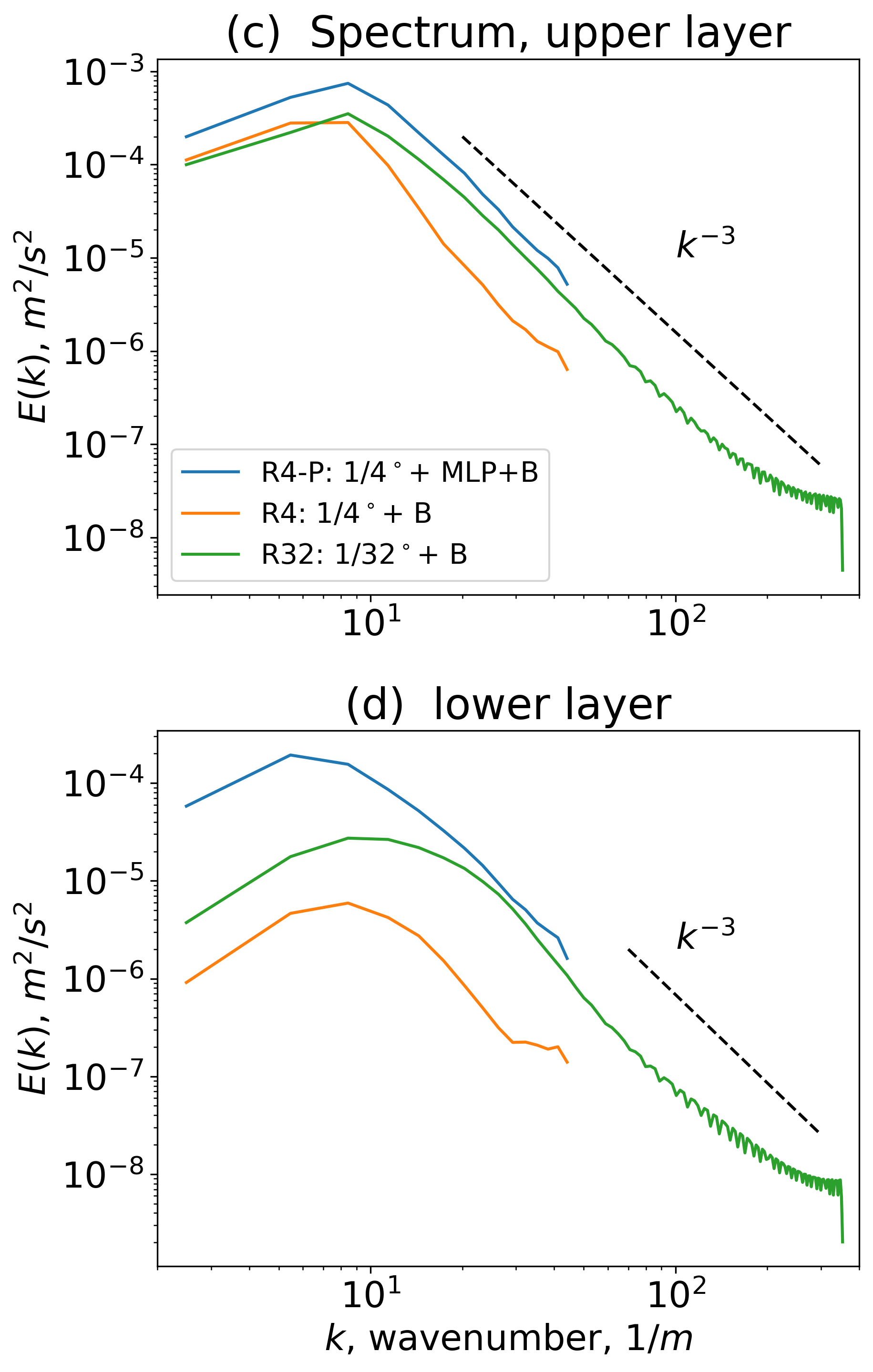}
\caption{Comparison of KE time series (a and b) and spectra (c and d) for perturbed-topography tests at the flow upper layer (top row) and the lower layer (bottom row) between the coarse resolution model R4 (orange), fine resolution R32 (green) and the coarse resolution model with ML parameterizations R4-P (blue). The dashed lines in (a and b) are mean values of KE over the last 5 years. The dashed lines in (c and d) are the spectral slope of kinetic energy spectrum corresponding to inertial interval of enstrophy. MLP is short for the ML parameterization.}
\label{fig4.2_7}
\end{figure}

\begin{figure}[htbp]
\centering
\includegraphics[width=1.0\textwidth]{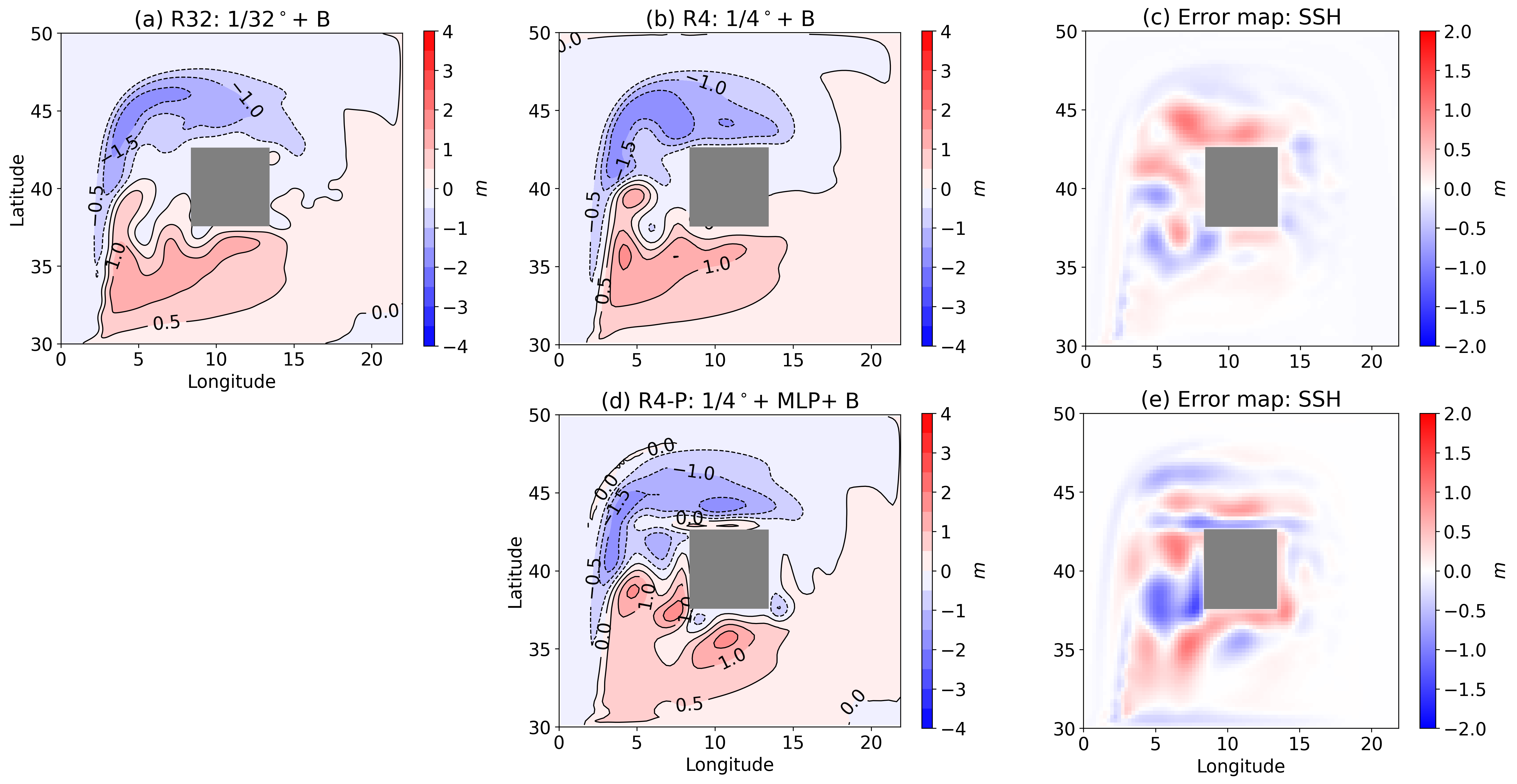}
\caption{Comparison of five-year averaged SSH for perturbed-topography tests between the coarse resolution model with (R4-P) and without (R4) the subgrid parameterization and the target fine resolution model (R32). The error maps (c and d) are obtained by subtracting low-resolution SSH (with or without parameterization) from coarse-grained high-resolution SSH.  MLP is short for the ML parameterization and B is short for the box (grey rectangle).}
\label{fig4.2_8}
\end{figure}

In Section \ref{sec3.2} we noted the CNN parameterization induced artifacts at the wall boundaries.
Strong zonally sheared eddies highlighted by the black box in the left plot of Figure \ref{fig3.3_1} are not realistic, with no counterpart in the fine resolution model results.
The training data used by GZ21 was from limited regions of the CM2.6 model and deliberately excluded any coastal waters or land.
Therefore, by construction the parameterization was not trained to "know" what to do near boundaries.
We hypothesize that the exclusion of coastal waters in the four selected regions contributes to the boundary artifacts.
To better illustrate the boundary artifacts near model coastlines after the CNN parameterization, we perturb the double gyre test by adding a box in the middle of the domain (positioned from $8.5^\circ$ to $13.5^\circ$ in longitude and $37.5^\circ$ to $42.5^\circ$ in latitude, see Figure \ref{fig4.2_5}) with vertical walls.
This is a severe topographic obstacle in the path of the wind-driven jet and we expect it to test the limits of the CNN parameterization.
A snapshot of the upper layer relative vorticity shows how much the new geometry affects for the coarse R4-P model using the parameterization.
Strong sheared structures can be seen both around the box island as well as at the southern boundary as before (Figure \ref{fig4.2_6}a).
Introducing the box island to the fine resolution R32 model does not develop any comparable structures (Figure \ref{fig4.2_6}b).
The kinetic energy time series and spectra (Figure \ref{fig4.2_7}) also suggest that the parameterization over-energetizes the flow close to the boundary.
As before without the box island, the CNN parameterization injects too much energy into the upper layer, but also now in the lower layer.
The limit of the parameterization near wall boundaries is also evident from the time-mean sea surface height (Figure \ref{fig4.2_8}).
The RMS difference between R4-P SSH and R32 SSH is increased to $0.2503$m from $0.1765$ for R4 SSH (without parameterization), which makes matters worse.

We believe that retraining the same CNN model using the velocity data from the entire globe might address the issue. 
However, the volume of data that will be used in the retaining process is roughly 40 times greater than that from the four subdomains used to train the current CNN model (GZ21).
The cost of the training process will be dramatically increased given the complex architecture of the current CNN model.
Extending to the global domain raises the question of how to handle land points. 
One option is to set the velocity components at the land points to either $NaN$ or 0.
However, the precision of training at the wet points that have land points in their $21 \times 21$ stencil will be reduced or lost entirely.
Another choice is to exclude from the training data anywhere that the stencil includes land points.
Using this approach, however, the network is losing many samples within 20 cells of the coasts (which is of order 120-200~km in distance).
There needs to be more discussion on the better method to use.

A natural next step to address the boundary artifacts will be to retrain the same CNN model with the global data so that it may interpolate between "known" states in the model inference process and avoid this possible "out of sample" issue.

\subsection{Cost model for CPU and GPU implementations} \label{sec4.4}

\begin{figure}[htbp]
\centering
\includegraphics[width=1.0\textwidth]{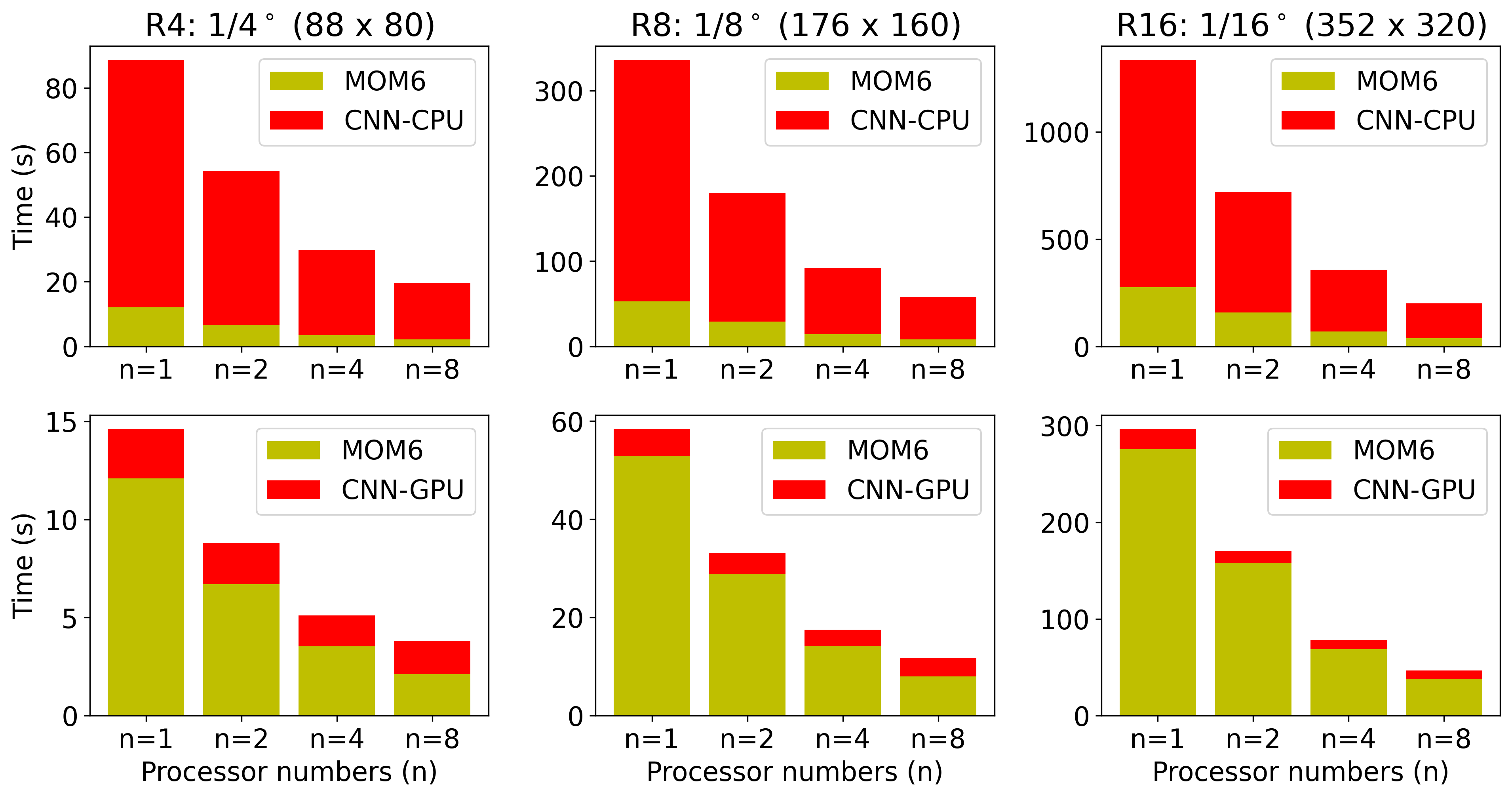}
\caption{Cumulative processing time of the CNN inference on CPU/GPU and MOM6 simulations at various grid resolutions and parallel processor counts.}
\label{fig4.3_1}
\end{figure}

The computation of the CNN model inference may involve many more floating point computations than the dynamical model itself.
Many conventional closed-form parameterizations typically cost a small fraction of the dynamical model so the potentially high cost may appear to be prohibitive to adopting neural network based parameterizations.
The total time complexity \cite{he2015CNNcost} of one time step inference is
\be
O \left(\sum_{l=1}^d ( w_{l-1}\cdot s_l^2\cdot + 1 ) w_l\right)
\label{eq4.3_1}
\ee
where $l$ is the index of a convolutional layer, $d$ is the depth (number of convolutional layers),  $w_l$ is the number of output channels (also known as “width”) for the $l$-th layer, $w_{l-1}$ is the number of input channels of the $l$-th layer, and $s_l$ is the spatial size of the filter.
This formula counts the numbers of weights needed to describe the neural network and allows us to estimate the approximate number of floating point operations (FLOPs) assuming for convenience that a multiply-add pair counts as a single operation.
The network of GZ21 we use has $s_l = 5, 5, 3, 3, 3, 3, 3, 3$ and $w_l = 128, 64, 32, 32, 32, 32, 32, 4$. The first layer has two inputs (namely the $u$ and $v$ components of flow) and the four outputs of the last layer correspond to the mean and standard deviation of the zonal and meridional momentum forcing.
The inference for our CNN model requires at least $268,005$ in FLOPs for each grid point of the dynamical model, which is significantly more operations than what is required by conventional parameterizations, and is even more than that of the dynamical model itself (typically on the order of hundreds to thousands).
The stacked bar charts in Figure \ref{fig4.3_1} show the measured processing time spent computing the CNN inference and dynamical core, for various spatial resolutions and parallel MPI processes.
The upper panels of Figure \ref{fig4.3_1} show, that for this simple two-layer double gyre case, the CNN inference processing time, on that same CPUs as the dynamical core is running, is around $O(10)$ times that of the dynamical core, and this ratio is essentially constant over a range of grid resolutions.

The above results for cost of inference on CPUs are prohibitive for most applications for global or regional simulations.
Most machine learning applications utilize GPUs which work well on the tensor-like operations within a neural network.
Typically, one GPU is only accessible by one CPU processor at a time, and the rest of the CPU processors must wait in queue.
CUDA provides the \href{https://docs.nvidia.com/deploy/mps/index.html}{Multi-Process Service} (MPS) which allows multiple CPU processors to access a GPU card.
This allows us to run the dynamical model on mutliple CPU processors and move the CNN inference to a shared GPU and can call CNN computation asynchronously.
With this strategy, we find the processing time for inference is dramatically decreased (around 1/5 in wall-clock time).
As shown in Figure \ref{fig4.3_1}, the cumulative processing time required for CNN inference on GPUs (lower panels) is considerably less than that of the dynamical core running on the CPUs, and the ratio of time on CNN decreases as the grid resolution is increased.

Although utilizing GPUs for the CNN inference is efficient, various challenges remain to prevent widespread adoption.
Currently, MPS only permits a maximum of $16$ CPU processors per GPU (\url{https://docs.nvidia.com/deploy/mps/index.html}).
This restriction complicates the implementation of data transfer because subdomains for MPI exchange on CPU would necessarily be different to the subdomains for CNN inference on GPUs when more that $16$ CPUs are used.
We have not tried using any configurations with more than 32 CPU processors and 2 GPU cards.
A practical matter is that not every computer or cluster has a GPU card directly attached, or indirectly available.
In this instance, inference has to occur on the CPUs and so the number of weights in the network (which most directly controls the computational cost) has to be limited.
We could make trade-offs between the depth, number of filters, and filter size within the CNN model in order to balance accuracy and implementation costs.
Such considerations are usually part of the hyper-parameter tuning during the training process but the restrictions imposed by the cost of inference on CPUs would significantly change the balance of factors.

Related to the number of weights is the volume of data needing to be communicated laterally between parallel processes so that the inputs to the CNN are all valid across the full stencil.
For the GZ21 network, each output point has a stencil of $21\times21$ input points.
This requires a halo of width $10$ to surround each computational subdomain which must be updated prior to passing to the CNN for inference.
In our implementation, we could have made the halo wider for all variables in the model but this would have increased the cost of communication for the whole model which generally has halo widths of $3$ or $4$.
Instead we made two temporary arrays (one for each of $u$ and $v$) with wide halos of $10$ and the cost of updating these halos proved to not be significant.


\section{Conclusions}\label{sec5}

We have described an investigation into how well a stochastic-deep learning parameterization of subgrid momentum forcing performs in an idealized ocean model.
We set out to explore how to use a pre-defined ML parameterization in a general use, global ocean circulation model written in Fortran.
We focused on one particular parameterization, GZ21, that targets the backscatter of energy from unresolved flows.
However, the tests, lessons learned, and recommendations apply broadly to any deep learning ocean or atmosphere parameterizations developed  \cite{krasnopolsky2010accurate, rasp2018deep, o2018using, maulik2019subgrid, Bolton2019, yuval2021GRL, beucler2021climate,christensen2022parametrization}.

The ML parameterization was originally trained on a geographic sub-sample of surface flow from a realistic, relatively fine-resolution, fully coupled climate model (CM2.6).
We applied the ML parameterization “as is” in a coarse-resolution, idealized wind-driven baroclinic model primitive equation model for which we could afford to run a fine-resolution “truth” simulation.
We employed several metrics, i.e., kinetic energy, spectra, and sea surface height error, to access the performance.
Out of the box, the ML parameterization did improve some aspects of the coarse-resolution solution.
However, some artifacts were apparent that were not evident in the original online testing in a barotropic model with flat bottom by \cite{Guillaumin1&Zanna-JAMES21}.
Despite these negative aspects, the network produced results that improve some of the model physics without generating infinities or nonsense, so our results are evidence of some underlying robustness of the parameterization.
We found the overall energization to be too efficient and that global tuning could be used to yield better results, similarly to \cite{zanna2020data}.
Our results are improved if we tune layer-by-layer, which re-enforces a notion that surface currents and interior currents have different dynamics.
Tuning was able to optimize one metric (in our case we used mean KE), and while separate metrics (such as SSH) improved they were not always optimal nor was it obvious they were robustly sensitive to the tuning.
The geographic sub-sample used for training seems to have selected sheared flow structures that led to sheared artifacts near boundaries in our tests.
This might be a classic example of the “out of sample” problem whereby a network should be trained with enough samples that it is interpolating between “known” states rather than extrapolating beyond.
However, as just stated, the network did not “blow up” which is a more common manifestation of “out of sample” problems.
The robustness may be connected to the use of the stochastic method with the network, since to exhibit an uncontrolled "blow up" both the network has to be out of sample and the random numbers need to be consistently large (which is statistically unlikely).
We propose that re-training with the surface currents across the whole globe and at different depths, including near boundaries, might eliminate sheared artifacts and potentially address the need for layer-by-layer tuning.
The local resolution was not explicitly encoded in the network and we found that the parameterization returned reduced forcing at finer resolutions so did not adversely modify the finer resolution solutions to a major degree.
We suspect that the network returns weaker forcing at finer and finer resolutions because it is operating on smaller amplitude flow anomalies congruent with finer scales.
This resolution-dependent behavior suggests that the network is ignoring the absolute values of the input velocities.
The network is thus recovering a property of traditional parameterizations that use spatial derivatives.
However, tuning at each resolution suggests a weak nonlinear response to the inputs at different resolutions since we had to moderately scale up the parameterization as we refined the resolution.
We found the optimal scaling as a function of resolution to be relatively predictable and so suspect that scale-awareness is achievable with this parameterization if the network were trained with multiple resolutions.

The network we used is deep ($8$ convolutional layers) and thus has a wide stencil ($21 \times 21$) relative to most lateral spatial operators found in a conventional ocean model.
This proved to not cause much overhead in our model but is nevertheless a consideration since some infrastructure frameworks may not work so easily.
The wide stencil means that the many near-coast ocean points could feel the choice of how “land values” are handled.
Our test results reveal obvious artifacts near the boundary.
Improvement is needed in the treatment of coastlines by this parameterization, and we propose that the parameterization would benefit if the network was trained with the global data with more flow regimes including data near coastlines.
It is the common practice in ocean models to stagger variables in space.
The MOM6 model uses the Arakawa C-grid with flow components normal to the cell used for the continuity budget.
The network was trained with co-located variables (B-grid) and so similarly to online tests in \cite{Guillaumin1&Zanna-JAMES21} we had to interpolate the MOM6 variables to the same point, then interpolate the momentum forcing back.
There is a null-space in this approach; structures near the grid-scale that are neither felt by the parameterization nor influenced by the parameterization.
We did not investigate the consequences of our interpolation choices but recognize there is potentially wasted resolution below the scales that are affected by the parameterization.
The wide stencil and the width of the network (number of channels in the hidden layers) was such that there were $268,005$ weights making the number of floating point operations per grid point per time step very large.
We were able to offload the network inference to GPUs which made the network affordable.
Nevertheless, the wall-clock time spent on the GPUs was still a finite fraction of the wall clock of the model (on CPUs) and so reducing the size of the network will very likely be beneficial.
Given the growing propensity of GPUs, and the challenges of porting existing models to GPUs, utilizing GPUs for machine learned parameterizations seems a viable opportunity \cite{partee2022SmartSim}.
We tackled the inter-language barrier with a lightweight Fortran module (\url{https://github.com/ylikx/forpy}).
There are various solutions available for inter-language coupling and using Forpy we found we had to understand various technical aspects of the hardware, e.g., CPU and GPU configurations.
Turn-key solutions such as {\it SmartSim} that handle much of the technical work will likely prove more and more useful in this arena.
We encountered a hardware constraint that prohibited us from evaluating the ML parameterization for the full-scale realistic model, OM4 \cite{adcroft2019gfdl}, which requires more processors and GPUs than we had available.


The neural network imported, with a quarter of a million weights, is treated as a "black box" in our study; we implicitly trust the parameterization and the calculated weights. 
We can make an analogy with the individual weights used in the polynomial expressions for the Gibbs free-energy of sea-water \cite{feistel2008teos10}; in this case as well, we implicitly trust the authors to have calculated those weights appropriately when we readily use their weights (and software).
When we import a new equation of state, we test the implementation in our model to both evaluate our implementation as well as the new equation of state itself.
Here, we conducted such tests with the neural network backscatter parameterization and made an assessment: the original network performed better than we might have anticipated given that it was trained only on surface data and for limited geographic regions, but there was room for improvement which might need to be assessed in future studies.
Choices made for the network architecture do leave open questions.
For instance, the parameterization calculates a momentum forcing written as a body force and not as the divergence of a stress tensor.
Model developers often rely on integral constraints or conservation principles to test and evaluate their models but this parameterization conserves neither momentum nor energy.
Constraints can be imposed during training as done in \cite{Beuclerconstraint,zanna2020data,Ross-et-al2022}, through a choice of architecture design.  
In addition, other strategies such as post-processing can achieve similar results \cite{Bolton2019}. 
Ensuring such conservation can help model developers during the implementation stage, since the properties of the terms in a conventional closed-form parameterization often lend themselves to analysis, which is undeniably harder here unless imposed.
Despite no direct imposition of property conservation, we find the network used and revised here to show considerable promise, and the exercise of importing into a conventional model to be manageable.
We fully expect to see more widespread use of machine learned parameterizations in the future.

\clearpage
\appendix

\setcounter{figure}{0}

\section{Model equations}\label{appa}
We use the model in an adiabatic limit with no buoyancy forcing which simplifies the equations of motion to the stacked shallow water equations. The equations are written in vector-invariant form as
\ba
 \frac{\partial \bar{\bfu}_k}{\partial t} + \frac{f+\bar{\zeta}_k}{\bar{h}_k} \hat{\bf{z}} \times \bar{h}_k \bar{\bfu}_k + \nabla \bar{K}_k + \nabla \bar{M}_k
&=&  \bar{\mathbf{F}}_k + \mathbf{S}_k \label{eq_appa_1}\\
\frac{\partial \bar{h}_k}{\partial t}+\bar{\nabla}\cdot(\bar{\bfu} \bar{h}_k) &=& 0
\ea
where the overbar is the horizontal filtering and coarse graining, $\bfu_k$ is the horizontal component of velocity, $h_k$ is the layer thickness, $f$ is Coriolis parameter, $\zeta_k$ is the vertical component of the relative vorticity, $K_k=(1/2) \bfu_k\cdot\bfu_k$ is the kinetic energy per unit mass in horizontal, $\hat{\bf{z}}$ is the unit vector pointing in the upward vertical direction, $k$ is the vertical layer index with $k=1$ at the top and $k=N$ at the bottom, $\nabla$ is the horizontal gradient and $\nabla \cdot$ is the horizontal divergence.
$M_k= \sum_{l=1}^kg'_{l-1/2}\eta_{l-1/2}$ is the Montgomery potential, where $g'_{k-1/2}$ is the reduced gravity of each layer, $\eta_{k-1/2}$ is the interface position.
$\mathbf{F}_k=\frac{1}{\rho_0 h_k}(\bft_{k-1/2}-\bft_{k+1/2}) - \nabla \cdot \nu_4 \nabla(\nabla^2\bfu)$ represents the accelerations due to the divergence of stresses including the lateral parameterizations that are not inferred from ML-based models.
$\mathbf{S}_k$, which is defined in equation \eqref{eq2.1_2}, is the subgrid momentum forcing from the machine learned parameterizations. $\rho_0$ is the reference density, $\bft_{k-1/2}$ is the vertical stress, and  $\nabla^2=\nabla \cdot \nabla$ is the horizontal Laplacian.
The turbulence model that we use is a biharmonic friction with a Smagorinsky eddy viscosity following \cite{griffies2000biharmonic}.
The eddy viscosity reads
\be
\nu_4 = C_S \Delta^4 \sqrt{D_T^2+D_S^2}
\ee
where $D_T=\partial_x u-\partial_y v$ and $D_S=\partial_y u+\partial_x v$ (in Cartesian coordinates) are horizontal tension and shearing strain, respectively, $\Delta = \sqrt{\frac{2 (\Delta x)^2 (\Delta y)^2}{(\Delta x)^2 + (\Delta y)^2}}$ is a measure of grid spacing.

\setcounter{figure}{0}

\section*{Software Availability Statement}
The source code of the MOM6 version used for implementing the ML parameterization is accessible through Zenodo \cite{MOM6Forpy_src}, while the CNN model files used for the online evaluation in this study (GZ21) can also be accessed via Zenodo \cite{ForpyCNNGZ21}. To facilitate the setup process for the wind-driven double gyre case in the study, we have made the setup files available online \cite{doublegyre_setup}. 

\acknowledgments
We thank all members of the M$^2$LInES team for helpful discussions and their support throughout this project. We thank Marshall Ward and Wenda Zhang for useful comments on a draft of this manuscript, and Arthur Guillaumin for assistance with the networks.
This research received support through the generosity of Eric and Wendy Schmidt by recommendation of the Schmidt Futures program.
AA was also supported by award NA18OAR4320123, from the National Oceanic and Atmospheric Administration (NOAA), U.S. Department of Commerce and which funded the Princeton Stellar computer resources used for the inference stage of the research.
The statements, findings, conclusions, and recommendations are those of the author(s) and do not necessarily reflect the views of the National Oceanic and Atmospheric Administration, or the U.S. Department of Commerce.
CG was supported by a MacCracken Fellowship.
CFG was partially supported by NSF DMS award 2009752.
This research was also supported in part through the NYU IT High Performance Computing resources, services, and staff expertise.

\clearpage

\bibliography{zhang-etal}

\begin{thebibliography}{}

\bibitem [\protect \citeauthoryear {%
Adcroft%
\ \protect \BOthers {.}}{%
Adcroft%
\ \protect \BOthers {.}}{%
{\protect \APACyear {2019}}%
}]{%
adcroft2019gfdl}
\APACinsertmetastar {%
adcroft2019gfdl}%
\begin{APACrefauthors}%
Adcroft, A.%
, Anderson, W.%
, Balaji, V.%
, Blanton, C.%
, Bushuk, M.%
, Dufour, C\BPBI O.%
\BDBL {}others%
\end{APACrefauthors}%
\unskip\
\newblock
\APACrefYearMonthDay{2019}{}{}.
\newblock
{\BBOQ}\APACrefatitle {The {GFDL} global ocean and sea ice model {OM4. 0}:
  Model description and simulation features} {The {GFDL} global ocean and sea
  ice model {OM4. 0}: Model description and simulation features}.{\BBCQ}
\newblock
\APACjournalVolNumPages{Journal of Advances in Modeling Earth
  Systems}{11}{10}{3167--3211}.
\newblock
\begin{APACrefDOI} \doi{https://doi.org/10.1029/2019MS001726} \end{APACrefDOI}
\PrintBackRefs{\CurrentBib}

\bibitem [\protect \citeauthoryear {%
Anstey%
\ \BBA {} Zanna%
}{%
Anstey%
\ \BBA {} Zanna%
}{%
{\protect \APACyear {2017}}%
}]{%
anstey2017deformation}
\APACinsertmetastar {%
anstey2017deformation}%
\begin{APACrefauthors}%
Anstey, J\BPBI A.%
\BCBT {}\ \BBA {} Zanna, L.%
\end{APACrefauthors}%
\unskip\
\newblock
\APACrefYearMonthDay{2017}{}{}.
\newblock
{\BBOQ}\APACrefatitle {A deformation-based parametrization of ocean mesoscale
  eddy Reynolds stresses} {A deformation-based parametrization of ocean
  mesoscale eddy reynolds stresses}.{\BBCQ}
\newblock
\APACjournalVolNumPages{Ocean Modelling}{112}{}{99--111}.
\newblock
\begin{APACrefDOI} \doi{https://doi.org/10.1016/j.ocemod.2017.02.004}
  \end{APACrefDOI}
\PrintBackRefs{\CurrentBib}

\bibitem [\protect \citeauthoryear {%
Balwada%
, Xie%
, Marino%
\BCBL {}\ \BBA {} Feraco%
}{%
Balwada%
\ \protect \BOthers {.}}{%
{\protect \APACyear {2022}}%
}]{%
balwada2022direct}
\APACinsertmetastar {%
balwada2022direct}%
\begin{APACrefauthors}%
Balwada, D.%
, Xie, J\BHBI H.%
, Marino, R.%
\BCBL {}\ \BBA {} Feraco, F.%
\end{APACrefauthors}%
\unskip\
\newblock
\APACrefYearMonthDay{2022}{}{}.
\newblock
{\BBOQ}\APACrefatitle {Direct observational evidence of an oceanic dual kinetic
  energy cascade and its seasonality} {Direct observational evidence of an
  oceanic dual kinetic energy cascade and its seasonality}.{\BBCQ}
\newblock
\APACjournalVolNumPages{arXiv preprint arXiv:2202.08637}{}{}{}.
\newblock
\begin{APACrefDOI} \doi{https://doi.org/10.48550/arXiv.2202.08637}
  \end{APACrefDOI}
\PrintBackRefs{\CurrentBib}

\bibitem [\protect \citeauthoryear {%
Beck%
\ \BBA {} Kurz%
}{%
Beck%
\ \BBA {} Kurz%
}{%
{\protect \APACyear {2021}}%
}]{%
beck2021CFDparameter}
\APACinsertmetastar {%
beck2021CFDparameter}%
\begin{APACrefauthors}%
Beck, A.%
\BCBT {}\ \BBA {} Kurz, M.%
\end{APACrefauthors}%
\unskip\
\newblock
\APACrefYearMonthDay{2021}{}{}.
\newblock
{\BBOQ}\APACrefatitle {A perspective on machine learning methods in turbulence
  modeling} {A perspective on machine learning methods in turbulence
  modeling}.{\BBCQ}
\newblock
\APACjournalVolNumPages{GAMM-Mitteilungen}{44}{1}{e202100002}.
\newblock
\begin{APACrefDOI} \doi{https://doi.org/10.1002/gamm.202100002}
  \end{APACrefDOI}
\PrintBackRefs{\CurrentBib}

\bibitem [\protect \citeauthoryear {%
Berner%
, Shutts%
, Leutbecher%
\BCBL {}\ \BBA {} Palmer%
}{%
Berner%
\ \protect \BOthers {.}}{%
{\protect \APACyear {2009}}%
}]{%
berner2009spectral}
\APACinsertmetastar {%
berner2009spectral}%
\begin{APACrefauthors}%
Berner, J.%
, Shutts, G.%
, Leutbecher, M.%
\BCBL {}\ \BBA {} Palmer, T.%
\end{APACrefauthors}%
\unskip\
\newblock
\APACrefYearMonthDay{2009}{}{}.
\newblock
{\BBOQ}\APACrefatitle {A spectral stochastic kinetic energy backscatter scheme
  and its impact on flow-dependent predictability in the ECMWF ensemble
  prediction system} {A spectral stochastic kinetic energy backscatter scheme
  and its impact on flow-dependent predictability in the ecmwf ensemble
  prediction system}.{\BBCQ}
\newblock
\APACjournalVolNumPages{Journal of the Atmospheric Sciences}{66}{3}{603--626}.
\newblock
\begin{APACrefDOI} \doi{https://doi.org/10.1175/2008jas2677.1} \end{APACrefDOI}
\PrintBackRefs{\CurrentBib}

\bibitem [\protect \citeauthoryear {%
Beucler%
, Pritchard%
, Rasp%
\BCBL {}\ \protect \BOthers {.}}{%
Beucler%
, Pritchard%
, Rasp%
\BCBL {}\ \protect \BOthers {.}}{%
{\protect \APACyear {2021}}%
}]{%
Beuclerconstraint}
\APACinsertmetastar {%
Beuclerconstraint}%
\begin{APACrefauthors}%
Beucler, T.%
, Pritchard, M.%
, Rasp, S.%
, Ott, J.%
, Baldi, P.%
\BCBL {}\ \BBA {} Gentine, P.%
\end{APACrefauthors}%
\unskip\
\newblock
\APACrefYearMonthDay{2021}{Mar}{}.
\newblock
{\BBOQ}\APACrefatitle {Enforcing Analytic Constraints in Neural Networks
  Emulating Physical Systems} {Enforcing analytic constraints in neural
  networks emulating physical systems}.{\BBCQ}
\newblock
\APACjournalVolNumPages{Phys. Rev. Lett.}{126}{}{098302}.
\newblock
\begin{APACrefURL}
  \url{https://link.aps.org/doi/10.1103/PhysRevLett.126.098302}
  \end{APACrefURL}
\newblock
\begin{APACrefDOI} \doi{10.1103/PhysRevLett.126.098302} \end{APACrefDOI}
\PrintBackRefs{\CurrentBib}

\bibitem [\protect \citeauthoryear {%
Beucler%
, Pritchard%
, Yuval%
\BCBL {}\ \protect \BOthers {.}}{%
Beucler%
, Pritchard%
, Yuval%
\BCBL {}\ \protect \BOthers {.}}{%
{\protect \APACyear {2021}}%
}]{%
beucler2021climate}
\APACinsertmetastar {%
beucler2021climate}%
\begin{APACrefauthors}%
Beucler, T.%
, Pritchard, M.%
, Yuval, J.%
, Gupta, A.%
, Peng, L.%
, Rasp, S.%
\BDBL {}others%
\end{APACrefauthors}%
\unskip\
\newblock
\APACrefYearMonthDay{2021}{}{}.
\newblock
{\BBOQ}\APACrefatitle {Climate-Invariant Machine Learning} {Climate-invariant
  machine learning}.{\BBCQ}
\newblock
\APACjournalVolNumPages{arXiv preprint arXiv:2112.08440}{}{}{}.
\newblock
\begin{APACrefDOI} \doi{https://doi.org/10.48550/arXiv.2112.08440}
  \end{APACrefDOI}
\PrintBackRefs{\CurrentBib}

\bibitem [\protect \citeauthoryear {%
Bolton%
\ \BBA {} Zanna%
}{%
Bolton%
\ \BBA {} Zanna%
}{%
{\protect \APACyear {2019}}%
}]{%
Bolton2019}
\APACinsertmetastar {%
Bolton2019}%
\begin{APACrefauthors}%
Bolton, T.%
\BCBT {}\ \BBA {} Zanna, L.%
\end{APACrefauthors}%
\unskip\
\newblock
\APACrefYearMonthDay{2019}{}{}.
\newblock
{\BBOQ}\APACrefatitle {{Applications of Deep Learning to Ocean Data Inference
  and Subgrid Parameterization}} {{Applications of Deep Learning to Ocean Data
  Inference and Subgrid Parameterization}}.{\BBCQ}
\newblock
\APACjournalVolNumPages{Journal of Advances in Modeling Earth
  Systems}{11}{1}{376--399}.
\newblock
\begin{APACrefDOI} \doi{10.1029/2018MS001472} \end{APACrefDOI}
\PrintBackRefs{\CurrentBib}

\bibitem [\protect \citeauthoryear {%
Brenowitz%
\ \BBA {} Bretherton%
}{%
Brenowitz%
\ \BBA {} Bretherton%
}{%
{\protect \APACyear {2018}}%
}]{%
brenowitz2018GRL}
\APACinsertmetastar {%
brenowitz2018GRL}%
\begin{APACrefauthors}%
Brenowitz, N\BPBI D.%
\BCBT {}\ \BBA {} Bretherton, C\BPBI S.%
\end{APACrefauthors}%
\unskip\
\newblock
\APACrefYearMonthDay{2018}{}{}.
\newblock
{\BBOQ}\APACrefatitle {Prognostic validation of a neural network unified
  physics parameterization} {Prognostic validation of a neural network unified
  physics parameterization}.{\BBCQ}
\newblock
\APACjournalVolNumPages{Geophysical Research Letters}{45}{12}{6289--6298}.
\newblock
\begin{APACrefDOI} \doi{https://doi.org/10.1029/2020GL091363} \end{APACrefDOI}
\PrintBackRefs{\CurrentBib}

\bibitem [\protect \citeauthoryear {%
Christensen%
\ \BBA {} Zanna%
}{%
Christensen%
\ \BBA {} Zanna%
}{%
{\protect \APACyear {2022}}%
}]{%
christensen2022parametrization}
\APACinsertmetastar {%
christensen2022parametrization}%
\begin{APACrefauthors}%
Christensen, H.%
\BCBT {}\ \BBA {} Zanna, L.%
\end{APACrefauthors}%
\unskip\
\newblock
\APACrefYearMonthDay{2022}{}{}.
\newblock
{\BBOQ}\APACrefatitle {Parametrization in Weather and Climate Models}
  {Parametrization in weather and climate models}.{\BBCQ}
\newblock
\BIn{} \APACrefbtitle {Oxford Research Encyclopedia of Climate Science.}
  {Oxford research encyclopedia of climate science.}
\newblock
\begin{APACrefDOI} \doi{https://doi.org/10.1093/acrefore/9780190228620.013.826}
  \end{APACrefDOI}
\PrintBackRefs{\CurrentBib}

\bibitem [\protect \citeauthoryear {%
Curcic%
}{%
Curcic%
}{%
{\protect \APACyear {2019}}%
}]{%
curcic2019neuralFortran}
\APACinsertmetastar {%
curcic2019neuralFortran}%
\begin{APACrefauthors}%
Curcic, M.%
\end{APACrefauthors}%
\unskip\
\newblock
\APACrefYearMonthDay{2019}{}{}.
\newblock
{\BBOQ}\APACrefatitle {A parallel Fortran framework for neural networks and
  deep learning} {A parallel fortran framework for neural networks and deep
  learning}.{\BBCQ}
\newblock
\BIn{} \APACrefbtitle {Acm sigplan fortran forum} {Acm sigplan fortran forum}\
  (\BVOL~38, \BPGS\ 4--21).
\newblock
\begin{APACrefDOI} \doi{https://doi.org/10.1145/3323057.3323059}
  \end{APACrefDOI}
\PrintBackRefs{\CurrentBib}

\bibitem [\protect \citeauthoryear {%
Delman%
\ \BBA {} Lee%
}{%
Delman%
\ \BBA {} Lee%
}{%
{\protect \APACyear {2021}}%
}]{%
delman2021global}
\APACinsertmetastar {%
delman2021global}%
\begin{APACrefauthors}%
Delman, A.%
\BCBT {}\ \BBA {} Lee, T.%
\end{APACrefauthors}%
\unskip\
\newblock
\APACrefYearMonthDay{2021}{}{}.
\newblock
{\BBOQ}\APACrefatitle {Global contributions of mesoscale dynamics to meridional
  heat transport} {Global contributions of mesoscale dynamics to meridional
  heat transport}.{\BBCQ}
\newblock
\APACjournalVolNumPages{Ocean Science}{17}{4}{1031--1052}.
\newblock
\begin{APACrefDOI} \doi{https://doi.org/10.5194/os-17-1031-2021}
  \end{APACrefDOI}
\PrintBackRefs{\CurrentBib}

\bibitem [\protect \citeauthoryear {%
Espinosa%
, Sheshadri%
, Cain%
, Gerber%
\BCBL {}\ \BBA {} DallaSanta%
}{%
Espinosa%
\ \protect \BOthers {.}}{%
{\protect \APACyear {2022}}%
}]{%
Espinosa_etal-GRL2022}
\APACinsertmetastar {%
Espinosa_etal-GRL2022}%
\begin{APACrefauthors}%
Espinosa, Z\BPBI I.%
, Sheshadri, A.%
, Cain, G\BPBI R.%
, Gerber, E\BPBI P.%
\BCBL {}\ \BBA {} DallaSanta, K\BPBI J.%
\end{APACrefauthors}%
\unskip\
\newblock
\APACrefYearMonthDay{2022}{apr}{}.
\newblock
{\BBOQ}\APACrefatitle {Machine Learning Gravity Wave Parameterization
  Generalizes to Capture the {QBO} and Response to Increased {CO}$_2$} {Machine
  learning gravity wave parameterization generalizes to capture the {QBO} and
  response to increased {CO}$_2$}.{\BBCQ}
\newblock
\APACjournalVolNumPages{Geophysical Research Letters}{49}{8}{}.
\newblock
\begin{APACrefDOI} \doi{https://doi.org/10.1029/2022gl098174} \end{APACrefDOI}
\PrintBackRefs{\CurrentBib}

\bibitem [\protect \citeauthoryear {%
Feistel%
}{%
Feistel%
}{%
{\protect \APACyear {2008}}%
}]{%
feistel2008teos10}
\APACinsertmetastar {%
feistel2008teos10}%
\begin{APACrefauthors}%
Feistel, R.%
\end{APACrefauthors}%
\unskip\
\newblock
\APACrefYearMonthDay{2008}{}{}.
\newblock
{\BBOQ}\APACrefatitle {A {Gibbs} function for seawater thermodynamics for $-6$
  to $80^\circ ${C} and salinity up to $120$ g kg$^{-1}$} {A {Gibbs} function
  for seawater thermodynamics for $-6$ to $80^\circ ${C} and salinity up to
  $120$ g kg$^{-1}$}.{\BBCQ}
\newblock
\APACjournalVolNumPages{Deep Sea Research Part I: Oceanographic Research
  Papers}{55}{12}{1639-1671}.
\newblock
\begin{APACrefDOI} \doi{https://doi.org/10.1016/j.dsr.2008.07.004}
  \end{APACrefDOI}
\PrintBackRefs{\CurrentBib}

\bibitem [\protect \citeauthoryear {%
Frederiksen%
\ \BBA {} Davies%
}{%
Frederiksen%
\ \BBA {} Davies%
}{%
{\protect \APACyear {1997}}%
}]{%
frederiksen1997eddy}
\APACinsertmetastar {%
frederiksen1997eddy}%
\begin{APACrefauthors}%
Frederiksen, J\BPBI S.%
\BCBT {}\ \BBA {} Davies, A\BPBI G.%
\end{APACrefauthors}%
\unskip\
\newblock
\APACrefYearMonthDay{1997}{}{}.
\newblock
{\BBOQ}\APACrefatitle {Eddy viscosity and stochastic backscatter
  parameterizations on the sphere for atmospheric circulation models} {Eddy
  viscosity and stochastic backscatter parameterizations on the sphere for
  atmospheric circulation models}.{\BBCQ}
\newblock
\APACjournalVolNumPages{Journal of the atmospheric
  sciences}{54}{20}{2475--2492}.
\newblock
\begin{APACrefDOI}
  \doi{https://doi.org/10.1175/1520-0469(1997)054<2475:evasbp>2.0.co;2}
  \end{APACrefDOI}
\PrintBackRefs{\CurrentBib}

\bibitem [\protect \citeauthoryear {%
Gent%
, Willebrand%
, McDougall%
\BCBL {}\ \BBA {} McWilliams%
}{%
Gent%
\ \protect \BOthers {.}}{%
{\protect \APACyear {1995}}%
}]{%
gent_parameterizing_1995}
\APACinsertmetastar {%
gent_parameterizing_1995}%
\begin{APACrefauthors}%
Gent, P\BPBI R.%
, Willebrand, J.%
, McDougall, T\BPBI J.%
\BCBL {}\ \BBA {} McWilliams, J\BPBI C.%
\end{APACrefauthors}%
\unskip\
\newblock
\APACrefYearMonthDay{1995}{{\APACmonth{04}}}{}.
\newblock
{\BBOQ}\APACrefatitle {Parameterizing {Eddy}-{Induced} {Tracer} {Transports} in
  {Ocean} {Circulation} {Models}} {Parameterizing {Eddy}-{Induced} {Tracer}
  {Transports} in {Ocean} {Circulation} {Models}}.{\BBCQ}
\newblock
\APACjournalVolNumPages{Journal of Physical Oceanography}{25}{4}{463--474}.
\newblock
\begin{APACrefDOI} \doi{10.1175/1520-0485(1995)025<0463:PEITTI>2.0.CO;2}
  \end{APACrefDOI}
\PrintBackRefs{\CurrentBib}

\bibitem [\protect \citeauthoryear {%
Gill%
\ \BBA {} Adrian%
}{%
Gill%
\ \BBA {} Adrian%
}{%
{\protect \APACyear {1982}}%
}]{%
gill1982atmosphere}
\APACinsertmetastar {%
gill1982atmosphere}%
\begin{APACrefauthors}%
Gill, A\BPBI E.%
\BCBT {}\ \BBA {} Adrian, E.%
\end{APACrefauthors}%
\unskip\
\newblock
\APACrefYear{1982}.
\newblock
\APACrefbtitle {Atmosphere-ocean dynamics} {Atmosphere-ocean dynamics}\
  (\BVOL~30).
\newblock
\APACaddressPublisher{}{Academic press}.
\PrintBackRefs{\CurrentBib}

\bibitem [\protect \citeauthoryear {%
Greatbatch%
, Zhai%
, Claus%
, Czeschel%
\BCBL {}\ \BBA {} Rath%
}{%
Greatbatch%
\ \protect \BOthers {.}}{%
{\protect \APACyear {2010}}%
}]{%
greatbatch2010transport}
\APACinsertmetastar {%
greatbatch2010transport}%
\begin{APACrefauthors}%
Greatbatch, R.%
, Zhai, X.%
, Claus, M.%
, Czeschel, L.%
\BCBL {}\ \BBA {} Rath, W.%
\end{APACrefauthors}%
\unskip\
\newblock
\APACrefYearMonthDay{2010}{}{}.
\newblock
{\BBOQ}\APACrefatitle {Transport driven by eddy momentum fluxes in the Gulf
  Stream Extension region} {Transport driven by eddy momentum fluxes in the
  gulf stream extension region}.{\BBCQ}
\newblock
\APACjournalVolNumPages{Geophysical Research Letters}{37}{24}{}.
\newblock
\begin{APACrefDOI} \doi{https://doi.org/10.1029/2010gl045473} \end{APACrefDOI}
\PrintBackRefs{\CurrentBib}

\bibitem [\protect \citeauthoryear {%
Griffies%
\ \protect \BOthers {.}}{%
Griffies%
\ \protect \BOthers {.}}{%
{\protect \APACyear {1998}}%
}]{%
griffies_isoneutral_1998}
\APACinsertmetastar {%
griffies_isoneutral_1998}%
\begin{APACrefauthors}%
Griffies, S\BPBI M.%
, Gnanadesikan, A.%
, Pacanowski, R\BPBI C.%
, Larichev, V\BPBI D.%
, Dukowicz, J\BPBI K.%
\BCBL {}\ \BBA {} Smith, R\BPBI D.%
\end{APACrefauthors}%
\unskip\
\newblock
\APACrefYearMonthDay{1998}{{\APACmonth{05}}}{}.
\newblock
{\BBOQ}\APACrefatitle {Isoneutral {Diffusion} in a z-{Coordinate} {Ocean}
  {Model}} {Isoneutral {Diffusion} in a z-{Coordinate} {Ocean} {Model}}.{\BBCQ}
\newblock
\APACjournalVolNumPages{Journal of Physical Oceanography}{28}{5}{805--830}.
\newblock
\APACrefnote{Publisher: American Meteorological Society Section: Journal of
  Physical Oceanography}
\newblock
\begin{APACrefDOI} \doi{10.1175/1520-0485(1998)028<0805:IDIAZC>2.0.CO;2}
  \end{APACrefDOI}
\PrintBackRefs{\CurrentBib}

\bibitem [\protect \citeauthoryear {%
Griffies%
\ \BBA {} Hallberg%
}{%
Griffies%
\ \BBA {} Hallberg%
}{%
{\protect \APACyear {2000}}%
}]{%
griffies2000biharmonic}
\APACinsertmetastar {%
griffies2000biharmonic}%
\begin{APACrefauthors}%
Griffies, S\BPBI M.%
\BCBT {}\ \BBA {} Hallberg, R\BPBI W.%
\end{APACrefauthors}%
\unskip\
\newblock
\APACrefYearMonthDay{2000}{}{}.
\newblock
{\BBOQ}\APACrefatitle {Biharmonic friction with a {Smagorinsky-like} viscosity
  for use in large-scale eddy-permitting ocean models} {Biharmonic friction
  with a {Smagorinsky-like} viscosity for use in large-scale eddy-permitting
  ocean models}.{\BBCQ}
\newblock
\APACjournalVolNumPages{Monthly Weather Review}{128}{8}{2935--2946}.
\newblock
\begin{APACrefDOI}
  \doi{https://doi.org/10.1175/1520-0493(2000)128<2935:bfwasl>2.0.co;2}
  \end{APACrefDOI}
\PrintBackRefs{\CurrentBib}

\bibitem [\protect \citeauthoryear {%
Griffies%
\ \protect \BOthers {.}}{%
Griffies%
\ \protect \BOthers {.}}{%
{\protect \APACyear {2015}}%
}]{%
Griffies_etal-JC15}
\APACinsertmetastar {%
Griffies_etal-JC15}%
\begin{APACrefauthors}%
Griffies, S\BPBI M.%
, Winton, M.%
, Anderson, W\BPBI G.%
, Benson, R.%
, Delworth, T\BPBI L.%
, Dufour, C\BPBI O.%
\BDBL {}Zhang, R.%
\end{APACrefauthors}%
\unskip\
\newblock
\APACrefYearMonthDay{2015}{feb}{}.
\newblock
{\BBOQ}\APACrefatitle {Impacts on Ocean Heat from Transient Mesoscale Eddies in
  a Hierarchy of Climate Models} {Impacts on ocean heat from transient
  mesoscale eddies in a hierarchy of climate models}.{\BBCQ}
\newblock
\APACjournalVolNumPages{Journal of Climate}{28}{3}{952--977}.
\newblock
\begin{APACrefDOI} \doi{https://doi.org/10.1175/jcli-d-14-00353.1}
  \end{APACrefDOI}
\PrintBackRefs{\CurrentBib}

\bibitem [\protect \citeauthoryear {%
Guillaumin%
\ \BBA {} Zanna%
}{%
Guillaumin%
\ \BBA {} Zanna%
}{%
{\protect \APACyear {2021}}%
}]{%
Guillaumin1&Zanna-JAMES21}
\APACinsertmetastar {%
Guillaumin1&Zanna-JAMES21}%
\begin{APACrefauthors}%
Guillaumin, A.%
\BCBT {}\ \BBA {} Zanna, L.%
\end{APACrefauthors}%
\unskip\
\newblock
\APACrefYearMonthDay{2021}{mar}{}.
\newblock
{\BBOQ}\APACrefatitle {Stochastic Deep Learning parameterization of Ocean
  Momentum Forcing} {Stochastic deep learning parameterization of ocean
  momentum forcing}.{\BBCQ}
\newblock
\APACjournalVolNumPages{Journal of Advances in Modeling Earth Systems}{}{}{}.
\newblock
\begin{APACrefDOI} \doi{https://doi.org/10.1002/essoar.10506419.1}
  \end{APACrefDOI}
\PrintBackRefs{\CurrentBib}

\bibitem [\protect \citeauthoryear {%
Hallberg%
}{%
Hallberg%
}{%
{\protect \APACyear {2013}}%
}]{%
Hallberg-2013}
\APACinsertmetastar {%
Hallberg-2013}%
\begin{APACrefauthors}%
Hallberg, R.%
\end{APACrefauthors}%
\unskip\
\newblock
\APACrefYearMonthDay{2013}{}{}.
\newblock
{\BBOQ}\APACrefatitle {Using a resolution function to regulate
  parameterizations of oceanic mesoscale eddy effects} {Using a resolution
  function to regulate parameterizations of oceanic mesoscale eddy
  effects}.{\BBCQ}
\newblock
\APACjournalVolNumPages{Ocean Modelling}{72}{}{92-103}.
\newblock
\begin{APACrefDOI} \doi{https://doi.org/10.1016/j.ocemod.2013.08.007}
  \end{APACrefDOI}
\PrintBackRefs{\CurrentBib}

\bibitem [\protect \citeauthoryear {%
Hallberg%
\ \protect \BOthers {.}}{%
Hallberg%
\ \protect \BOthers {.}}{%
{\protect \APACyear {2023}}%
}]{%
MOM6Forpy_src}
\APACinsertmetastar {%
MOM6Forpy_src}%
\begin{APACrefauthors}%
Hallberg, R.%
, Adcroft, A.%
, Marques, G.%
, Ward, M.%
, Hedstrom, K.%
, Shao, A.%
\BDBL {}Stern, A.%
\end{APACrefauthors}%
\unskip\
\newblock
\APACrefYearMonthDay{2023}{{\APACmonth{02}}}{}.
\newblock
\APACrefbtitle {chzhangudel/MOM6: Initial Release
  (forpy2/2023.02.22)[Software].} {chzhangudel/mom6: Initial release
  (forpy2/2023.02.22)[software].}
\newblock
\APACaddressPublisher{}{Zenodo}.
\newblock
\begin{APACrefDOI} \doi{10.5281/zenodo.7663075} \end{APACrefDOI}
\PrintBackRefs{\CurrentBib}

\bibitem [\protect \citeauthoryear {%
Hallberg%
\ \BBA {} Rhines%
}{%
Hallberg%
\ \BBA {} Rhines%
}{%
{\protect \APACyear {2000}}%
}]{%
hallberg_rhines_2000}
\APACinsertmetastar {%
hallberg_rhines_2000}%
\begin{APACrefauthors}%
Hallberg, R.%
\BCBT {}\ \BBA {} Rhines, P\BPBI B.%
\end{APACrefauthors}%
\unskip\
\newblock
\APACrefYearMonthDay{2000}{}{}.
\newblock
{\BBOQ}\APACrefatitle {Boundary Sources of Potential Vorticity in Geophysical
  Circulations} {Boundary sources of potential vorticity in geophysical
  circulations}.{\BBCQ}
\newblock
\BIn{} R\BPBI M.~Kerr\ \BBA {} Y.~Kimura\ (\BEDS), \APACrefbtitle {{IUTAM}
  {Symposium} on {Developments} in {Geophysical} {Turbulence}} {{IUTAM}
  {Symposium} on {Developments} in {Geophysical} {Turbulence}}\ (\BPGS\
  51--65).
\newblock
\APACaddressPublisher{Dordrecht}{Springer Netherlands}.
\newblock
\begin{APACrefDOI} \doi{https://doi.org/10.1007/978-94-010-0928-7_5}
  \end{APACrefDOI}
\PrintBackRefs{\CurrentBib}

\bibitem [\protect \citeauthoryear {%
He%
\ \BBA {} Sun%
}{%
He%
\ \BBA {} Sun%
}{%
{\protect \APACyear {2014}}%
}]{%
he2015CNNcost}
\APACinsertmetastar {%
he2015CNNcost}%
\begin{APACrefauthors}%
He, K.%
\BCBT {}\ \BBA {} Sun, J.%
\end{APACrefauthors}%
\unskip\
\newblock
\APACrefYearMonthDay{2014}{}{}.
\newblock
\APACrefbtitle {Convolutional Neural Networks at Constrained Time Cost.}
  {Convolutional neural networks at constrained time cost.}
\newblock
\begin{APACrefDOI} \doi{https://arxiv.org/abs/1412.1710} \end{APACrefDOI}
\PrintBackRefs{\CurrentBib}

\bibitem [\protect \citeauthoryear {%
Hewitt%
\ \protect \BOthers {.}}{%
Hewitt%
\ \protect \BOthers {.}}{%
{\protect \APACyear {2020}}%
}]{%
hewitt2020resolving}
\APACinsertmetastar {%
hewitt2020resolving}%
\begin{APACrefauthors}%
Hewitt, H\BPBI T.%
, Roberts, M.%
, Mathiot, P.%
, Biastoch, A.%
, Blockley, E.%
, Chassignet, E\BPBI P.%
\BDBL {}others%
\end{APACrefauthors}%
\unskip\
\newblock
\APACrefYearMonthDay{2020}{}{}.
\newblock
{\BBOQ}\APACrefatitle {Resolving and parameterising the ocean mesoscale in
  earth system models} {Resolving and parameterising the ocean mesoscale in
  earth system models}.{\BBCQ}
\newblock
\APACjournalVolNumPages{Current Climate Change Reports}{6}{4}{137--152}.
\newblock
\begin{APACrefDOI} \doi{https://doi.org/10.5281/zenodo.3685918}
  \end{APACrefDOI}
\PrintBackRefs{\CurrentBib}

\bibitem [\protect \citeauthoryear {%
Jansen%
\ \BBA {} Held%
}{%
Jansen%
\ \BBA {} Held%
}{%
{\protect \APACyear {2014}}%
}]{%
jansen2014parameterizing}
\APACinsertmetastar {%
jansen2014parameterizing}%
\begin{APACrefauthors}%
Jansen, M\BPBI F.%
\BCBT {}\ \BBA {} Held, I\BPBI M.%
\end{APACrefauthors}%
\unskip\
\newblock
\APACrefYearMonthDay{2014}{}{}.
\newblock
{\BBOQ}\APACrefatitle {Parameterizing subgrid-scale eddy effects using
  energetically consistent backscatter} {Parameterizing subgrid-scale eddy
  effects using energetically consistent backscatter}.{\BBCQ}
\newblock
\APACjournalVolNumPages{Ocean Modelling}{80}{}{36--48}.
\newblock
\begin{APACrefDOI} \doi{https://doi.org/10.1016/j.ocemod.2014.06.002}
  \end{APACrefDOI}
\PrintBackRefs{\CurrentBib}

\bibitem [\protect \citeauthoryear {%
Juricke%
, Danilov%
, Koldunov%
, Oliver%
\BCBL {}\ \BBA {} Sidorenko%
}{%
Juricke%
\ \protect \BOthers {.}}{%
{\protect \APACyear {2020}}%
}]{%
juricke2020ocean}
\APACinsertmetastar {%
juricke2020ocean}%
\begin{APACrefauthors}%
Juricke, S.%
, Danilov, S.%
, Koldunov, N.%
, Oliver, M.%
\BCBL {}\ \BBA {} Sidorenko, D.%
\end{APACrefauthors}%
\unskip\
\newblock
\APACrefYearMonthDay{2020}{}{}.
\newblock
{\BBOQ}\APACrefatitle {Ocean kinetic energy backscatter parametrization on
  unstructured grids: Impact on global eddy-permitting simulations} {Ocean
  kinetic energy backscatter parametrization on unstructured grids: Impact on
  global eddy-permitting simulations}.{\BBCQ}
\newblock
\APACjournalVolNumPages{Journal of Advances in Modeling Earth
  Systems}{12}{1}{e2019MS001855}.
\newblock
\begin{APACrefDOI} \doi{https://doi.org/10.1029/2019ms001855} \end{APACrefDOI}
\PrintBackRefs{\CurrentBib}

\bibitem [\protect \citeauthoryear {%
Juricke%
, Palmer%
\BCBL {}\ \BBA {} Zanna%
}{%
Juricke%
\ \protect \BOthers {.}}{%
{\protect \APACyear {2017}}%
}]{%
juricke2017stochastic}
\APACinsertmetastar {%
juricke2017stochastic}%
\begin{APACrefauthors}%
Juricke, S.%
, Palmer, T\BPBI N.%
\BCBL {}\ \BBA {} Zanna, L.%
\end{APACrefauthors}%
\unskip\
\newblock
\APACrefYearMonthDay{2017}{}{}.
\newblock
{\BBOQ}\APACrefatitle {Stochastic subgrid-scale ocean mixing: impacts on
  low-frequency variability} {Stochastic subgrid-scale ocean mixing: impacts on
  low-frequency variability}.{\BBCQ}
\newblock
\APACjournalVolNumPages{Journal of Climate}{30}{13}{4997--5019}.
\PrintBackRefs{\CurrentBib}

\bibitem [\protect \citeauthoryear {%
Kjellsson%
\ \BBA {} Zanna%
}{%
Kjellsson%
\ \BBA {} Zanna%
}{%
{\protect \APACyear {2017}}%
}]{%
kjellsson2017impact}
\APACinsertmetastar {%
kjellsson2017impact}%
\begin{APACrefauthors}%
Kjellsson, J.%
\BCBT {}\ \BBA {} Zanna, L.%
\end{APACrefauthors}%
\unskip\
\newblock
\APACrefYearMonthDay{2017}{}{}.
\newblock
{\BBOQ}\APACrefatitle {The impact of horizontal resolution on energy transfers
  in global ocean models} {The impact of horizontal resolution on energy
  transfers in global ocean models}.{\BBCQ}
\newblock
\APACjournalVolNumPages{Fluids}{2}{3}{45}.
\newblock
\begin{APACrefDOI} \doi{https://doi.org/10.3390/fluids2030045} \end{APACrefDOI}
\PrintBackRefs{\CurrentBib}

\bibitem [\protect \citeauthoryear {%
Krasnopolsky%
, Fox-Rabinovitz%
, Hou%
, Lord%
\BCBL {}\ \BBA {} Belochitski%
}{%
Krasnopolsky%
\ \protect \BOthers {.}}{%
{\protect \APACyear {2010}}%
}]{%
krasnopolsky2010accurate}
\APACinsertmetastar {%
krasnopolsky2010accurate}%
\begin{APACrefauthors}%
Krasnopolsky, V.%
, Fox-Rabinovitz, M.%
, Hou, Y.%
, Lord, S.%
\BCBL {}\ \BBA {} Belochitski, A.%
\end{APACrefauthors}%
\unskip\
\newblock
\APACrefYearMonthDay{2010}{}{}.
\newblock
{\BBOQ}\APACrefatitle {Accurate and fast neural network emulations of model
  radiation for the {NCEP} coupled climate forecast system: Climate simulations
  and seasonal predictions} {Accurate and fast neural network emulations of
  model radiation for the {NCEP} coupled climate forecast system: Climate
  simulations and seasonal predictions}.{\BBCQ}
\newblock
\APACjournalVolNumPages{Monthly Weather Review}{138}{5}{1822--1842}.
\newblock
\begin{APACrefDOI} \doi{https://doi.org/10.1175/2009mwr3149.1} \end{APACrefDOI}
\PrintBackRefs{\CurrentBib}

\bibitem [\protect \citeauthoryear {%
Legg%
\ \protect \BOthers {.}}{%
Legg%
\ \protect \BOthers {.}}{%
{\protect \APACyear {2009}}%
}]{%
legg_improving_2009}
\APACinsertmetastar {%
legg_improving_2009}%
\begin{APACrefauthors}%
Legg, S.%
, Briegleb, B.%
, Chang, Y.%
, Chassignet, E\BPBI P.%
, Danabasoglu, G.%
, Ezer, T.%
\BDBL {}Yang, J.%
\end{APACrefauthors}%
\unskip\
\newblock
\APACrefYearMonthDay{2009}{{\APACmonth{05}}}{}.
\newblock
{\BBOQ}\APACrefatitle {Improving Oceanic Overflow Representation in Climate
  Models: The Gravity Current Entrainment Climate Process Team} {Improving
  oceanic overflow representation in climate models: The gravity current
  entrainment climate process team}.{\BBCQ}
\newblock
\APACjournalVolNumPages{Bulletin of the American Meteorological
  Society}{90}{5}{657--670}.
\newblock
\begin{APACrefDOI} \doi{10.1175/2008BAMS2667.1} \end{APACrefDOI}
\PrintBackRefs{\CurrentBib}

\bibitem [\protect \citeauthoryear {%
Liu%
\ \protect \BOthers {.}}{%
Liu%
\ \protect \BOthers {.}}{%
{\protect \APACyear {2021}}%
}]{%
liu2021gas}
\APACinsertmetastar {%
liu2021gas}%
\begin{APACrefauthors}%
Liu, C.%
, Zhang, H.%
, Cheng, Z.%
, Shen, J.%
, Zhao, J.%
, Wang, Y.%
\BDBL {}Cheng, Y.%
\end{APACrefauthors}%
\unskip\
\newblock
\APACrefYearMonthDay{2021}{}{}.
\newblock
{\BBOQ}\APACrefatitle {Emulation of an atmospheric gas-phase chemistry solver
  through deep learning: Case study of Chinese Mainland} {Emulation of an
  atmospheric gas-phase chemistry solver through deep learning: Case study of
  chinese mainland}.{\BBCQ}
\newblock
\APACjournalVolNumPages{Atmospheric Pollution Research}{12}{6}{101079}.
\newblock
\begin{APACrefDOI} \doi{https://doi.org/10.1016/j.apr.2021.101079}
  \end{APACrefDOI}
\PrintBackRefs{\CurrentBib}

\bibitem [\protect \citeauthoryear {%
MacKinnon%
\ \protect \BOthers {.}}{%
MacKinnon%
\ \protect \BOthers {.}}{%
{\protect \APACyear {2017}}%
}]{%
mackinnon_climate_2017}
\APACinsertmetastar {%
mackinnon_climate_2017}%
\begin{APACrefauthors}%
MacKinnon, J\BPBI A.%
, Zhao, Z.%
, Whalen, C\BPBI B.%
, Waterhouse, A\BPBI F.%
, Trossman, D\BPBI S.%
, Sun, O\BPBI M.%
\BDBL {}Alford, M\BPBI H.%
\end{APACrefauthors}%
\unskip\
\newblock
\APACrefYearMonthDay{2017}{{\APACmonth{11}}}{}.
\newblock
{\BBOQ}\APACrefatitle {Climate Process Team on Internal Wave–Driven Ocean
  Mixing} {Climate process team on internal wave–driven ocean mixing}.{\BBCQ}
\newblock
\APACjournalVolNumPages{Bulletin of the American Meteorological
  Society}{98}{11}{2429--2454}.
\newblock
\APACrefnote{Publisher: American Meteorological Society Section: Bulletin of
  the American Meteorological Society}
\newblock
\begin{APACrefDOI} \doi{10.1175/BAMS-D-16-0030.1} \end{APACrefDOI}
\PrintBackRefs{\CurrentBib}

\bibitem [\protect \citeauthoryear {%
Maulik%
, San%
, Rasheed%
\BCBL {}\ \BBA {} Vedula%
}{%
Maulik%
\ \protect \BOthers {.}}{%
{\protect \APACyear {2019}}%
}]{%
maulik2019subgrid}
\APACinsertmetastar {%
maulik2019subgrid}%
\begin{APACrefauthors}%
Maulik, R.%
, San, O.%
, Rasheed, A.%
\BCBL {}\ \BBA {} Vedula, P.%
\end{APACrefauthors}%
\unskip\
\newblock
\APACrefYearMonthDay{2019}{}{}.
\newblock
{\BBOQ}\APACrefatitle {Subgrid modelling for two-dimensional turbulence using
  neural networks} {Subgrid modelling for two-dimensional turbulence using
  neural networks}.{\BBCQ}
\newblock
\APACjournalVolNumPages{Journal of Fluid Mechanics}{858}{}{122--144}.
\newblock
\begin{APACrefDOI} \doi{https://doi.org/10.1017/jfm.2018.770} \end{APACrefDOI}
\PrintBackRefs{\CurrentBib}

\bibitem [\protect \citeauthoryear {%
O'Gorman%
\ \BBA {} Dwyer%
}{%
O'Gorman%
\ \BBA {} Dwyer%
}{%
{\protect \APACyear {2018}}%
}]{%
o2018using}
\APACinsertmetastar {%
o2018using}%
\begin{APACrefauthors}%
O'Gorman, P\BPBI A.%
\BCBT {}\ \BBA {} Dwyer, J\BPBI G.%
\end{APACrefauthors}%
\unskip\
\newblock
\APACrefYearMonthDay{2018}{}{}.
\newblock
{\BBOQ}\APACrefatitle {Using machine learning to parameterize moist convection:
  Potential for modeling of climate, climate change, and extreme events} {Using
  machine learning to parameterize moist convection: Potential for modeling of
  climate, climate change, and extreme events}.{\BBCQ}
\newblock
\APACjournalVolNumPages{Journal of Advances in Modeling Earth
  Systems}{10}{10}{2548--2563}.
\newblock
\begin{APACrefDOI} \doi{https://doi.org/10.1029/2018ms001351} \end{APACrefDOI}
\PrintBackRefs{\CurrentBib}

\bibitem [\protect \citeauthoryear {%
Ott%
\ \protect \BOthers {.}}{%
Ott%
\ \protect \BOthers {.}}{%
{\protect \APACyear {2020}}%
}]{%
ott2020FKB}
\APACinsertmetastar {%
ott2020FKB}%
\begin{APACrefauthors}%
Ott, J.%
, Pritchard, M.%
, Best, N.%
, Linstead, E.%
, Curcic, M.%
\BCBL {}\ \BBA {} Baldi, P.%
\end{APACrefauthors}%
\unskip\
\newblock
\APACrefYearMonthDay{2020}{}{}.
\newblock
{\BBOQ}\APACrefatitle {A {Fortran-Keras} deep learning bridge for scientific
  computing} {A {Fortran-Keras} deep learning bridge for scientific
  computing}.{\BBCQ}
\newblock
\APACjournalVolNumPages{Scientific Programming}{2020}{}{}.
\newblock
\begin{APACrefDOI} \doi{https://doi.org/10.1155/2020/8888811} \end{APACrefDOI}
\PrintBackRefs{\CurrentBib}

\bibitem [\protect \citeauthoryear {%
Partee%
\ \protect \BOthers {.}}{%
Partee%
\ \protect \BOthers {.}}{%
{\protect \APACyear {2022}}%
}]{%
partee2022SmartSim}
\APACinsertmetastar {%
partee2022SmartSim}%
\begin{APACrefauthors}%
Partee, S.%
, Ellis, M.%
, Rigazzi, A.%
, Shao, A\BPBI E.%
, Bachman, S.%
, Marques, G.%
\BCBL {}\ \BBA {} Robbins, B.%
\end{APACrefauthors}%
\unskip\
\newblock
\APACrefYearMonthDay{2022}{}{}.
\newblock
{\BBOQ}\APACrefatitle {Using Machine Learning at scale in numerical simulations
  with {SmartSim}: An application to ocean climate modeling} {Using machine
  learning at scale in numerical simulations with {SmartSim}: An application to
  ocean climate modeling}.{\BBCQ}
\newblock
\APACjournalVolNumPages{Journal of Computational Science}{62}{}{101707}.
\newblock
\begin{APACrefDOI} \doi{https://doi.org/10.1016/j.jocs.2022.101707}
  \end{APACrefDOI}
\PrintBackRefs{\CurrentBib}

\bibitem [\protect \citeauthoryear {%
Rasp%
, Pritchard%
\BCBL {}\ \BBA {} Gentine%
}{%
Rasp%
\ \protect \BOthers {.}}{%
{\protect \APACyear {2018}}%
}]{%
rasp2018deep}
\APACinsertmetastar {%
rasp2018deep}%
\begin{APACrefauthors}%
Rasp, S.%
, Pritchard, M\BPBI S.%
\BCBL {}\ \BBA {} Gentine, P.%
\end{APACrefauthors}%
\unskip\
\newblock
\APACrefYearMonthDay{2018}{}{}.
\newblock
{\BBOQ}\APACrefatitle {Deep learning to represent subgrid processes in climate
  models} {Deep learning to represent subgrid processes in climate
  models}.{\BBCQ}
\newblock
\APACjournalVolNumPages{Proceedings of the National Academy of
  Sciences}{115}{39}{9684--9689}.
\newblock
\begin{APACrefDOI} \doi{https://doi.org/10.1073/pnas.1810286115}
  \end{APACrefDOI}
\PrintBackRefs{\CurrentBib}

\bibitem [\protect \citeauthoryear {%
Ross%
, Li%
, Perezhogin%
, Fernandez-Granda%
\BCBL {}\ \BBA {} Zanna%
}{%
Ross%
\ \protect \BOthers {.}}{%
{\protect \APACyear {{\protect \bibnodate {}}}}%
}]{%
Ross-et-al2022}
\APACinsertmetastar {%
Ross-et-al2022}%
\begin{APACrefauthors}%
Ross, A\BPBI S.%
, Li, Z.%
, Perezhogin, P.%
, Fernandez-Granda, C.%
\BCBL {}\ \BBA {} Zanna, L.%
\end{APACrefauthors}%
\unskip\
\newblock
\APACrefYearMonthDay{{\protect \bibnodate {}}}{}{}.
\newblock
{\BBOQ}\APACrefatitle {Benchmarking of machine learning ocean subgrid
  parameterizations in an idealized model} {Benchmarking of machine learning
  ocean subgrid parameterizations in an idealized model}.{\BBCQ}
\newblock
\APACjournalVolNumPages{Journal of Advances in Modeling Earth
  Systems}{n/a}{n/a}{e2022MS003258}.
\newblock
\begin{APACrefDOI} \doi{https://doi.org/10.1029/2022MS003258} \end{APACrefDOI}
\PrintBackRefs{\CurrentBib}

\bibitem [\protect \citeauthoryear {%
Scott%
\ \BBA {} Arbic%
}{%
Scott%
\ \BBA {} Arbic%
}{%
{\protect \APACyear {2007}}%
}]{%
scott2007spectral}
\APACinsertmetastar {%
scott2007spectral}%
\begin{APACrefauthors}%
Scott, R\BPBI B.%
\BCBT {}\ \BBA {} Arbic, B\BPBI K.%
\end{APACrefauthors}%
\unskip\
\newblock
\APACrefYearMonthDay{2007}{}{}.
\newblock
{\BBOQ}\APACrefatitle {Spectral energy fluxes in geostrophic turbulence:
  Implications for ocean energetics} {Spectral energy fluxes in geostrophic
  turbulence: Implications for ocean energetics}.{\BBCQ}
\newblock
\APACjournalVolNumPages{Journal of physical oceanography}{37}{3}{673--688}.
\newblock
\begin{APACrefDOI} \doi{https://doi.org/10.1175/jpo3027.1} \end{APACrefDOI}
\PrintBackRefs{\CurrentBib}

\bibitem [\protect \citeauthoryear {%
Thuburn%
, Kent%
\BCBL {}\ \BBA {} Wood%
}{%
Thuburn%
\ \protect \BOthers {.}}{%
{\protect \APACyear {2014}}%
}]{%
thuburn2014cascades}
\APACinsertmetastar {%
thuburn2014cascades}%
\begin{APACrefauthors}%
Thuburn, J.%
, Kent, J.%
\BCBL {}\ \BBA {} Wood, N.%
\end{APACrefauthors}%
\unskip\
\newblock
\APACrefYearMonthDay{2014}{}{}.
\newblock
{\BBOQ}\APACrefatitle {Cascades, backscatter and conservation in numerical
  models of two-dimensional turbulence} {Cascades, backscatter and conservation
  in numerical models of two-dimensional turbulence}.{\BBCQ}
\newblock
\APACjournalVolNumPages{Quarterly Journal of the Royal Meteorological
  Society}{140}{679}{626--638}.
\newblock
\begin{APACrefDOI} \doi{https://doi.org/10.1002/qj.2166} \end{APACrefDOI}
\PrintBackRefs{\CurrentBib}

\bibitem [\protect \citeauthoryear {%
Yuval%
, O'Gorman%
\BCBL {}\ \BBA {} Hill%
}{%
Yuval%
\ \protect \BOthers {.}}{%
{\protect \APACyear {2021}}%
}]{%
yuval2021GRL}
\APACinsertmetastar {%
yuval2021GRL}%
\begin{APACrefauthors}%
Yuval, J.%
, O'Gorman, P\BPBI A.%
\BCBL {}\ \BBA {} Hill, C\BPBI N.%
\end{APACrefauthors}%
\unskip\
\newblock
\APACrefYearMonthDay{2021}{}{}.
\newblock
{\BBOQ}\APACrefatitle {Use of neural networks for stable, accurate and
  physically consistent parameterization of subgrid atmospheric processes with
  good performance at reduced precision} {Use of neural networks for stable,
  accurate and physically consistent parameterization of subgrid atmospheric
  processes with good performance at reduced precision}.{\BBCQ}
\newblock
\APACjournalVolNumPages{Geophysical Research Letters}{48}{6}{e2020GL091363}.
\newblock
\begin{APACrefDOI} \doi{https://doi.org/10.1029/2020GL091363} \end{APACrefDOI}
\PrintBackRefs{\CurrentBib}

\bibitem [\protect \citeauthoryear {%
Zanna%
\ \BBA {} Bolton%
}{%
Zanna%
\ \BBA {} Bolton%
}{%
{\protect \APACyear {2020}}%
}]{%
zanna2020data}
\APACinsertmetastar {%
zanna2020data}%
\begin{APACrefauthors}%
Zanna, L.%
\BCBT {}\ \BBA {} Bolton, T.%
\end{APACrefauthors}%
\unskip\
\newblock
\APACrefYearMonthDay{2020}{}{}.
\newblock
{\BBOQ}\APACrefatitle {Data-driven equation discovery of ocean mesoscale
  closures} {Data-driven equation discovery of ocean mesoscale
  closures}.{\BBCQ}
\newblock
\APACjournalVolNumPages{Geophysical Research Letters}{47}{17}{e2020GL088376}.
\newblock
\begin{APACrefDOI} \doi{https://doi.org/10.1002/essoar.10503535.1}
  \end{APACrefDOI}
\PrintBackRefs{\CurrentBib}

\bibitem [\protect \citeauthoryear {%
Zanna%
\ \protect \BOthers {.}}{%
Zanna%
\ \protect \BOthers {.}}{%
{\protect \APACyear {2018}}%
}]{%
zanna_uncertainty_2018}
\APACinsertmetastar {%
zanna_uncertainty_2018}%
\begin{APACrefauthors}%
Zanna, L.%
, Brankart, J\BPBI M.%
, Huber, M.%
, Leroux, S.%
, Penduff, T.%
\BCBL {}\ \BBA {} Williams, P\BPBI D.%
\end{APACrefauthors}%
\unskip\
\newblock
\APACrefYearMonthDay{2018}{{\APACmonth{10}}}{}.
\newblock
{\BBOQ}\APACrefatitle {Uncertainty and scale interactions in ocean ensembles:
  {From} seasonal forecasts to multidecadal climate predictions} {Uncertainty
  and scale interactions in ocean ensembles: {From} seasonal forecasts to
  multidecadal climate predictions}.{\BBCQ}
\newblock
\APACjournalVolNumPages{Quarterly Journal of the Royal Meteorological
  Society}{0}{0}{}.
\newblock
\begin{APACrefDOI} \doi{10.1002/qj.3397} \end{APACrefDOI}
\PrintBackRefs{\CurrentBib}

\bibitem [\protect \citeauthoryear {%
Zanna%
, Mana%
, Anstey%
, David%
\BCBL {}\ \BBA {} Bolton%
}{%
Zanna%
\ \protect \BOthers {.}}{%
{\protect \APACyear {2017}}%
}]{%
zanna2017scale}
\APACinsertmetastar {%
zanna2017scale}%
\begin{APACrefauthors}%
Zanna, L.%
, Mana, P\BPBI P.%
, Anstey, J.%
, David, T.%
\BCBL {}\ \BBA {} Bolton, T.%
\end{APACrefauthors}%
\unskip\
\newblock
\APACrefYearMonthDay{2017}{}{}.
\newblock
{\BBOQ}\APACrefatitle {Scale-aware deterministic and stochastic
  parametrizations of eddy-mean flow interaction} {Scale-aware deterministic
  and stochastic parametrizations of eddy-mean flow interaction}.{\BBCQ}
\newblock
\APACjournalVolNumPages{Ocean Modelling}{111}{}{66--80}.
\newblock
\begin{APACrefDOI} \doi{https://doi.org/10.1016/j.ocemod.2017.01.004}
  \end{APACrefDOI}
\PrintBackRefs{\CurrentBib}

\bibitem [\protect \citeauthoryear {%
Zhang%
}{%
Zhang%
}{%
{\protect \APACyear {2023}}%
{\protect \APACexlab {{\protect \BCnt {1}}}}}]{%
doublegyre_setup}
\APACinsertmetastar {%
doublegyre_setup}%
\begin{APACrefauthors}%
Zhang, C.%
\end{APACrefauthors}%
\unskip\
\newblock
\APACrefYearMonthDay{2023{\protect \BCnt {1}}}{{\APACmonth{02}}}{}.
\newblock
\APACrefbtitle {chzhangudel/Double\_Gyre: Initial Release (v1.0.1)[Software].}
  {chzhangudel/double\_gyre: Initial release (v1.0.1)[software].}
\newblock
\APACaddressPublisher{}{Zenodo}.
\newblock
\begin{APACrefDOI} \doi{10.5281/zenodo.7663128} \end{APACrefDOI}
\PrintBackRefs{\CurrentBib}

\bibitem [\protect \citeauthoryear {%
Zhang%
}{%
Zhang%
}{%
{\protect \APACyear {2023}}%
{\protect \APACexlab {{\protect \BCnt {2}}}}}]{%
ForpyCNNGZ21}
\APACinsertmetastar {%
ForpyCNNGZ21}%
\begin{APACrefauthors}%
Zhang, C.%
\end{APACrefauthors}%
\unskip\
\newblock
\APACrefYearMonthDay{2023{\protect \BCnt {2}}}{{\APACmonth{02}}}{}.
\newblock
\APACrefbtitle {chzhangudel/Forpy\_CNN\_GZ21: Initial Release
  (v1.0.0)[Software].} {chzhangudel/forpy\_cnn\_gz21: Initial release
  (v1.0.0)[software].}
\newblock
\APACaddressPublisher{}{Zenodo}.
\newblock
\begin{APACrefDOI} \doi{10.5281/zenodo.7663062} \end{APACrefDOI}
\PrintBackRefs{\CurrentBib}

\end{thebibliography}

\end{document}


\title{Supporting Information for ''Implementation and Evaluation of a Machine Learned Mesoscale Eddy Parameterization into a Numerical Ocean Circulation Model''}

DOI: xxxxxx

\authors{Cheng Zhang\affil{1}, Pavel Perezhogin\affil{2}, Cem Gultekin\affil{2}, Alistair Adcroft\affil{1}, Carlos Fernandez-Granda\affil{2,3}, Laure Zanna\affil{2,3}}

\affiliation{1}{Program in Atmospheric and Oceanic Sciences, Princeton University, Princeton, NJ 08542, USA}
\affiliation{2}{Courant Institute of Mathematical Sciences, New York University, New York, NY 10012, USA}
\affiliation{3}{Center for Data Science, New York University, New York, NY 10011, USA}


\begin{article}
This supplement provides figures to show the dynamic sensitivity of the machined-learned parameterization used in the main text. It also provides figures to demonstrate how similar the ensemble members perform in the tests.

\noindent\textbf{Contents of this file}
\begin{enumerate}
\item Figures S1 to S6
\end{enumerate}
\noindent\textbf{Dynamic sensitivity of CNN parameterization}

The stochastic Convolutional Neural Network (CNN) used in this study returns the mean and standard deviation of a Gaussian probability distribution of the subgrid momentum forcing. The subgrid momentum forcing used for parameterization can be written as
\be
S_{C,i,j}=S_{C,i,j}^{(mean)}+\epsilon_{C,i,j} \cdot S_{C,i,j}^{(std)}; \hspace{0.2in}  C = x,y;\hspace{0.1in} i=1,...,m;\hspace{0.1in} j=1,...,n 
\label{eqS1_1}
\ee
where $i$ and $j$ are the ocean model spatial indices, $C$ indicates the component of momentum forcing (zonal "$x$" or meridional "$y$"), and $\epsilon_{C,i,j}$ are random 2D fields sampled from the standard normal distribution, independent for each grid cell, zonal/meridional component, vertical layer, and time step. Both the mean component $S_{C,i,j}^{(mean)}$ and standard deviation component $S_{C,i,j}^{(std)}$ are dynamic because they are functions of the flow. To examine if the dynamic behavior of the parameterization is important, we alternatively parameterize the subgrid forcing using time-invariant values of $S_{C,i,j}^{(mean)}$ and $S_{C,i,j}^{(std)}$. The subgrid forcing for this test is written as
\be
S_{C,i,j}=\overline{S_{C,i,j}^{(mean)}}+\epsilon_{C,i,j} \cdot \overline{S_{C,i,j}^{(std)}}; \hspace{0.2in}  C = x,y;\hspace{0.1in} i=1,...,m;\hspace{0.1in} j=1,...,n 
\label{eqS1_2}
\ee
where the over-bar denotes the time average over the last five years. 

\begin{figure}[htbp]
\centering
\includegraphics[width=1.0\textwidth]{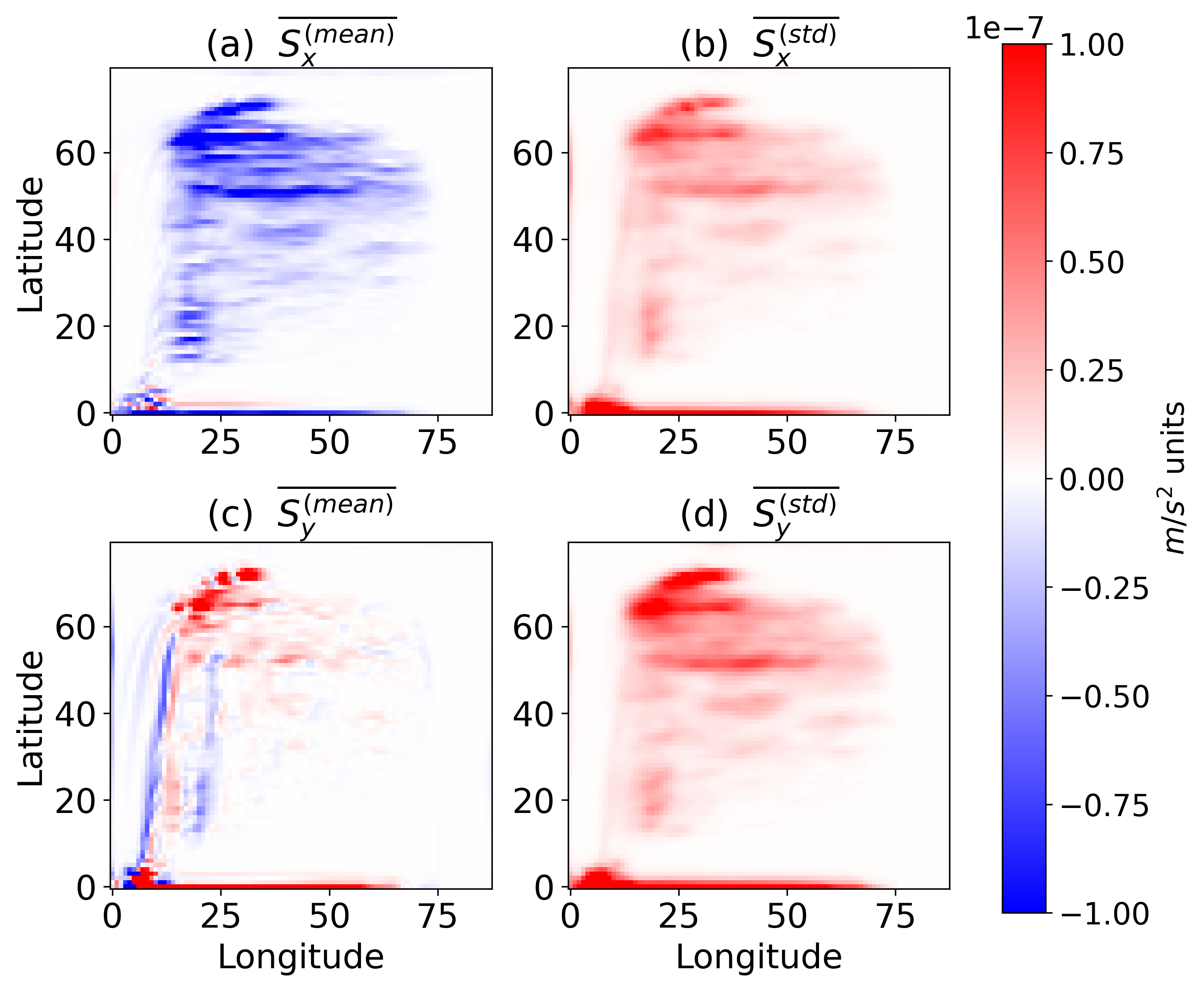}
\caption{Time-invariant spatially-varying map of two components (mean and standard deviation components) of the subgrid momentum forcing averaging over the last five years from a dynamic parameterization test. The components correspond to the variables in Equation \eqref{eqS1_2}.}
\label{figS1_1}
\end{figure}

\begin{figure}[htbp]
\centering
\includegraphics[width=1.0\textwidth]{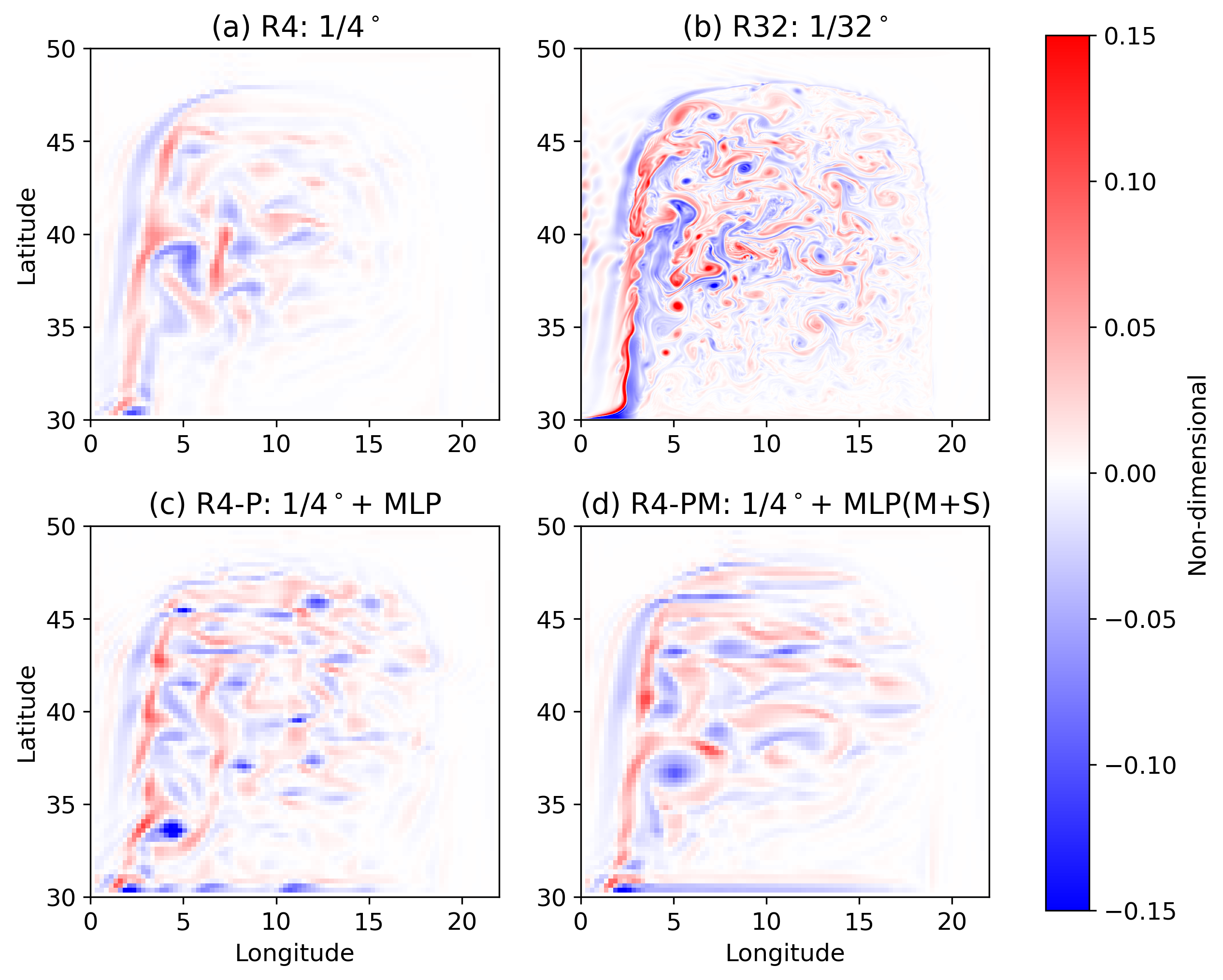}
\caption{Snapshots of relative vorticity (normalized by the planetary vorticity) at the upper layer flow without subgrid parameterizations (a and b), or with subgrid parameterizations (c and d). The grid sizes of the simulations are $1/4 ^\circ$ (R4, a, c and d) and $1/32 ^\circ$ (R32, b).  MLP is short for dynamic parameterization and MLP(M+S) is short for parameterization with time-invariant mean and standard deviation components.}
\label{figS1_2}
\end{figure}

\begin{figure}[htbp]
\centering
\includegraphics[width=0.495\textwidth]{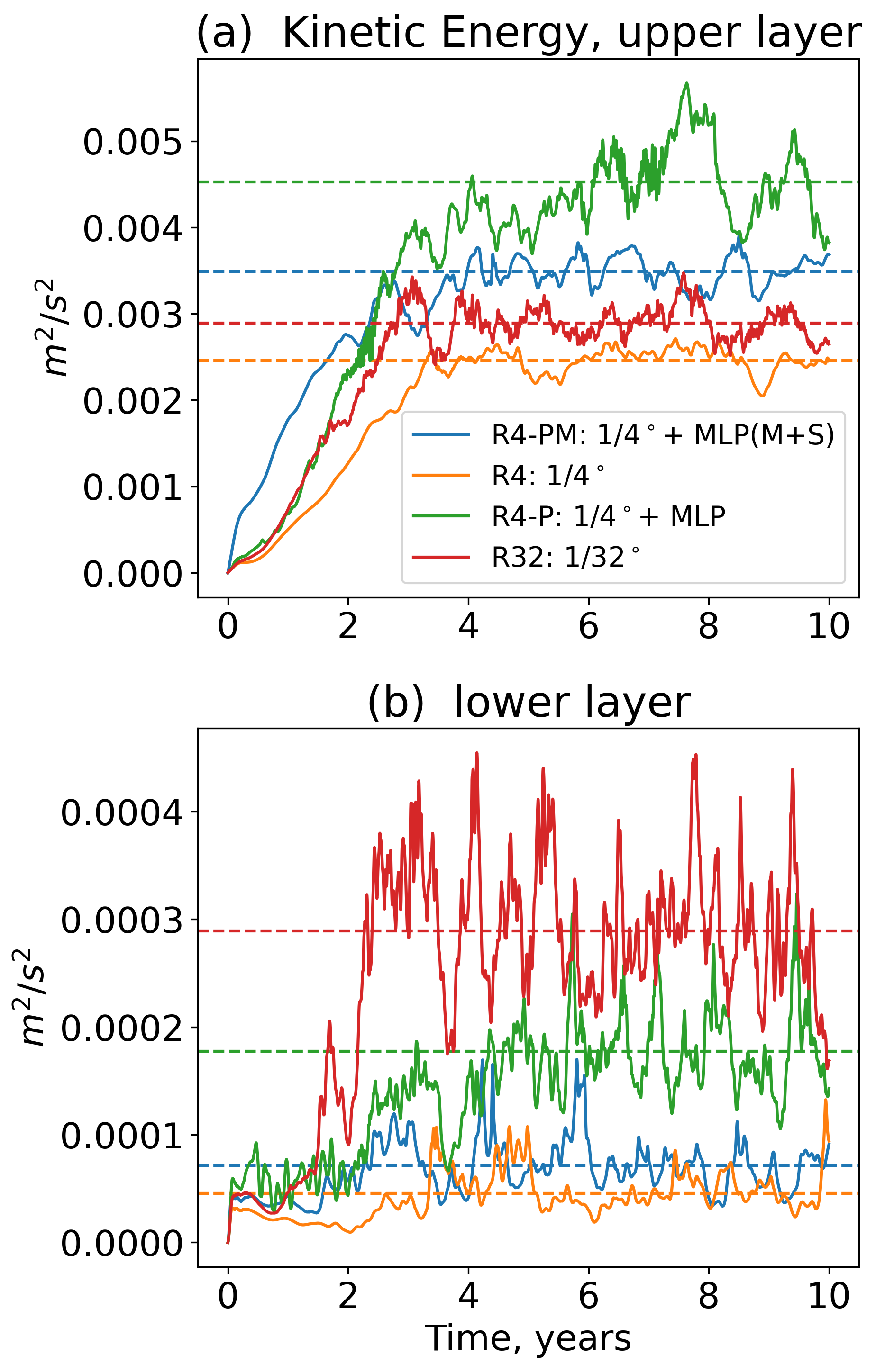}
\includegraphics[width=0.495\textwidth]{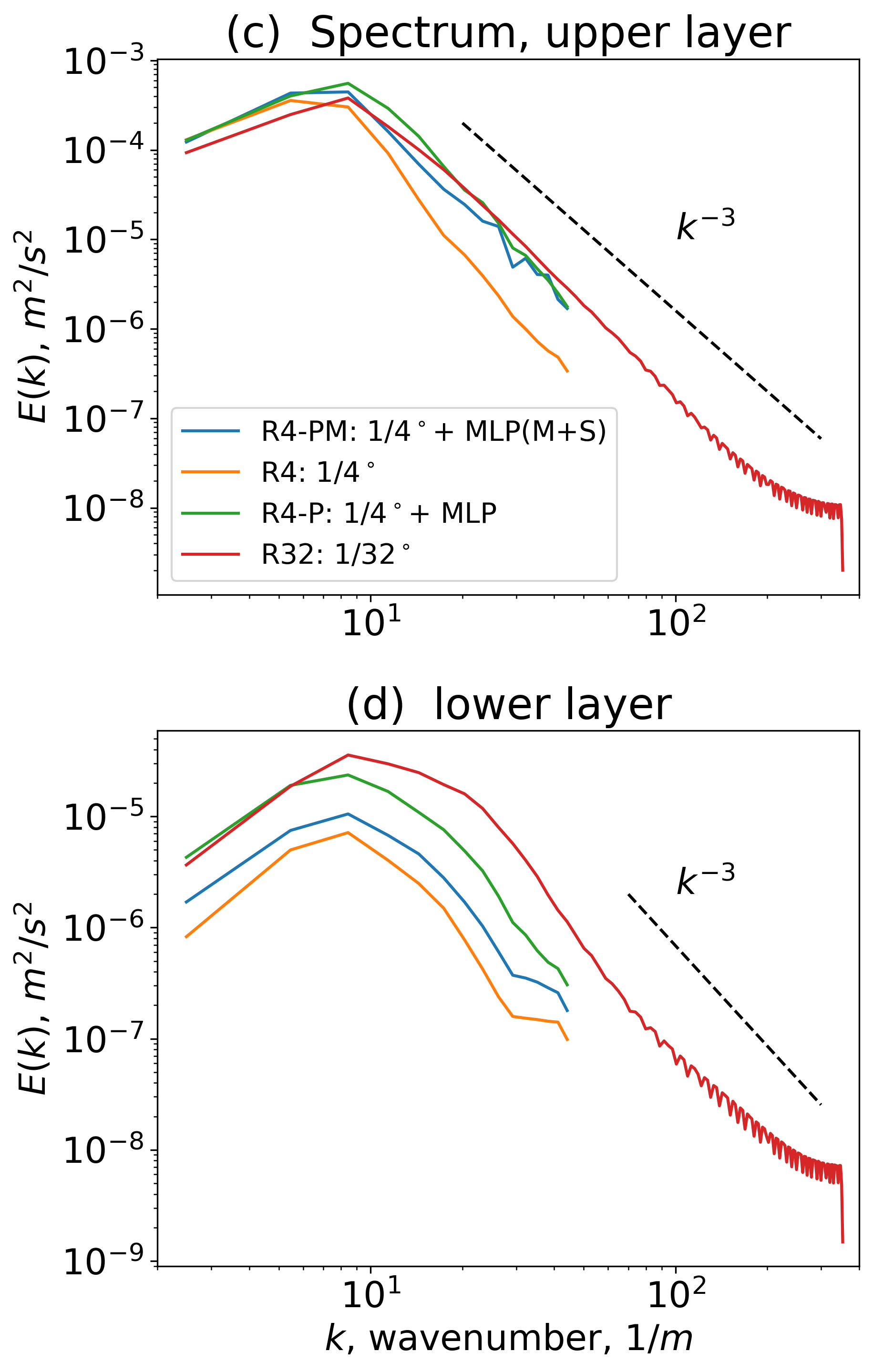}
\caption{Comparison of KE time series (a and b) and spectra (c and d) for the flow upper layer and the lower layer between the coarse resolution model R4(orange), fine resolution R32 (red) and the coarse resolution model with dynamic parameterizations R4-P (green) or time-invariant mean and standard deviation components R4-PM (blue). The dashed lines in (a and b) are mean values of KE over the last 5 years. The dashed lines in (c and d) are the spectral slope of kinetic energy spectrum corresponding to inertial interval of enstrophy. }
\label{figS1_3}
\end{figure}

Figure \ref{figS1_1} shows the time-invariant spatially-varying map of two components (mean and standard deviation components) of the subgrid momentum forcing averaging over the last five years from a dynamic parameterization test.  We find that all four components from the CNN model are of the same order. The strongest forcing is near the flow separation and the south wall boundary. We then use these static fields in a case using equation \eqref{eqS1_2} as the subgrid forcing formula (R4-PM), where the components are now time-invariant. From the snapshots of the upper layer relative vorticity in Figure \ref{figS1_2}(d), we see that the static parameterization does inject energy into the flow, but we do not see the small scale eddies that are present in the dynamic parameterization case (R4-P, Figure \ref{figS1_2}(c)). This is also illustrated by the kinetic energy (KE) spectrum in Figure \ref{figS1_3}(c), where the case R4-PM has a lower energy density at the small scale than the case R4-P does. In addition, the time series shows that the KE from the static parameterization is less than the KE from the dynamic parameterization, for both upper and lower layers.
Using the static parameterization has no clear improvement in time-mean sea surface height (SSH, Figure \ref{figS1_3}). The RMSE between R4-PM SSH and R32 SSH is increased to $0.3445$m from $0.2780$ for R4 SSH (without parameterization), while the RMSE between R4-P SSH and R32 SSH is $0.2202$.

Therefore, the dynamic response of the machine-learned parameterization is important in improving the solutions of the main text.

\noindent\textbf{Similarity of ensemble members}

To stochastic nature of the parameterization allows us to run an ensemble with multiple members with different random seeds.
The randomness is from the random 2D fields $\epsilon_{C,i,j}$ in equation \eqref{eqS1_1}.
The main text sometimes illustrates results from only one of the ensemble members. 
Here we show the similarity of the statistics among the ensemble members.
Figure \ref{figS2_1} shows snapshots of the relative vorticity (a-c) and KE (d-f), and five-year averaged SSH (g-i), of the first three ensemble runs with ML parameterizations. Although the flow may vary somewhat across snapshots, the patterns in the plots have the same scale. We compare KE time series and spectra of the flow between five different ensemble runs in Figure \ref{figS2_2}. The times series and spectra curves from the different members exhibit good agreement.


\end{article}





\begin{figure}[htbp]
\centering
\includegraphics[width=1.0\textwidth]{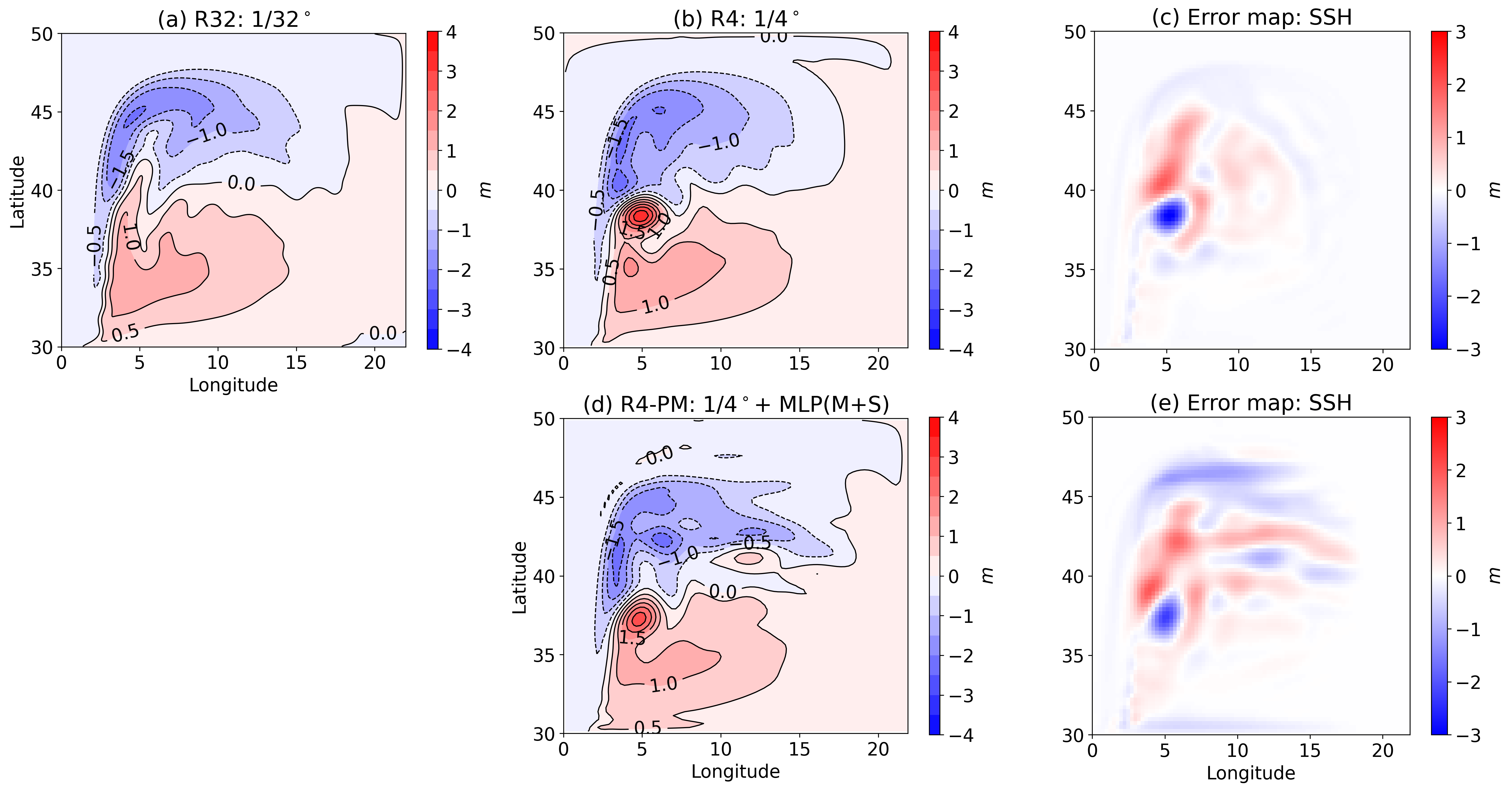}
\caption{Comparison of five-year averaged SSH between the coarse resolution model with (R4-PM) and without (R4) the static parameterization and the target fine resolution model (R32). The error maps (c and d) are obtained by subtracting low-resolution SSH (with or without parameterization) from coarse-grained high-resolution SSH. 
The R4-P SSH (d) is averaged from $50$ ensemble members.}
\label{figS1_4}
\end{figure}

\begin{figure}[htbp]
\centering
\includegraphics[width=1.0\textwidth]{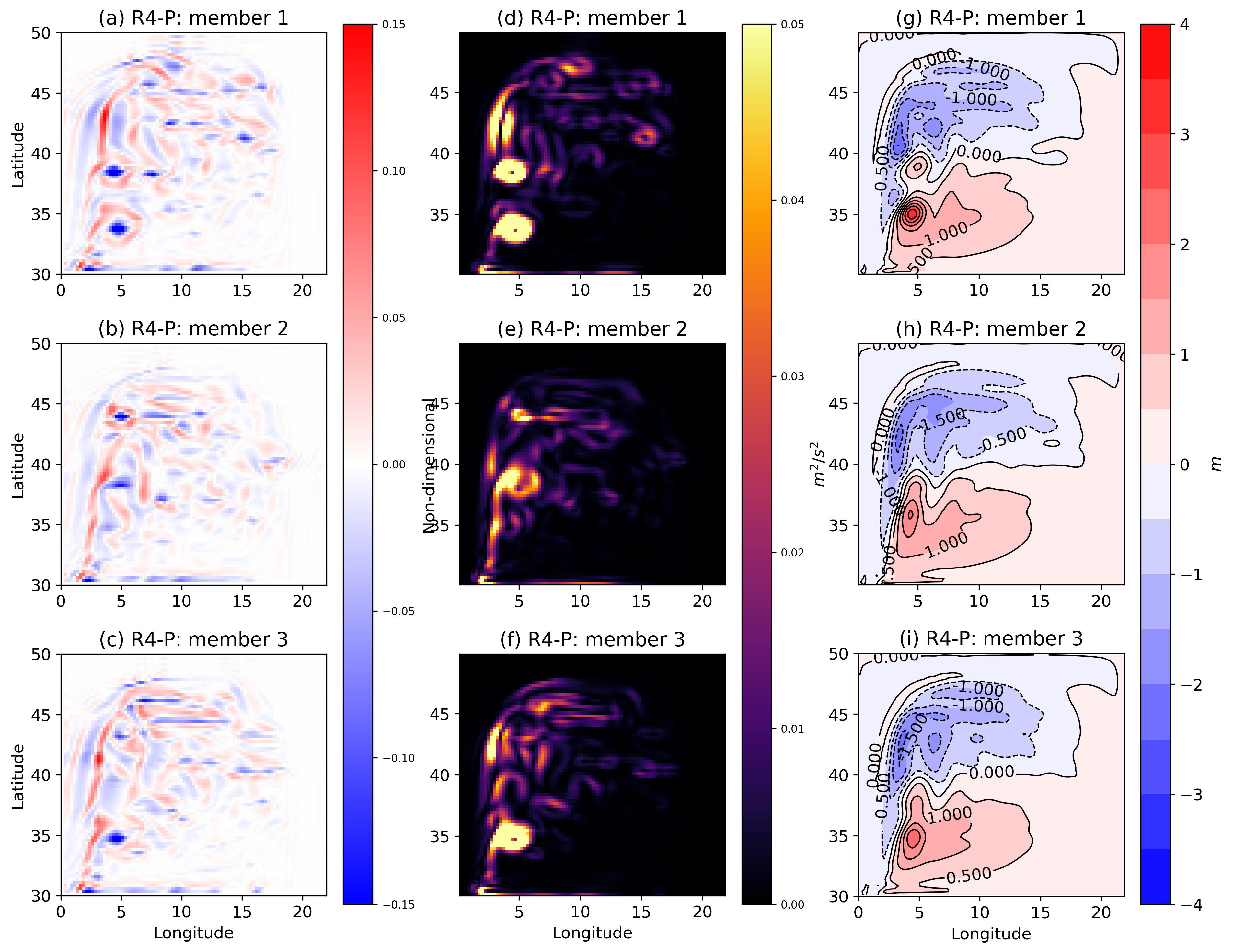}
\caption{Snapshots of the upper layer relative vorticity (normalized by the planetary vorticity, a-c) and KE in [m$^2$/s$^2$] (d-f), and five-year averaged SSH in [m] (g-i), of the first three ensemble runs for the coarse resolution model with ML parameterizations (R4-P).}
\label{figS2_1}
\end{figure}

\begin{figure}
\centering
\includegraphics[width=0.495\textwidth]{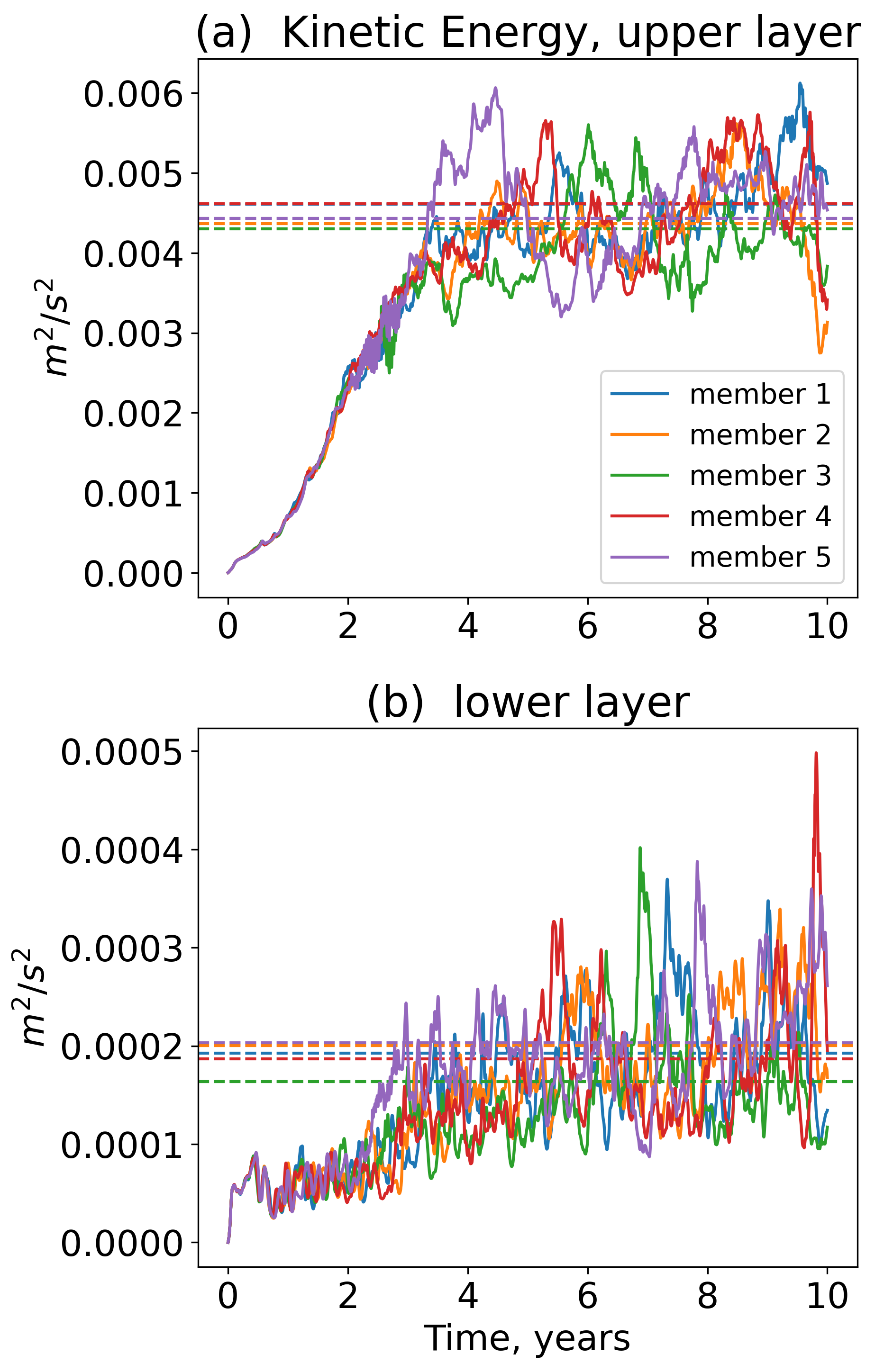}
\includegraphics[width=0.495\textwidth]{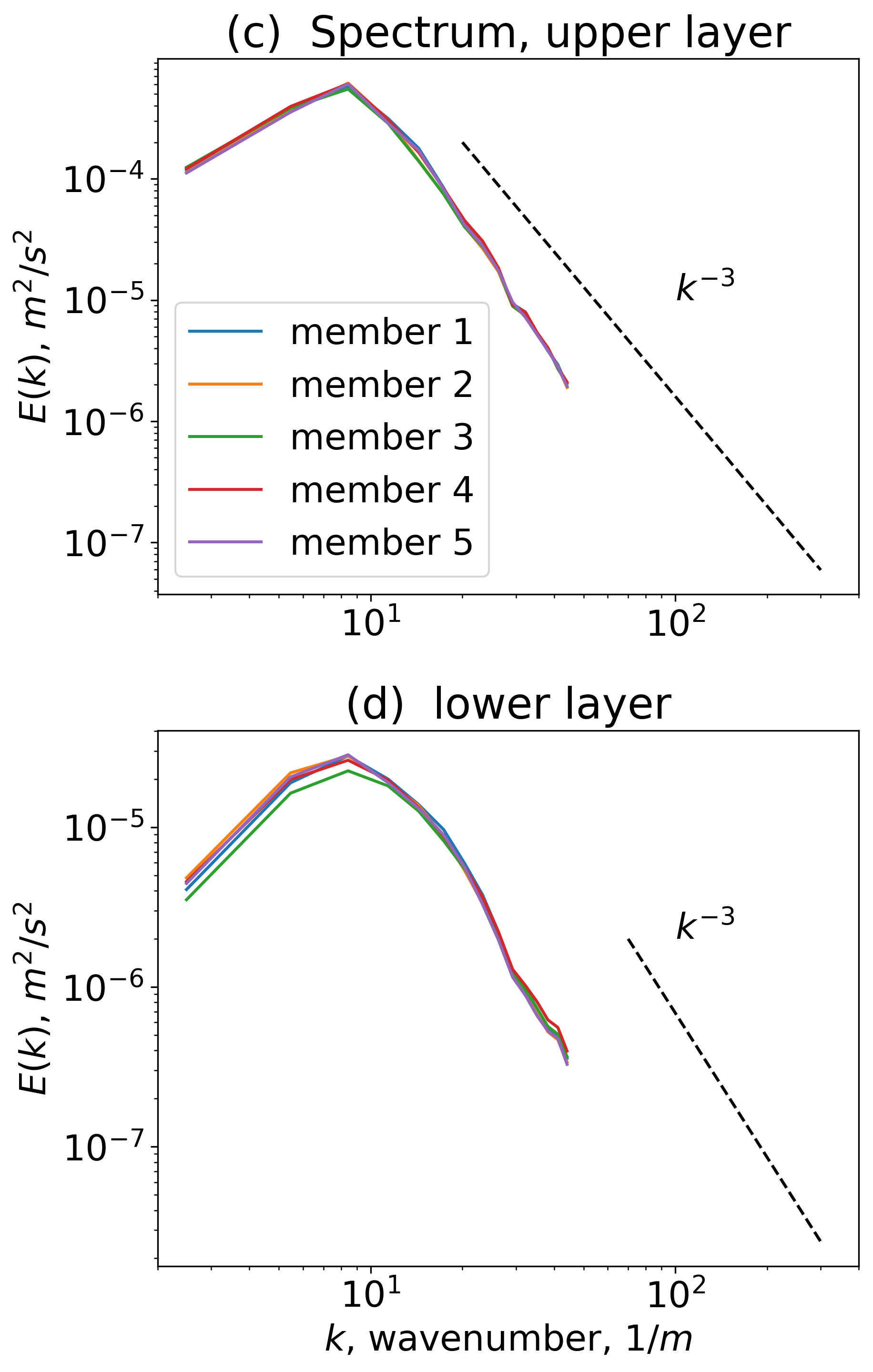}
\caption{Comparison of KE time series (a and b) and spectra (c and d) for the flow upper layer and the lower layer from five different ensemble runs. The dashed lines in (a and b) are mean values of KE over the last 5 years. The dashed lines in (c and d) are the spectral slope of kinetic energy spectrum corresponding to inertial interval of enstrophy. }
\label{figS2_2}
\end{figure}